\documentclass[8 pt]{article}

\usepackage[utf8x]{inputenc}
\usepackage{dsfont}
\usepackage{amsthm}
\usepackage{amsfonts}
\usepackage{amssymb}
\usepackage{mathtools}
\usepackage[T1]{fontenc}
\usepackage[cm]{fullpage}
\usepackage{graphicx}
\usepackage{bm}
\usepackage{setspace}
\usepackage{enumitem}
\usepackage{mdwlist}
\usepackage{parskip}
\usepackage{listings}
\usepackage{color}
\usepackage{tikz,datatool}
\usepackage{hyperref}
\usepackage{mathabx}
\usepackage{multicol}
\usepackage{caption}
\usepackage{amsthm}

\AtBeginDocument{
	\let\myThePage\thepage
	\renewcommand{\thepage}{\oldstylenums{\myThePage}}
}

\newcommand{\ket}[1]{\left| #1 \right\rangle}
\newcommand{\bra}[1]{\left\langle #1 \right|}
\newcommand{\bket}[2]{\left\langle #1 \middle| #2 \right\rangle}
\newcommand{\obket}[3]{\left\langle #1 \middle| #2 \middle| #3\right\rangle}
\newcommand{\ldefeq}{\mathrel{\rlap{%
			\raisebox{0.3ex}{$\m@th\cdot$}}%
		\raisebox{-0.3ex}{$\m@th\cdot$}}%
	=}
\makeatletter
\newcommand*{\rdefeq}{\mathrel{\rlap{%
			= 
			\raisebox{0.3ex}{$\m@th\cdot$}}%
		\raisebox{-0.3ex}{$\m@th\cdot$}}%
}
\makeatother

\usepackage[top=3cm,bottom=3cm,left=3cm,right=3cm]{geometry}

	\renewcommand{\baselinestretch}{1.3} 

\begin{document}
	
	\pagenumbering{roman}
	
	\begin{titlepage} 
		
		\centering 
		
		\scshape 
		
		\vspace*{\baselineskip} 
		
		
		\rule{\textwidth}{1.6pt}\vspace*{-\baselineskip}\vspace*{2pt} 
		\rule{\textwidth}{0.4pt} 
		
		\vspace{0.75\baselineskip}
		
		{\LARGE Black Hole Information \\ and \\  Thermodynamics \\} 
		
		\vspace{0.75\baselineskip} 
		
		\rule{\textwidth}{0.4pt}\vspace*{-\baselineskip}\vspace{3.2pt} 
		\rule{\textwidth}{1.6pt} 
		
		\vspace{1\baselineskip}

		
		{\scshape\Large Dieter L\"ust \\ Ward Vleeshouwers \\} 
		
		\vspace{1.5\baselineskip}
		
		\textit{Ludwig-Maximilians-Universität \\ München}

		\begin{abstract}
This SpringerBrief is based on a masters course on black hole thermodynamics and the black hole information problem taught by Dieter Lüst during the summer term 2017 at the Ludwig-Maximilians-Universität in Munich; it was written by Ward Vleeshouwers. It provides a short introduction to general relativity, which describes gravity in terms of the curvature of space-time, and examines the properties of black holes. These are central objects in general relativity which arise when sufficient energy is compressed into a finite volume, so that even light cannot escape its gravitational pull. We will see that black holes exhibit a profound connection with thermodynamic systems. Indeed, by quantizing a field theory on curved backgrounds, one can show that black holes emit thermal (Hawking) radiation, so that the connection with thermodynamics is more than a formal similarity. Hawking radiation gives rise to an apparent conflict between general relativity and quantum mechanics known as the black hole information problem. If a black hole formed from a pure quantum state evaporates to form thermal radiation, which is in a mixed state, then the unitarity postulate of quantum mechanics is violated. We will examine the black hole information problem, which has plagued the physics community for over four decades, and consider prominent examples of proposed solutions, in particular, the string theoretical construction of the Tangherlini black hole, and the infinite number of asymptotic symmetries given by BMS-transformations.
		\end{abstract}
		
	\end{titlepage}

		\clearpage
	
	\setcounter{tocdepth}{1}
	
		\renewcommand{\baselinestretch}{.9} 
	
	\tableofcontents

	\clearpage
	
	\renewcommand{\baselinestretch}{1.1} 
	
	\pagenumbering{arabic}

	\section{Special relativity}
	
	In non-relativistic settings, the symmetry group of space-time is the \textit{Galilean group}, which consists of rotations and translations. These transformations leave spatial distances (as well as temporal intervals) invariant. For example, in two spatial dimensions with coordinates $(x,y)$, the squared distance $s^2 = (\Delta x)^2 + (\Delta y)^2 $ is invariant under rotations, which are of the form
	
	$$ \begin{pmatrix}x \\ y \end{pmatrix} \mapsto \begin{pmatrix}x' \\ y' \end{pmatrix} = \begin{pmatrix} \cos \alpha & \sin \alpha \\  -\sin \alpha & \cos \alpha \end{pmatrix}\begin{pmatrix}x \\ y \end{pmatrix} ~.
	 $$
	
	The invariance of spatial (Euclidean) distance is then given by $s'^2 = (\Delta x')^2 + (\Delta y')^2 = s^2$. 
	
	In the context of special relativity, our notion of invariant distance changes. Namely, if we consider a $(3+1)$-dimensional space i.e. three spatial and one temporal dimension parametrized by $(t,x,y,z)$, we define the \textit{Minkowski distance}
	
	\begin{equation}
s^2 = - (c \Delta t)^2 + (\Delta x)^2 + (\Delta y)^2 + (\Delta z)^2~,
	\end{equation}
	
	where $c$ is the speed of light. The group that leaves this distance invariant is called the \textit{Poincar\'{e} group}, which extends the Galilean group to include boosts. Distances in special relativity are no longer positive semi-definite, namely, we distinguish
	
	\begin{enumerate}
		\item $s^2 > 0 ~~~ (\text{III})$           Space-like distance ,
		\item $s^2 = 0 ~~~ (II)$           Light-like distance ,
		\item $s^2 < 0 ~~~ (I)$           Time-like distance .
	\end{enumerate}

	\begin{figure}[h]
	\begin{center}
		\includegraphics[width=8cm]{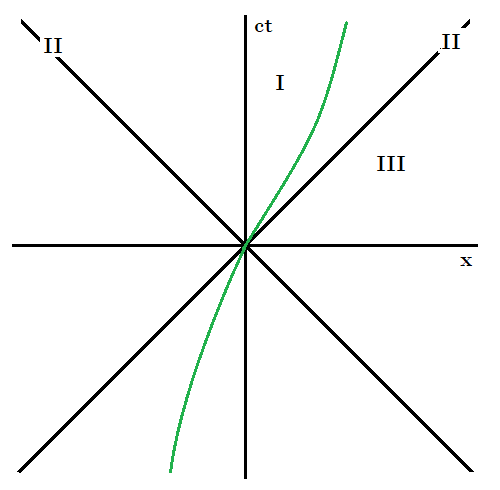}
		\caption{An example of a space-time diagram. The region indicated by Roman numeral $I$ $(III)$ consists of points at time-like (space-like) distance from the origin at $x^{\mu} = 0$, that is, points with $s^2 < 0$ $(s^2 > 0)$. Region $II$ is the light-cone of the origin, which separates time-like and space-like regions. The green line indicates an example of a time-like trajectory with $x(t=0) = 0 $. \label{stdiag}}
	\end{center}
\end{figure}
	
	If we consider a line segment parametrized in $(t,x)$, the Roman numerals above correspond to the regions in the space-time given in figure \ref{stdiag}. For intervals with $s^2<0$, the (squared) \textit{proper time} $\tau $ is given by
	
	\begin{equation}
	\tau^2 \coloneqq - \frac{s^2 }{c^2} ~.
	\end{equation}
	
	We will use \textit{four-vector notation} $x^{ \mu} = (x^0 , x^1 , x^2 , x^3 ) = (ct, x, y ,z )$ and denote spatial coordinates as $x^i  = (x^1,y^2,z^3)$. The \textit{Minkowski metric} is then given by
	
	\begin{equation}
	\eta_{\mu \nu} = \begin{pmatrix}
	-1 & 0&0& 0\\
	0&1&0&0\\
	0&0&1&0\\
	0&0&0&1
	\end{pmatrix}~.
	\end{equation} 
	
	The Minkowski distance can then be conveniently be written as
	
	\begin{equation}
	s^2 = \eta_{\mu \nu} \Delta x^{\mu} \Delta x^{\nu}~.
	\end{equation}

	We also employ \textit{Einstein summation convention}
	
	$$ x_{\mu}x^{\mu} \coloneqq  \sum_{\mu}  x_{\mu}x^{\mu} ~. $$

	I.e. we omit the summation symbol for repeated indices. The symmetry transformations that make up the Poincar\'{e} group are then written as
	
	\begin{enumerate*}
		\item $x^{\mu} \mapsto \tilde{x}^{\mu} = x^{\mu} + a^{\mu}$   (translations) \\
		\item $x^{\mu} \mapsto \tilde{x}^{\mu} = \Lambda^{\mu}_{\nu} x^{\nu}$   (boosts \& rotations)
	\end{enumerate*}
	
	From the invariance of Minkowskian distance under boosts and rotations, we find

	\begin{align}
	s^2 & = (\Delta x)^T \eta (\Delta x) \notag \\
		& = (\Delta \tilde{x})^T \eta (\Delta \tilde{x}) \notag \\
		& = (\Delta {x})^T \Lambda^T \eta  \Lambda (\Delta {x}) ~~~ , ~~  \eta' \coloneqq \Lambda^T \eta  \Lambda ~.
		\end{align}
	
The condition above is written in tensor notation as

\begin{equation}
\eta_{\rho \sigma } = \Lambda^{\mu}_{\rho} \Lambda^{\nu}_{\sigma} \eta_{\mu \nu}~.
\end{equation}

	We thus conclude that $\Lambda \in O(3,1)$, the $(3+1)$-dimensional orthogonal group. The Poincar\'{e} group is the semi-direct product of the \textit{Lorentz group} and $(3+1)$-dimensional translations. Elements of the Lorentz group include

	\begin{enumerate}
		\item Rotations (in $x,y$-plane): $\Lambda^{\mu}_{\nu} = \begin{pmatrix} 1 & 0 & 0 & 0 \\ 0 & \cos \theta & \sin \theta & 0 \\  0  & - \sin \theta & \cos \theta & 0 \\ 0 & 0 & 0 & 1 \end{pmatrix}  ~~~, ~~  0 \leq \theta < 2 \pi  ~,$
	\item Boosts (in $x$-direction): $\Lambda^{\mu}_{\nu} = \begin{pmatrix} \cosh \phi & -\sinh \phi & 0 & 0\\ - \sinh \phi & \cosh \phi & 0 & 0 \\ 0 & 0 & 1 & 0 \\ 0 & 0 & 0 & 1 \end{pmatrix} ~~~, ~~ -\infty < \phi < \infty ~.$
		\end{enumerate}

	\vspace{.3cm}
	
	One can easily check that these matrices satisfy $ \eta = \Lambda^T \eta  \Lambda$, hence they leave the Minkowskian distance invariant. We see that a boost transforms our coordinates as
	
	\begin{align}
	\tilde{t} & = t \cosh \phi - x \sinh \phi \notag \\ 
	\tilde{x} & = x \cosh \phi - t \sinh \phi ~.
	\end{align}
	
	For paths with $\tilde{x} = 0$, the speed in our original reference frame is
	
	\[ v = \frac{x}{t} = \frac{ \sinh \phi}{ \cosh \phi} = \tanh \phi ~~ \Rightarrow ~~ \phi = \tanh^{-1} v~, \]
	
	so that we find
	
	\begin{align}
	\tilde{t} & = \gamma ( t - v x) \notag \\ 
	\tilde{x} & = \gamma (x - vt ) ~~~, ~~  \gamma \coloneqq \frac{1}{\sqrt{1 - v^2/c^2}} ~.
	\end{align}
	
	We see that $s^2 = 0 $ for a tajectory with $v=c$. From now on, we set $c=1$, so that light-like trajectories are given by $x = \pm t$. Note that the light-cone is mapped to itself under Lorentz transformations. Time-like trajectories have $v<c$, while space-like trajectories have $v>c$. Hypothetical particles which have $v>c$ are called \textit{tachyons}, which are generally considered not to exist as propagating degrees of freedom.
	
	\clearpage
	
	\section{Riemannian geometry}

	We consider (d+1)-dimensional \textit{smooth manifolds} $\mathcal{M}$, which are topological manifold that look locally like $\mathbb{R}^n$. $\mathcal{M}$ can be covered by open sets $U_i$, $i \in I$, where $I$ is some indexing set. The \textit{charts} are then defined as bijective maps $\phi : U_i \rightarrow \mathbb{R}^{1,d}$ with the requirement that, for $U_i \bigcap U_j \neq 0$, the \textit{transition function} $\phi_i \circ \phi_j^{-1}$ is $C^{\infty}$. The collection of all $U_i$ is then called an \textit{atlas}. 
	
	At each point $p \in \mathcal{M}$, we can define a \textit{tangent space} $T_p \mathcal{M}$, which consists of all the tangent vectors at point $p$. The basis of $T_p \mathcal{M}$ is written as $\{ \hat{e}_{\left(\mu\right)} \}$, so that any vector can be written as $V= V^{\mu}\left(x^{\nu}\right) \hat{e}_{\left(\mu\right)}$. As we saw in  the previous lecture, Lorentz transformation acts as $x^{\mu} = \Lambda_{\nu}^{\mu}x^{\nu}$. Since $\partial_{\mu'} = \frac{\partial x^{\mu}}{\partial x'^{\mu}} \partial_{\mu}$ and $V^{\mu}\partial_{\mu} = V^{\mu'}\partial_{\mu'}$, it follows that $V^{\mu'} = \frac{\partial x^{\mu'}}{\partial x^{\mu}} V^{\mu}$. We thus see that that $\hat{e}_{\left( \mu \right)} = \partial_{\mu}$, i.e. the tangent space is spanned by partial derivatives defined at $p$. The dual space to the tangent space is called the \textit{cotangent space}, which is denoted by $T^*_p\mathcal{M}$. It is dual to $T_p \mathcal{M}$ in the sense that its basis covectors $\hat{\theta}^{\left(\mu\right)}$ satisfy $\hat{\theta}^{\left(\mu\right)}\hat{e}_{\left(\nu\right)}=\delta_{\nu}^{\mu}$. Covectors are then expressed as $w= w_{\mu}\left( x^{\nu} \right) \hat{\theta}^{\left(\mu\right)}$. The basis covectors are given by the differential forms, which we write as  see that $\hat{\theta}^{\left(\mu\right)} = dx^{\mu}$. The union of all $T_p \mathcal{M}$ over all $p \in \mathcal{M}$ is called the \textit{tangent bundle} over $\mathcal{M}$ and is denoted by $T \mathcal{M}$. The collection of all $T_p \mathcal{M}$ over all $p \in \mathcal{M}$ is then called the \textit{cotangent bundle} over $\mathcal{M}$, denoted by $T^* \mathcal{M}$.
	
	\subsection{Tensors} 
	
	Tensors are basically higher-dimensional generalizations of vectors and covectors. A $(k,l)$-tensor is denoted as $T^{(k)}_{(l)}= T^{\mu_1, \mu_2  \dots \mu_k}_{~~~~~~~~~~~ \mu_1, \mu_2 \dots \mu_l} \partial_{\mu_1} \otimes \partial_{\mu_2} \otimes  \dots \otimes \partial_{\mu_k}\otimes  dx^{\nu_1} \otimes dx^{\nu_2} \otimes \dots \otimes dx^{\nu_l}  $. These are elements of $\underbrace{T_p \mathcal{M}\otimes \dots \otimes T_p \mathcal{M}}_\text{k} \otimes \underbrace{ T_p^* \mathcal{M}\otimes \dots \otimes T_p^* \mathcal{M}}_\text{l}$. Important examples include the Minkowski metric we encountered before, which is a $(0,2)$-tensor, as well as the Riemann and Ricci tensors, to be considered in due time. 
	
	\subsection{Differential forms} 
	
	Differential forms are completely antisymmetric $(0, p)$-tensors. They can thus be written as 
	
	\[  A_{(p)} =  A_{ \left[ \mu_1  \mu_2 \dots \mu_p \right] }  dx_1 \wedge dx_2 \wedge \dots \wedge dx_p \]
	
The lowest-dimensional examples of p-forms are:
	
	\begin{enumerate*}
		\item  0-form: scalar $\phi\left( x^{\mu}\right)$
	
\item 1-form: vectors $A_{(1)}= A_{\mu}dx^{\mu}$ e.g. the electromagnetic potential
	
\item 2-form: $F_{(2)} = F_{\mu \nu } dx^{\mu}\wedge dx^{\nu}$ e.g. the electromagnetic field strength.
\end{enumerate*}

We can construct (p+1)-forms  out of p-forms by applying the \textit{exterior derivative} $F_{(p+1)} = dA_{(p)}$, which acts as

\begin{equation}
d_p A_{(p)} = \frac{1}{p!} \left( \partial_\nu A_{ \left[ \mu_1  \mu_2 \dots \mu_p \right] } \right) dx^\nu \wedge dx^{\mu_1} \wedge \dots \wedge dx^{\mu_p}
\end{equation}

 This allows us to construct the electromagnetic field strength out of the vector potential as $F_{(2)} = d A_{(1)} = \partial_{\mu} A_{\nu}  dx^{\mu} \wedge dx^{\nu} $.

	\subsection{de Rham cohomology}
	
	There are two important types of forms which we will consider further, namely, \textit{closed} and \textit{exact} forms. Closed forms $\omega$ satisfy $ d \omega= 0 $, whereas exact $p$-forms $F_{(p)}$ satisfy $ F_{(p)}= dA_{(p-1)}$, for some $(p-1)$-form $A_{(p-1)}$. It is easy to show that $d d A = 0$ for any $A$ because of the antisymmetrization of the differential forms; one therefore writes this as $d^2 =0$. Hence $dA$ is automatically closed for any differential form $A$. An important question is whether there exist closed forms which are not exact. Forms with this property form the basis of the cohomology group of our manifold, namely, we consider a $p$-form $ F_{(p)}$ topologically equivalent to some $ F_{(p)}'$ if they differ by an \textit{exact} form. We denote the space of closed p-forms by $Z_p(M) = \{ F_{(p)} \vert dF_{(p)}=0 \}$ and the space of exact p-forms by $B_p(M) = \{ F_{(p)} \vert F_{(p)}= dA_{(p-1)} \}$. The \textit{De Rham cohomology} is then defined as 
	
	\begin{equation}
H_p(M)= Z_p(M) / B_p(M) = \{ F_{(p)} \mid \text{closed but not exact} ~. \}
	\end{equation}
	
	We consider an example of a physical application of de Rham cohomology. We saw that the electromagnetic field strangth $F_{\mu \nu}$ can be written in form notation as $F_{(2)} = F_{\mu \nu } dx^{\mu} \wedge dx^{\nu}$, of which the matrix form is
	
	\begin{equation*}
	F_{\mu\nu}= \begin{pmatrix} 
	0 & E_1 & E_2 & E_3 \\
	-E_1 & 0 & B_3 & -B_2\\
	-E_2 & - B_3 & 0 & B_1 \\ 
	-E_3 & B_2 & -B_1 & 0\\
	\end{pmatrix}
	\end{equation*}
	
	The Bianchi identity is then given by  $dF_{(2)}= 0$. Writing the gauge potential in form notation, $A_{(1)} = A_{\mu}dx^{\mu}$, we see that $F_{(2)}=dA_{(1)}= \partial_{\mu}A_{\nu} dx^{\mu} \wedge dx^{\nu}$, hence $F^{(2)}$ is exact in the vacuum. The Bianchi identity is written in terms of the vector potential as $dF_{(2)}=d^2 A_{(1)}= \partial_{\rho}\partial_{\mu}A_{\nu} dx^{\mu} \wedge dx^{\nu}\wedge dx^{\rho} =0$. Gauge transformations are parametrized by a scalar, that is, a 0-form, which we denote by $\chi_{(0)}$. Gauge equivalent potentials then differ by an exact form, i.e. $A_{\mu}' \sim A_{\mu} - \partial_{\mu} \chi$. In form notation, this is written as $A_{(1)} = A_{(1)} - d \chi_{(0)}$. We now consider two examples of toplogically non-trivial situations.

	\begin{enumerate}
		
			\item \textbf{Aharonov-Bohm effect}: We consider an infinite solenoid in an otherwise $\mathbb{R}^3$. The relevant manifold is $\mathcal{M}^{(1,3)} = \mathbb{R}^{(1,3)}/\text{cylinder}$, where the cylinder is the site of a solenoid. Outside this solenoid the magnetic field is zero, inside the solenoid the magnetic field need not be zero. Then, $A_{(1)} = d\chi_{(0)}$, but, by Stokes' theorem, $ \oint_{\mathcal{C}} d\chi_{(0)} = \int_{\mathcal{S}} F  = \Phi $, where $\mathcal{C}$ is a closed curve which encloses the cylinder, $\mathcal{S}$ is a surface such that $\mathcal{C}=\partial\mathcal{S}$, and $\Phi$ is the magnetic flux through $\mathcal{S}$. If we take an electron around any closed path $\mathcal{C}$, its wave-function changes as
			
			\[\psi \to \psi e^{i \frac{e}{\hbar} \oint_{\mathcal{C} = \partial \mathcal{S}} \vec{A} \cdot d \vec{\ell} } = \psi e^{i \frac{e}{\hbar} \int_\mathcal{S} \vec{B} \cdot d \vec{S } }=\psi e^{i \frac{e}{\hbar} \Phi } ~. \]
			
			Demanding that the wave function is single-valued leads to magnetic flux quantization:
			
			\[ \Phi = \frac{\hbar}{e} 2 \pi n = \frac{hn}{e}~, ~~~ n \in \mathbb{Z}~.  \]

		\item \textbf{Magnetic monopole}: This is a configuration with finite magnetic charge density $\rho(x)$ at $r=0$, so that $ \vec{B}(\vec{x}) \sim \frac{g \vec{r}}{r^2}$, where $g$ is the magnetic charge. Gauss' law then tells us that $dF_{(2)} = \rho(\vec{x})$. This configuration cannot be covered by a single chart due to the presence of a \textit{Dirac string} where $F_{(2)} \neq dA_{(1)}$. We therefore have to introduce multiple coordinate patches so that $F_{(2)} = dA_{(1)}$ on each patch, but we cannot introduce a single patch that covers the entire space on which $F_{(2)} = dA_{(1)}$ holds. Similar to the Aharonov-Bohm effect, moving an electron in a circular path that encloses the magnetic monopole gives an exchange phase given by $ \frac{-4\pi eg}{\hbar}$, so that demanding single-valuedness leads to the \textit{Dirac quantiation condition}
		
		\begin{equation}
		g = \frac{\hbar n}{2e} ~, ~~~ n \in \mathbb{Z}~.
		\end{equation}
		
	That is, in the context of electric charges, all magnetic charges are quantized, and vice versa.

		
	\end{enumerate}
	

		\clearpage

	\section{Introduction to general relativity}

	
	General relativity (GR) can be seen as the gauge theory of local coordinate transformations. The basis of GR is the \textit{equivalence principle}, which, in some form, states that physics should be independent of the choice of coordinate system. This is already present  in Newtonian dynamics, where that the inertial mass of a particle is equal to its gravitational mass.
	
	Consider a thought experiment due to Einstein where we have an observer in a box. The observer should not be able to distinguish between gravitational and inertial acceleration by means of local experiments i.e. if the box is small enough no observer will be able to distinguish between these two types of acceleration. This is known as the \textit{weak equivalence principle}. However, if we put two observers in a large enough box, they will be able to distinguish between gravitational and inertial acceleration, since, assuming that the metric is independent of time, they can measure a gradient in the acceleration in the case of gravitational acceleration but, not in the case of inertial acceleration. This obviously constitutes a non-local experiment and is therefore not a violation of the weak equivalence principle.
	
	We now consider changes of the coordinate system due to constant accelerations. Assume the trajectory of an observer $\mathcal{O}$ is given by a path in Minkowski space given by $x^{\mu}(\tau)$, where $\tau$ is the proper time of the observer. Hence we have the velocity vector $u^\mu (\tau) = \frac{d}{d\tau} x^{\mu}$ and acceleration $a^\mu (\tau) = \frac{d^2}{d\tau^2} x^{\mu}$. We then have $a_{\mu} u^{\mu} = \eta_{\mu \nu} a^{\mu} u^{\nu} = 0$. In the case of constant acceleration, we have $a^{\mu} = ( 0, a, 0,0)$, $a \in \mathbb{R}$. Taking boundary condition $x^{\mu}(\tau = 0) = (0, a^{-1},0,0)$, a simple calculation gives
	
	 \[ x^{\mu}(\tau) = (a^{-1} \sinh (a\tau) , a^{-1}\cosh(a\tau) ,0,0) ~.\]
	 
	 This is colloquially referred to as `fake gravity', since the force is entirely due to the inertial acceleration of the particle and not due to a mass distribution. We can obtain the same result by performing a coordinate transformation on Minkowski space-time. Starting from flat Minkowski coordinates $(x^0, x^1, x^2, x^3)$ we go to hyperbolic coordinates, which we write as $( \eta, \rho, x^2, x^3)$. The coordinate systems are related as
	
	\begin{align}
	x^0 =& \rho \sinh (\eta) \notag \\
	x^1 =& \rho \cosh (\eta)~.
	\end{align}
	
	The line element is then written as
	
	\begin{align}
	ds^2 & = - (dx^0)^2 + d\vec{x}^2 \notag \\
	& = -\rho^2 d\eta^2 - d\rho^2 + (dx^2)^2 + (dx^3)^2~.
	\end{align}
	
	This space is known as Rindler space. As stated above, we require that the physics is invariant under such coordinate transformations, so the physics of Rindler and Minkowski space are in some way equivalent. We will further consider this point in due time. 
		
	Consider instead a physical gravitational force, namely, one induced by an energy distributions. Starting from some metric denoted as $g_{\mu \nu}(x)$, the line element is given by $ds^2 = g_{\mu \nu } dx^{\mu}dx^{\nu}$. Since the metric is symmetric, we have $\frac{d}{2}(d+1)$ independent components in $d$ dimensions, i.e. 10 independent components in $(3+1)$ dimensional space. In the case of physical gravitation, there does not exist a coordinate transformation of the form $x^{\mu} \rightarrow x'^{\mu} = x^{\mu} + \xi^{\mu}$ such that we retrieve the flat Minkowski metric. This distinguishes physical (`real') gravity from `fake' gravity.
	
	

		\clearpage

	\section{General relativity}
	
	\subsection{Equivalence principles}
	
	We distinguish the following equivalence principles:

	\begin{enumerate}
		\item	\textbf{Newton's equivalence principle} (which we retroactively denote as such for consistency in nomenclature): Inertial mass is the same as gravitational mass
	
\item	\textbf{Weak equivalence principle}: Gravitational and inertial acceleration are locally indistinguishable
	
	\item \textbf{Strong equivalence principle}: Particles travel along geodesics on a curved manifold (ignoring back-reaction of such particles on the metric as well as non-gravitational forces).
		\end{enumerate}
	
	We have not previously encountered the strong equivalence principle before, which states that particles move on the geodesics of a curved space-time manifold $\mathcal{M}$ regardless of the nature of the particle under consideration (in the limit where we ignore the back-reaction of the particle energy on the metric). This can be stated in words by saying that gravity is equivalent to curvature of space-time. 
	
	As an example we again consider Rindler space, given by
	
	\begin{align}
	t & = \rho \sinh(a\eta) \notag \\
	x& = \rho  \cosh (a\eta)
	\end{align}
	
	Where $\eta= \pm \frac{1}{a}   \text{arctanh} \left( \frac{t}{x} \right)$ $\rho = \pm \sqrt{x^2 - t^2} $. The plus and minus in $\rho$ correspond to the two `external' regions of the corresponding Penrose diagram, which are connected by space-like geodesics. We will return to this point in a later lecture. The above coordinate transformation gives
	
	 $$ ds^2 =   ( -a^2 \rho^2 d\eta^2 + d\rho^2) + (dx^2)^2 +(dx^3)^2~. $$
	
	Observers at rest in Rindler space, that is, at constant $\rho$, are constantly accelerated in Minkowski space. In a curved space-time, there does not exist a globally defined coordinate transformation which transforms $g_{\mu\nu}(x) \rightarrow \eta_{\mu\nu}$, namely, one cannot define a global chart for a curved manifold. The light cones that emanate from the origin are called \textit{Rindler horizons}. They are given by $\rho = 0, \eta = \infty$ and $\rho = 0, \eta = - \infty$. Hence they are never reached by locally inertial observers in Rindler coordinates i.e. observers at rest in Rindler space.

	\subsection{Curved manifolds}
	
	From the Riemannian metric $g_{\mu\nu}(x)$ we construct the Christoffel connection $\Gamma^{\rho}_{\mu \nu}$ and from that the Riemann and Ricci tensor as well as the Ricci scalar. The Christoffel connection gives us covariant derivative for parallel transport, Ricci tensor is the field strength of our connection, Ricci scalar gives a measure of curvature. We will briefly consider more familiar gauge theories before moving on to general relativity.
	
	In general gauge theories, such as electromagnetism, we demand that the action is invariant under local transformations of the form $\psi \rightarrow e^{ i e \Lambda(x)} \psi(x)$.  This requires us to replace $\partial_{\mu} \rightarrow D_{\mu} \coloneqq \partial_{\mu} - ie A_{\mu}(x)$ with $A_{\mu }(x) \rightarrow A_{\mu }(x) + \partial_{\mu}\Lambda(x)$.  The field strength is then given by $\left[ D_{\mu} , D_{\nu} \right]  $. 	
	In general relativity, the `gauge transformation' is of the form $x^{\mu} \rightarrow x^{\mu} + \xi^{\mu}(x)$ i.e. a local transformation of our coordinates. We then have $\nabla_{\mu}V^{\nu} = \partial_{\mu}V^{\nu} + \Gamma^{\nu}_{\mu\lambda}V^{\lambda}$ and $\left[ \nabla_{\mu} , \nabla_{\nu} \right] \sim R_{\mu \nu}$, where $R_{\mu \nu}$ is the Ricci tensor, as we will see below. 	
	For usual gauge theories, we consider gauge fields $A^i_{\mu}$, where $i$ is a group index. In GR the analogous object is $\Gamma^{\rho}_{\mu \nu}$, the group is (local) $SO(1,3)$. We can raise one index on $\Gamma$ to get $\Gamma^{\rho \nu}_{\mu}$, the indices $\rho$ and $\nu$ together form a group index of $SO(1,3)$.
	
	We consider $S^2$, which is an example of a curved space. Starting from $ds^2 = dx^2 + dy^2 + dz^2$, we embed $S^2$ in $\mathbb{R}^3$ AS $x = r \cos \theta \sin \varphi~, ~~ y= r \sin \theta \sin \varphi~,  ~~ z= r\cos \theta$. This gives

	\[ ds^2 = dr^2 + r^2 d\theta^2 + r^2 \sin^2 \theta d\varphi^2 ~. \]
	
	 To get $S^2$, we fix $r=1$. The unit normal vector on $S^2$ is $\hat{x}_n(\theta, \varphi ) = (\sin\theta \cos\varphi ,  \sin\theta \sin\varphi, \cos \theta)$. 	We now construct the tangent vector. Two basis vectors of tangent space $T_p S^2$ are
	
	\begin{align*}
	\vec{u}_1 &= \partial_{\theta}\hat{x} = ( \cos\theta \cos \varphi , \cos \theta \sin \varphi , - \sin \theta )  ~,\\
	\vec{u}_2 & = \partial_{\varphi}\hat{x} = ( -\sin\theta \sin \varphi , \sin \theta \cos \varphi , 0 ) ~.
	\end{align*}
	
	One can easily check by taking the inner product that these are orthogonal to the normal vector. We see from the line element that the metric is given by

	\[g_{\mu \nu} (\theta, \varphi ) = \vec{u}_{\mu}\vec{u}_{\nu}= 
	\begin{pmatrix} 1& 0 \\
	0& \sin^2 \theta
	\end{pmatrix} ~. \]
	
	We would like to define covariant derivatives which are intrinsic to manifold i.e. which can be found without referring to the embedding of our manifold in some other so. We first calculate
	
	\begin{align*}
	\partial_{\theta} \vec{u}_1 =& (- \sin \theta \cos \varphi , - \sin \theta \sin\varphi, - \cos\theta) = - \vec{x}_n  ~, \\
	\partial_{\varphi} \vec{u}_1 = & \partial_{\theta} \vec{u}_2 = (-\cos\theta \sin \varphi , \cos\theta\cos\varphi ,0 ) = \cot \theta \vec{u}_2  ~, \\
	\partial_{\varphi}\vec{u}_2 =& (-\sin\theta\cos\varphi , -\sin\theta\sin\varphi , 0) = - \sin^2 \theta \vec{x}_n - \cos\theta \sin\theta \vec{u}_1  ~.
	\end{align*}
	
	We now drop the normal components to the derivatives of the vectors we just calculated to find the covariant derivatives
	
	\[\nabla_{\theta} \vec{u}_1 = 0 ~~, ~~~ \nabla_{\varphi}\vec{u}_2 = \cot \theta \vec{u}_2  = \nabla_{\varphi}\vec{u}_1  ~~, ~~~ \nabla_{\varphi}\vec{u}_2 = -\cos\theta \sin \theta \vec{u}_1  ~. \]
	
We thus find
	
	 \[ \Gamma^2_{12}= \Gamma^2_{21}= \cot \theta ~~, ~~~\Gamma^1_{22 } = - \cos \theta \sin \theta ~,\]

so that
	
	\begin{align}
	\label{covdiv}
	\nabla_{\mu}V^{\nu} = & \partial_{\mu}V^{\nu} + \Gamma^{\nu}_{\mu \lambda}V^{\lambda} ~, \notag\\
	\nabla_{\mu}w_{\nu}= & \partial_{\mu}w_{\nu} - \Gamma^{\lambda}_{\mu \nu } w_{\lambda} ~
	\end{align}
	
	is satisfied. This is the general action of the covariant derivative $\nabla_{\mu}$ on a vector $V^{\nu}$ and a form $w_{\nu}$, which we added for completeness. 
	
	\vspace{.6cm}
	
		\begin{figure}[h!]
		\begin{center}
			\includegraphics[width=6cm]{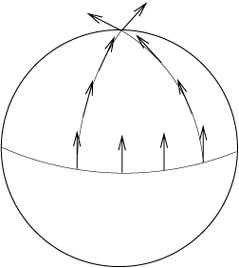}
			\caption{Parallel transport of a vector along a sphere. Image adapted from \cite{PTSimg}. \label{PTSimg}}
		\end{center}
	\end{figure}

	\subsubsection{Parallel transport, geodesics, and curvature}

	If $V^{\lambda}$  transform as a vector under local Lorentz transformations, then $\nabla_{\mu}V^{\lambda}$ transforms as a tensor under coordinate transformations while $\partial_{\mu}V^{\lambda}$ does not. We define the \textit{torsion tensor} as $ T^{\lambda}_{\mu\nu} \coloneqq \Gamma^{\lambda}_{\left[ \mu \nu \right]}$ A connection is called \textit{torsion-free} if $ T^{\lambda}_{\mu\nu} = 0 $. Further, a connection is \textit{metric-compatible} if $\nabla_{\rho} g_{\mu \nu}(x) = 0$. The \textit{Christoffel connection} is given by 
	
	\begin{equation}
	\Gamma^{\rho}_{\mu \nu } = \frac{1}{2} g^{\rho\kappa} \left( \partial_{\mu}g_{\nu \kappa} + \partial_{\nu}g_{\nu \kappa} - \partial_{\kappa}g_{\mu \nu} \right) ~.
		\end{equation}
		
		One can show that this connection is torsion-free and metric-compatible. The central object that characterizes space-time curvature is the \textit{Riemann tensor}. The Riemann tensor describes the parallel transport of a vector $V^{\mu}$ along an infinetesimally small path. Consider the translating a vector $V^\mu$ along a path $x^\mu(\lambda)$. This constitutes parallel transport if $\frac{D}{d\lambda} V^\mu \coloneqq \frac{dx^\nu}{d\lambda} \nabla_\nu V^\mu = 0$, which can be rewritten to give
		
	\[ \frac{d}{d\lambda} V^{\mu} + \Gamma^{\mu}_{\sigma \rho}\frac{dx^{\sigma}}{d\lambda}V^{\rho} = 0~.\]	
		
		As an example, we parallel $V^{\mu}$ along a curve given by $x^{\nu}(\lambda)$, which is a closed curve from $(1,0,0)$ to $(0,1,0)$ to $(0,0,1)$ back to $(1,0,0)$. The path is divided up into smooth sections $A$, $B$, $C$ as
	
\[ x_A(\lambda ) = (\cos \lambda, \sin \lambda , 0)~~,~~~ x_B(\lambda) = ( 0 , \sin \lambda , - \cos\lambda) ~~, ~~~ x_C(\lambda) = (-\sin\lambda, 0 , - \cos\lambda) ~~ \text{with} ~\lambda_A \in  \left[0, \frac{\pi}{2} \right]~. \]

This is illustrated in figure \ref{PTSimg}. One clearly sees that parallel transporting a vector along a closed path on a curved manifold, in this case a two-sphere, can lead to a change in direction, give by the angle between the two arrows at the North pole of the two-sphere. Infinetesimally, we have
	
	\[\left[ \nabla_{\mu}, \nabla_{\nu}\right] V^{\rho} = R^{\rho}_{ \sigma\mu \nu } V^{\sigma} - \Gamma^{\lambda}_{\mu \nu }\nabla_{\lambda}V^{\rho} ~. \]

	The second term is zero because of antisymmetry of the Christoffel connection.
	

		\clearpage

	\section{Einstein's equations}
	
	\subsection{Christoffel connection}
	
	The Christoffel connection is associated to a covariant derivative acting on tensors. In familiar gauge theories, the partial derivative is replaced by a covariant derivative as $\partial_{\mu} \rightarrow D_{\mu} = \partial_{\mu}+ ie A_{\mu}(x)$. In general relativity, the covariant derivative acts as $\partial_{\mu} \rightarrow \nabla_{\mu} = \left(\partial_{\mu}+ \Gamma^{\nu}_{\mu\lambda}\right)\circ$, where $\circ$ indicates that multiplication is tensorially non-trivial, see \ref{covdiv}.

	\subsubsection{Parallel transport}
	
	Parallel transport of a vector $V^{\mu}$ satisfies
	
	\[ \frac{d}{d \lambda} V^{\mu} + \Gamma^{\mu}_{\rho \sigma} \frac{dx^{\sigma}}{d\lambda } V^{\rho} = 0 ~.\]
	
	A special case of parallel transport involves the geodesics on $\mathcal{M}$. Expressing $V^\mu$ as a tangent vector on $\mathcal{M}$ given by $V^{\mu } = \frac{dx^{\mu}}{d\lambda}$, we find
	
	\begin{equation}
	\frac{d^2x^\mu(\lambda)}{d\lambda^2} + \Gamma^{\mu}_{\rho \sigma } \frac{dx^{\rho}}{d\lambda}\frac{dx^{\sigma}}{d\lambda} =0 ~.
	\end{equation}
	
	This is known as the \textit{geodesic equation}.
	
	\subsubsection{Curvature: the `field strength' of $\Gamma$}
	
	As we saw, in a typical gauge theory the field strength tensor is $F_{\mu \nu} = \partial_{\mu } A_{\nu} -\partial_{\nu } A_{\mu} = -i \left[ D_{\mu}, D_{\nu} \right]$. In general relativity, the analogous object is the Riemann (curvature) tensor, given by
	
	$$ \left[ \nabla_{\mu}, \nabla_{\nu} \right] V^{\rho} = \underbrace{R^{\rho}_{\sigma \mu \nu } V^{\sigma}}_{\text curvature} -  \underbrace{\Gamma^{\lambda}_{\mu \nu} \nabla_{\lambda} V^{\rho}}_{\text {torsion}} ~. $$
	
	
For connections that are symmetic in their lower indices, such as the Christoffel connection, the torsion is equal to zero. Plugging in the covariant derivative associated to the Christoffel connection gives
	
	\begin{equation}
	R^{\rho}_{\sigma \mu \nu} = \partial_{\mu}\Gamma^{\rho}_{\nu \sigma} - \partial_{\nu}\Gamma^{\rho}_{\mu \sigma} + \Gamma^{\rho}_{\mu \lambda} \Gamma^{\lambda}_{\nu \sigma} - \Gamma^{\rho}_{\nu \lambda} \Gamma^{\lambda}_{\mu \sigma} ~.
	\end{equation}

	The Riemann tensor becomes easier to manipulate if we lower an index using the metric as $R_{\rho \sigma \mu \nu} = g_{\rho \lambda} R^{\lambda}_{\sigma \mu \nu}$. The symmetry properties of this tensor are
	
	\begin{align}
	R_{\rho \sigma \mu \nu} =& R_{\mu \nu \rho \sigma}\notag \\ 
	R_{\rho \sigma \mu \nu} =& R_{[\rho \sigma] [\mu \nu]}~.
	\end{align}
	
	The Bianchi identity is given by
	
	\begin{equation}
	\nabla_{[ \lambda}  R_{\rho \sigma ] \mu \nu} = 0~.
	\end{equation}
	
	Note that the Riemann tensor transforms non-trivially under coordinate transformations e.g. Lorentz transformations. We then construct the \textit{Ricci tensor} as
	
	\begin{equation}
	R_{\mu \nu} = R^{\lambda}_{ \mu \lambda \nu}~~, ~~~ R_{\mu \nu } = R_{\nu \mu}~.
	\end{equation}
	
	The \textit{Ricci scalar} is then defined as 
	
	\begin{equation}
	R=R^{\mu}_{~ \mu}~.
	\end{equation}
	
	We combine these with the metric to construct the \textit{Einstein tensor}
	
	\begin{equation}
	G_{\mu \nu} = R_{\mu \nu } - \frac{1}{2} R g_{\mu \nu} ~~,~~~ \nabla^{\mu}G_{\mu \nu} = 0 ~.
	\end{equation}
	
	Where the second equality is the defining property of the Einstein tensor. As a simple example, consider the metric of $S^2$ with radius $a$, which is given by 
	
	\[ ds^2 = a^2\left( d\theta^2 + \sin^2 \theta d \varphi^2 \right) ~. \]
	
	For this metric, the Ricci scalar equals
	
\[R_{\rho \sigma \mu \nu} =  \frac{1}{a^2} \left( g_{\rho \mu} g_{\sigma \nu } - g_{\rho \nu} g_{\sigma \mu } \right)~.\]
	
	Space-times with a metric that can be written in this form are called \textit{maximally symmetric spaces} or \textit{Einstein spaces}. For an $n$-dimensional space(-time), they are easily seen to satisfy $R_{\mu \nu}= \frac{n-1}{a^2} g_{\mu \nu}$ and $R= \frac{n(n-1)}{a^2} $, e.g. for $S^2$ we have $R= \frac{2}{a^2}$.
	
	\subsection{Einstein equations}
	
	The Einstein equations tell us how matter or energy affects the curvature of a space-time, as well as how matter or energy is affected by the curvature of space-time. This is reminiscent of the equivalence between inertial mass and gravitational mass. Newton's equation of gravitation can be written as $m_I a = - m_G \nabla \Phi$, where $\Phi$ is the gravitational potential and $m_I$ and $m_G$ are the intertial and gravitational mass, respectively. Consider the geodesic equation of a massive particle with $\lvert\frac{dx^i}{d\tau}\rvert \ll  \lvert \frac{dt}{d\tau} \rvert$ i.e. in the limit of small velocities. Then
	
	\[ \frac{d^2x^{\mu}}{d\tau^2} + \Gamma^{\mu}_{00} \left( \frac{dt}{d\tau}\right)^2 = 0~.\]
	
	Using $\Gamma^{\mu}_{00} = -\frac{1}{2} g^{\mu \lambda}\partial_{\lambda} g_{00}$ and making a \textit{weak field approximation} $g_{\mu \nu} = \eta_{\mu \nu} + h_{\mu \nu}$, $\lvert h_{\mu \nu} \rvert \ll 1 $, we find
	
	\[ \frac{d^2x^{\mu}}{d\tau^2} = \frac{1}{2} \eta^{\mu \lambda} \partial_{\lambda}h_{00} \left( \frac{dt}{d\tau}\right)^2~.\]
	
	Since $\frac{dt}{d\tau} \approx 1$, we have
	
\[ \frac{d^2x^i}{d\tau^2} = \frac{1}{2} \partial_i h_{00} ~.\]
	
	For small $\Phi$, we have $g_{00} = -\left( 1+2\Phi\right)$, so that we retrieve the Newtonian limit 
	
	\[ \frac{d^2x^i}{d\tau^2} =  -\partial^i \Phi ~.\]

	Introducing a matter density $\rho(\vec{x})$, we find the Poisson equation
	
	\[ \nabla^2 \Phi = 4\pi G \rho\]
	
	Where $G$ is Newton's constant. On the left hand side, we have a geometric quantity akin to curvature, and on the right hand side we have an matter distribution. Einstein's equations of general relativity can be seen as a fully covariant generalization of this expression. Observe that $\rho = T^{00}$, where $T^{\mu \nu} = - \frac{2}{\sqrt{-g}} \frac{\delta S_{\text{matter}}}{\delta g _{\mu \nu}}$ is the \textit{energy-momentum tensor}. Here, $g = \det g_{\mu \nu}$, the determinant of the metric, and $S_{\text{matter}}$ is specific to the matter system we are considering e.g. the standard model. We thus find 
	
	\[ \nabla^2 h_{00} = -8 \pi G T_{00}~.\] 
	
	These can be seen to arise as a limit of the \textit{Einstein field equations}, which are given by
	
	\begin{equation}
	\label{EEQ}
	G_{\mu \nu} = R_{\mu \nu} - \frac{1}{2} R g_{\mu \nu} = 8\pi G T_{\mu \nu} - \frac{1}{2} \Lambda g_{\mu \nu} ~,
	\end{equation}
	
	where $\Lambda$ is know as the \textit{cosmological constant}, which constitutes a constant energy density in our space-time. From energy-momentum conservation, $\nabla^{\mu}T_{\mu \nu } = 0 = \nabla^{\mu}G_{\mu \nu }$, which explains the relevance of the Einstein tensor. 
	
A natural question is which action gives rise to the Einstein field equations. The answer is the \textit{Einstein-Hilbert action}, which is given by
	
	\begin{equation}
	S_{EH}= \frac{1}{8 \pi G} \int d^4 x \sqrt{-g} \left(R- \Lambda \right) + \mathcal{L }_{matter} 
	\end{equation}
	
	  Using 
	  \[ T_{\mu \nu} =  -\frac{2}{\sqrt{-g}} \frac{\delta S_{\text{matter}}}{\delta g^{\mu \nu}}\]
	
	and demanding $\frac{\delta S_{EH}}{\delta g_{\mu \nu }}=0$, one can derive the Einstein equation. Since the energy-momentum tensor depends on the metric, the equations we end up with are typically very complicated.
	
	\subsection{Remarks}
	
	Einstein equations are a system of coupled non-linear second order partial differential equations. This means that analytic solutions are rather difficult to find, and very few are known. To find a solution we have to choose initial conditions. In Newton, we choose $(p_i, x_i)(t_0)$, where $t_0$ is some initial time. In GR, we choose a space-like hypersurface $\Sigma$, which is called a \textit{Cauchy surface} if the union of its past and future domains of dependence cover the entire space-time. For such a $\Sigma$ we can formulate a well-defined initial value problem.
	
	Approximate solutions include two important classes
	
	\begin{enumerate}
		\item \textbf{Gravitational waves}: These are approximate solutions of the vacuum Einstein equations in the limit $\lvert h_{\mu \nu} \rvert \ll 1 $, $v=c$. This gives the following equation
	
	\[ \Box h_{\mu \nu} - \frac{1}{2} \eta_{\mu \nu} \Box h = - 16 \pi G T_{\mu \nu } \]
	
	We are in a vacuum, $T_{\mu \nu} = 0$, so that 
	
	\begin{equation}
	\Box h_{\mu \nu} = 0 ~.
	\end{equation}
	
	This describes a fully relativistic wave with two polarization, namely $h_{++}$ and $h_{--}$, so that this wave corresponds to a spin-2 particle.
	
\item \textbf{Newtonian limit}: In this case we take the limit $\lvert h_{\mu \nu} \rvert \ll 1 $, $v \ll c$, which we also considered in the previous section. For an object of mass $M$ at position $\vec{x}=0$, we then have
	
	\begin{equation}
	h_{00 } = -2 \Phi ~~,~~~ \rho(\vec{x}) = M \delta(\vec{x}) ~ \Rightarrow \Phi = -\frac{GM}{r} ~.
	\end{equation}
	
	Then
	
	\begin{align}
	\
	ds^2 & = - \left( 1+ 2\Phi \right)dt^2 + \left( 1 - 2\Phi \right)\left( dx^2 + dy^2 + dz^2 \right) \notag \\
	& = - \left( 1 - \frac{2GM}{r} \right) dt^2 + \left( 1 + \frac{2GM}{r} \right) \left( dr^2 + r^2 d\Omega^2 \right) ~.
	\end{align}
\end{enumerate}
	
	The attentive reader may wonder why a non-zero mass would give rise to a vacuum solution of Einstein's equations. This will be clarified in the next section where we consider the Schwarzschild black hole solution, which will be reminiscent of the metric given above.

	\clearpage

	\section{Black holes}
	
	We will now introduce the central objects of interest, namely, black holes. These arise from gravitational collapse of an object with some mass which is compressed into a small region of space-time. It is caracterized by a curvature singularity at the origin which is `screened' to outside observers by a coordinate singularity at finite radial distance. This coordinate singularity is known as the \textit{event horizon}, which will be seen to exhibit deep connections with thermodynamic systems.

	\subsection{Schwarzschild solution}
	
	The metric of a star of mass $M$ which we derived last time is the large $r$, small $M$ limit of the Schwarzschild solution, which we consider here. We look for an exact solution of $R_{\mu \nu} - \frac{1}{2} R g_{\mu \nu} = 8 \pi G T_{\mu \nu} + \frac{1}{2} \Lambda g_{\mu \nu}$ under the following assumptions.

	\begin{enumerate}
	\item \textbf{Vacuum solution}: $T_{\mu \nu} = 0$
	\item \textbf{Spherical symmetry}: $\partial_{\phi} g_{\mu \nu}=0$
\item \textbf{Asymptotic flatness}: $\Lambda = 0$, where $\Lambda$ is the cosmological constant
	\end{enumerate}
	
The solution which satisfies these assumptions was found by Schwarzschild only a few months after Einstein published his theory of general relativity. It reads
	
	\begin{equation}
	ds^2 = -\left(1-\frac{2M}{r}\right) dt^2 + \left(1-\frac{2M}{r}\right)^{-1}dr^2 + \underbrace{r^2(d\theta^2 + \sin^2 \theta d\varphi^2)}_{r^2 d\Omega^2}~. 
	\end{equation}

This is known as the Schwarzschild metric. We have one parameter, mass $M$, while electric charge $Q$ and angular momentum $L$ are zero. Birkhoff's theorem then tells us that the Schwarzschild black hole is the only spherically symmetric solution of Einstein's equations. We see that for large $r$, we can expand $\left(1-\frac{2M}{r}\right)^{-1} \approx 1+\frac{2M}{r}$, which gives the metric for a star derived last time. For $r \rightarrow \infty$ as well as $M\rightarrow 0$, we retrieve Minkowski space.

We now compute the curvature of this geometry. In this metric, we have $R_{\mu \nu} = 0 $, so that this is a so-called \textit{Ricci flat} space. Hence, we consider another quantity than the Ricci tensor, namely, the \textit{Kretschmann scalar}, which, for the Schwarzschild metric, is given by

\[ R_{\mu \nu \rho \sigma} R^{\mu \nu \rho \sigma} = \frac{48M^2}{r^6} \overset{r \to 0 }{\rightarrow} \infty ~.\]

 The $r\to 0$ behaviour of the Kretschmann scalar expresses the fact that $r=0$ is the locus of a \textit{space-time singularity}, namely, a singularity which cannot be resolved by a coordinate transformation. This is different from the \textit{coordinate singularity} at $r=2M$, where the Kretschmann scalar is finite. This coordinate transformation can be removed by going to a different coodrinate system, as we will see later.

\subsection{Event horizon}

The event horizon has the topology of $S^2$ and is located at $r=2M$. Here, the metric becomes divergent. However, all components of $R^{\lambda}_{\mu \nu \rho}$ are finite. Hence we see (as stated before) that this is not a true (space-time) singularity but merely a coordinate singularity. However, in spite of the fact that this singularity is removable, a lot of interesting physics takes place here. We distinguish between three different parts of a black hoe space-time

\begin{itemize}

\item For $r > 2M$, called region $I$ or the \textit{exterior} of the black hole, the metric has signature $(-,+,+,+)$.

\item For $r< 2M$, called region $II$ or the \textit{interior} of the black hole, the metric has signature $(+,-,+,+)$.

\item At $r=2M$, we have the event horizon, which is a \textit{null surface} since it has signature $(0,0,+,+)$.
\end{itemize}

We thus see that the event horizon is the point at which $x$ and $t$ exchange their respective signatures (positive and negative). In region $II$ we have to move forward in $r$ i.e. to $r=0$, which means one inevitably hits the singularity. This is is reminiscent of particles always moving forward in time (after setting our conventions of time direction) in region $I$. 

\begin{figure}[h]
	\begin{center}
		\includegraphics[width=8cm]{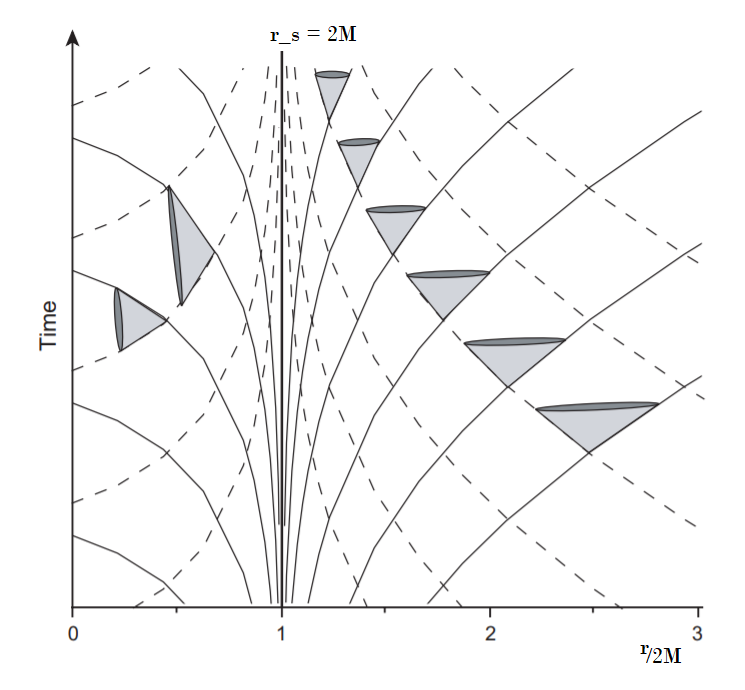}
		\caption{ Space-time diagram of the near-horizon region of the Schwarzschild metric. We see that the future directed light cones, given by the shaded areas, `close up' when we approach the Schwarzschild radius $r_s$ from $r>r_s$. At $r=r_s$, the direction of the light cones changes discontinuously as $r$ and $t$ exchange (positive and negative) signature.}
	\end{center}
\end{figure}


An outside observer will never see a signal reach the event horizon since signals close to the event horizon get infinitely red-shifted i.e. signals slow down until they become stationary close to the event horizon. Infinite redshift means that light loses all its energy as it climbs out of the gravitational potential of the mass at $r=0$, which is why black hole is an appropriate name for this object. At $r=2M$, the light-cone flips since metric signature changes from $(-,+,+,+)$ to $(+,-,+,+)$. 

We now consider a co-moving observer that is propagating toward a black hole. Such an observer crosses the event horizon in finite proper time $\tau$ and does not experience anything special when crossing the event horizon. This observer also reaches the space-time singularity in finite time. Consider a radial null geodesic i.e. $ds^2=0$ and $\frac{d\theta}{dt} = 0 = \frac{d\varphi}{dt}$. This gives

\begin{equation}
\label{radnull}
\frac{dr}{dt} = \pm \left( 1- \frac{2M}{r}\right)~.
\end{equation}

Equation \ref{radnull} tells us that the light cone indeed `closes up' when approaching the event horizon. We now change to a coordinate system which does not contain the coordinate singularity we have in the Schwarzschild solution. We first go to \textit{tortoise coordinates}. We first solve \ref{radnull}, which gives

\[ t= \pm \left[ r+2M \ln \left( \frac{r}{2M} -1 \right) \right] ~. \]

We then introduce $r^* = r + 2M \ln \left( \frac{r}{2M} -1 \right) \xrightarrow{r \to 2M } - \infty $. Our metric then becomes

\begin{equation}
ds^2 = \left( 1- \frac{2M}{r} \right) \left( -dt^2 + (dr^*)^2 \hspace{.05cm} \right) + r^2 d\Omega^2~.
\end{equation}

This metric is conformally equivalent to the Minkowski metric, which means that light cones are given by $r^* = \pm t$. However, this metric is not quite what we are looking for since $g_{tt},g_{rr} \xrightarrow{r \rightarrow 2M} 0 $.

We then introduce $v=t+r^*$ and $u=t-r^*$ such that $v=$constant and $u=$constant describe outgoing and infalling radial null curves, respectively.  We can then express our metric as follows

\begin{equation} 
ds^2 = - \left(1-\frac{2M}{r}\right) dv^2 + 2dv dr + r^2 d\Omega^2  ~.
\end{equation}

This is the so-called \textit{Eddington-Finkelstein metric}. For a null curve, we have $ds^2=0$. For $d\Omega = 0$ (i.e. radial null curves), this gives $ds^2 = -\left( 1- \frac{2M}{r}\right) dv^2 + 2dr dv =0 $. The solutions are 

\[\frac{dv}{dr}=\left\{
\begin{array}{ll}
0~,\\
\frac{2}{1-\frac{2M}{r}}\\
\end{array}
\right. ~.\]

Hence for $r<2M$ all future-directed paths are in the direction of decreasing $r$, which means that a signal at $r<2M$ will inevitably hit the space-time singularity at $r=0$. 

In the next lecture, we introduce the \textit{Kruskal-Szekeres} coordinates, which helps us completely get rid of coordinate singularities. This will make our space-time \textit{geodesically complete}, but we also get additional regions $III$ and $IV$ that are copies of $I$ and $II$.

\clearpage

\section{Kruskal-Szekeres coordinates and geodesics of the Schwarzschild black hole}

Last lecture, we saw that the Schwarzschild solution is characterized by a \textit{curvature singularity} at $r=0$ and a \textit{metric singularity} at $r=\frac{2GM}{c^2}$, where we re-introduced the gravitational constant $G$ and speed of light $c$. The space-time has an exterior and an interior region characterized by $r>2M$ and $r<2M$ and denoted by $I$ and $II$, respectively. Today we will consider a new coordinate system which will allow us to double the space-time and find copies of regions $I$ and $II$. So far we have introduced the following coordinate systems

\begin{enumerate}
	\item Schwarzschild coordinates: $(t,r, \theta,\varphi)$
\item Tortoise coordinates: $(t, r^*,\theta,\varphi )$ with $r^* = r+2M \ln \left( \frac{r}{2M} -1 \right) $
\item Eddington-Finkelstein coordinates: $(v,u,\theta , \varphi)$, where $u$, $v$ are light-cone coordinates given by $v=t+r^*$, $u=t-r^*$.
\end{enumerate}

We now introduce \textit{Kruskal-Szekeres coordinates} as

\begin{align}
\begin{rcases}
R&= \lvert \frac{r}{2M} -1 \rvert^{1/2} e^{r/4M} \cosh \left( \frac{t}{4M} \right) \notag \\
T&= \lvert \frac{r}{2M} -1 \rvert^{1/2} e^{r/4M} \sinh \left( \frac{t}{4M} \right)
\end{rcases}
\text{region $I$ ,}
\end{align}

\begin{align}
\begin{rcases}
R&= \lvert \frac{r}{2M} -1 \rvert^{1/2} e^{r/4M} \sinh \left( \frac{t}{4M} \right) \notag \\
T&= \lvert \frac{r}{2M} -1 \rvert^{1/2} e^{r/4M} \cosh \left( \frac{t}{4M} \right)
\end{rcases}
\text{region $II$ .}
\end{align}

The metric is then given by

\begin{equation}
ds^2 = \frac{32M^3}{r} e^{-r/2M} \left( - dT^2 + dR^2 \right) + r^2d\Omega_{(2)}^2~.
\end{equation}

A useful expression is $ T^2 - R^2 = \left( 1- \frac{r}{2M} \right) e^{r/2M}$, which holds in both regions $I$ and $II$. Kruskal-Szekeres coordinates are perhaps the most useful coordinates for describing a black hole since they are geodesically complete and they do not exhibit a metric singularity. They cover the entire space-time, in fact it is the double cover of our original space-time, as we will see. Some remarks are in order.

\begin{enumerate}
	\item We still have a curvature singularity at $r=0 \leftrightarrow T^2 = R^2 +1$. 
\item However, the metric is regular at $r=2M \leftrightarrow T= \pm R$. 
\item Radial null curves are given by $T = R + \text{constant}$.
\end{enumerate}

We can allow $T$ and $R$ to range over $-\infty < R < \infty$ and $ -\infty < T < \infty$  subject to $T^2 \leq R^2 +1$, which will give a second copy of our space-time. 

\begin{figure}[h]
	\begin{center}
		\includegraphics[width=10cm]{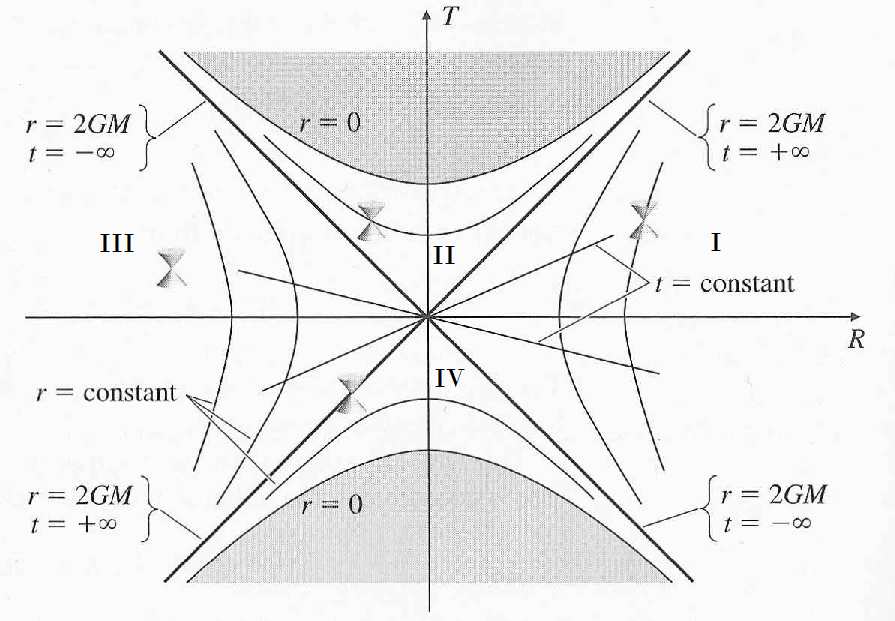}
		\caption{A diagram of the Schwarzschild black hole in Kruskal-Szekeres coordinates, where null curves travel at 45 degrees. Note that the even horizons are also at 45 degrees, which signals the fact that they are null-like hypersurfaces. We have four regions indicated by Roman numerals as described in the text. Image adapted from \cite{krusk}}
		\end{center}
\end{figure}

The four regions of the Kruskal-Szekeres diagram are\\

\begin{itemize*}
	\item $\begin{rcases}
I: r>2M: ~ \text{exterior of black hole}\\
II: r<2M: ~ \text{interior of black hole} 
\end{rcases}
\text{our original space-time~,}$

\item $\begin{rcases}
III: r>2M: ~\text{parallel universe, causally disconnected from I }\\
IV: r>2M: ~ \text{white hole} 
\end{rcases}
\text{a copy of our original space-time~.}
$\\
\end{itemize*}

The \textit{white hole} is a region of space-time from which signals can reach any point in space-time, but not the other way around. The white hole is shielded by a horizon, as per the \textit{cosmic censorship conjecture}. This holds that all curvature singularities are `screened' by an event horizon i.e. there are no so called \textit{naked singularities}.

\subsection{Geodesics and effective potential of the Schwarzschild black hole}


We consider paths outside the horizon and ask ourselves whether we can escape the gravitational pull of the black hole as long are we outside the event horizon. We will see that the answer is yes. Recall the geodesic equation

\[ \frac{d^2x^{\mu}}{d\tau^2} = - \Gamma^{\mu}_{\nu \lambda} \frac{dx^{\nu}}{d\tau}\frac{dx^{\lambda}}{d\tau}  ~.\]

The symmetries of our system allow us to isolate a single differential equation with an effective potential. The metric is invariant under rotations as well as time translations. This gives two \textit{Killing vectors}, namely 

\begin{align*}
K^{\mu}_{\phi} = (0,0,0,1)~,\\
K^{\mu}_t = (1,0,0,0)~.
\end{align*}

These satisfy the \textit{Killing equation}

\begin{equation}
K_{\mu} \frac{dx^{\mu}}{d\tau} = \text{constant} ~ .
\end{equation}

 This entails that angular momentum $L=r^2 \frac{d\phi}{d\tau}$ and energy $E=\left(1-\frac{2M}{r} \right) \frac{dt}{d\tau}$ are conserved along geodesics, corresponding to spherical symmetry and time invariance, respectively. The geodesic equation then gives

\begin{equation}
\frac{1}{2} \left(\frac{dr}{d\tau}\right)^2 + V(r) = \frac{1}{2}E^2  ~.
\end{equation}

Setting $\theta = \frac{\pi}{2}$, the effective potential $V(r)$ is given by

\begin{equation}
V(r) = \frac{1}{2}\left( \epsilon + \frac{L^2}{r^2} \right) \left( 1- \frac{2M}{r}\right) ~,
\end{equation}

with $\epsilon = +1, -1, 0$ for time-like, space-like, and null geodesics, respectively. We look for roots in the derivative of the effective potential

$$ \frac{dV(r)}{dr} = \epsilon \frac{M}{r^2} - \frac{L^2}{r^2} + \frac{3ML^2}{r^4} = 0  ~.$$

We only find a solution for $L>12M^2$. We have circular orbits at

\begin{equation}
r_{\pm} = \frac{L^2 \pm \sqrt{L^4 - 12M^2 L^2 }}{2M}   ~.
\end{equation}

A more detailed derivation can be found in \cite{mblau}.


\subsection{Gravitational energy}

 We use the standard definition of energy and Einstein's equations to find the energy in a space-time as

\[ E= \int_{\Sigma}d^3x T_{00} = \frac{1}{8\pi G} \int_{\Sigma}d^3x G_{00}  ~,\]

where $G_{\mu \nu}$ is the familiar Einstein tensor and $\Sigma$ is an achronal surface which covers our entire space. We perform a weak field approximation by writing $g_{\mu \nu } = \bar{g}_{\mu \nu} + h_{\mu \nu}$, where $h_{\mu \nu}$ is a perturbation around our background metric $\bar{g}_{\mu \nu}$, which we set to $\eta_{\mu \nu}$. At a linearized level, we have

\begin{align*}
G_{00} &= - \frac{1}{2} \Delta \left( \delta^{ik} h_{ik} \right) + \frac{1}{2}\partial_i \partial_k h^{ik}  \\
& = \partial_i \left( -\frac{1}{2} \partial^i h^k_k + \frac{1}{2}\partial_k h^{ik} \right)~.
\end{align*}

We can then use Stokes'theorem to find

\[ E= \frac{1}{16\pi G} \oint_{S^2_{\infty}}dS_i \left( \partial_k h^{ik} - \partial^i h^k_k \right) ~. \]

Covariantizing gives 

\begin{equation}
E= \frac{1}{16\pi G} \oint_{S^2_{\infty}}dS^i \left( \nabla^k h_{ik} - \nabla_i h \right) ~.
\end{equation}

This quantity is know as the ADM mass (energy) of a gravitational system. We look at the ADM mass for the Schwarzschild metric. We have $ \partial_k h_{ik} - \partial_i h_{kk} = \frac{4GM}{r^3}x_i$. 

\[ \oint dS^i x^i = \oint d\Omega r^2 n_i x^i = 4 \pi r^3.\]

We can then calculate

\begin{equation}
E_{ADM} = \displaystyle{\lim_{r \to \infty}} \frac{1}{16 \pi G} \frac{4GM}{r^3} 4 \pi  r^3 = M ~.
\end{equation}

We thus find that the energy of the Schwarzschild space-time is equal to the mass of the black hole, as we would naturally expect.

\clearpage

\section{ Conformal compactifications and Penrose diagrams}

Typical space-time metrics, e.g. $\mathbb{R}^{1,3}$ or Schwarzschild space, are infinite in coordinate extension. This means that there are boundaries of our space-time at infinite coordinate distance in this coordinate system. To make such space-times more manageable we perform so-called \textit{conformal compactifications}, which is a transformation of our original coordinate system such that:

\begin{enumerate}
	\item Space-time boundaries typically at infinite coordinate distance are mapped to lines, points, or hypersurfaces at finite distance
	\item The conformal structure is kep intact, in particular, we require that light rays travel at 45 degrees. 
\end{enumerate}

\subsection{Examples}

\subsubsection{Two-dimensional flat space $\mathbb{R}^{2} $}

We have $ds^2 = dx^2 + dy^2 = dr^2 + r^2 d\phi^2$ with $x = r \cos \phi, ~ y=r \sin \phi,~ 0 \leq r < \infty $. We then transform as $r=\tan \frac{\theta}{2},~ 0 \leq \theta < \pi$. The metric then becomes

$$ds^2 = \frac{1}{4\cos^4 \frac{\theta}{2} } \left( d\theta^2 + \sin^2 \theta d\phi^2\right) = f(\theta) \left( d\theta^2 +\sin^2 \theta d\phi^2 \right) \eqqcolon f(\theta) d \tilde{s}^2~.$$

This conformal rescaling allows us to relate the metric of a non-compact space $(ds^2)$ to the metric of a compact space $(d\tilde{s}^2)$. Generally, a conformal rescaling is a coordinate transformation of the form $g_{\alpha \beta}(x) \rightarrow \tilde{g}_{\alpha \beta}(x)= \Omega(x)^2 g_{\alpha \beta}$. Since it is simply a local rescaling of the metric, the causal structure remains the same i.e. light rays still travel at 45 degrees. The Penrose map is a combination of a coordinate transformation that maps infinity to finite coordinate distance and a conformal rescaling.

 \subsubsection{(1+3)-dimensional Minkowski space $\mathbb{R}^{1,3}$}
	
	The $(1+3)$-dimensional Minkowski metric is given by
	
	\begin{equation}
	ds^2 = -dt^2 + dr^2 + r^2 d\Omega_{(2)}^2 = -du dv + \left( \frac{v-u}{4}\right)^2d\Omega_{(2)}^2 ~, 
	\end{equation}
	
	with $v=t+r,~ u=t-r$. We now define $V \coloneqq \arctan v , ~ U \coloneqq  \arctan u$. The range of the coordinates is $ - \frac{\pi}{2} < U \leq V < \frac{\pi}{2}$. We can also define $T \coloneqq U+V,~ R=V-U$, which then gives $ \lvert T \rvert + R < \pi,~ 0 \leq R < \pi$. $T$ and $R$ thus cover a triangle i.e. a compact space. We then have
	
	\[ds^2 = \frac{1}{4 \cos^2 V \cos^2 U} \left(-4dUdV + \sin^2 (V-U) d\Omega_{(2)}^2 \right) \eqqcolon \frac{1}{4 \cos^2 V \cos^2 U} d \tilde{s}^2\]
	
or 
\[d\tilde{s}^2 = \underbrace{-dT^2 +dR^2}_{\text{flat metric}  + \sin^2 R d\Omega_{(2)}^2} ~. \]
	
	This is the metric of the Einstein static universe, its topology is given by $I \otimes S^3$, with $I$ a finite interval.
	
		\begin{figure}[h!]
		\begin{center}
			\includegraphics[width=4.5cm]{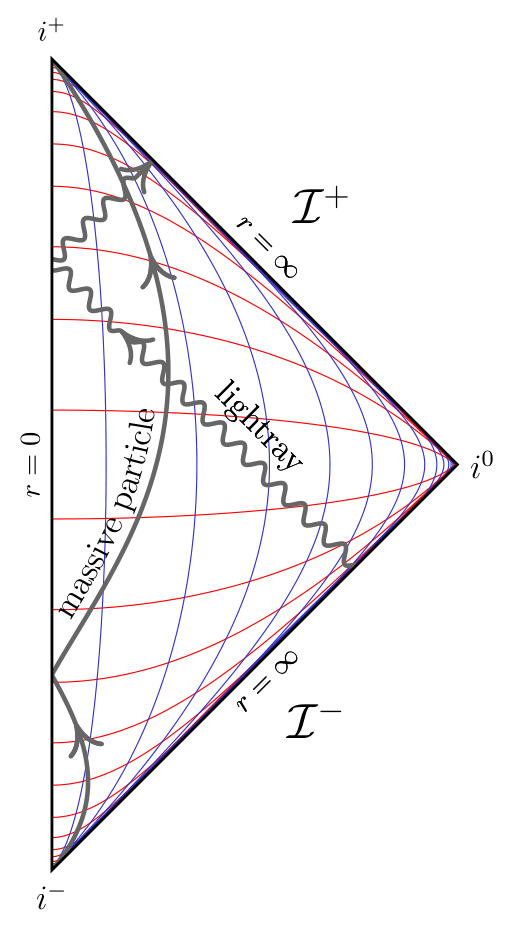}
			\begin{center}
			\caption{The conformal diagram of Minkowski space with time-, null-, and space-like coordinate infinities indicated as explained below. Image adapted from \cite{strominfr}.}
		\end{center}
		\end{center}
	\end{figure}

	We have
	
	\begin{itemize}
		\item  $i^+$: \textbf{Future timelike infinity} i.e. $t=\infty$, fixed $r$,
		\item 	$i^-$: \textbf{Past timelike infinity} i.e. $t=\infty$, fixed $r$,
	\item 	$i^0$: \textbf{Spatial infinity} i.e. $r=\infty$,
	\item 	$\mathcal{I}^+$: \textbf{Future null infinity} i.e. the $u=$fixed, $v\to \infty$  asymptotics of outgoing radial null curves,
	\item 	$\mathcal{I}^-$: \textbf{Past null infinity} i.e. the $v=$fixed, $u \to - \infty$ asymptotics of ingoing radial null curves.
		\end{itemize}

\vspace{.6cm}
	
	We make the following remarks
	
	\begin{enumerate}
		\item Radial null geodesics i.e. light cones are at $\pm 45$ degrees. 
	\item All points in the Penrose diagram represent a 2-sphere except $i^0$ and $i^{\pm}$.
	\item $i^0$ and $i^{\pm}$ are really points
		\item $ \mathcal{I}^{\pm}$ are null hypersurfaces $\mathbb{R} x S^2$
	\item All infinitely extended timelike geodesics begin at $i^-$ and end at $i^+$.
	\item All infinitely extended spacelike geodesics begin at $i^0$, are reflected at $r=0$ and come back $i^0$.
	\item All infinitely extended null geodesics begin at $\mathcal{I}^-$, are reflected at $r=0$, and at $\mathclap{I}^+$.
\end{enumerate}
	
	 Any time-like geodesic observer will eventually be able to see all of Minkowski space i.e. at $i^+$ the past light-cones of all observers cover all of Minkowski space.The past and future light-cones of any two events have a non-empty intersection. In particular, any two events in Minkowski were causally connected at some point in the past. This entails that there is no horizon in Minkowski space. We will see that this will not be true for the following example.
	
	\subsubsection{Two-dimensional Rindler space}

	We start from the $(1+1)$-dimensional Minkowski metric given by $ds^2 = -dt^2 + dx^2,~ -\infty<t<\infty,~ -\infty<x<\infty$. We define $x^{\pm} \coloneqq t\pm x$ and $X^{\pm} \coloneqq \arctan x^{\pm} $. This gives the following compactified metric
	
	\[ d\tilde{s}^2 = - dX^+ dX^- ~~, ~~ \frac{\pi}{2} \leq X^+, X^- \leq \frac{\pi}{2}~. \]
	
\begin{figure}[h!]
		\begin{center}
			\includegraphics[width=8cm]{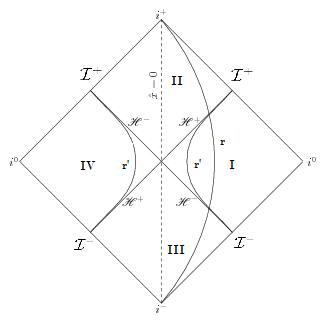}
			\caption{The maximally extended conformal (Penrose) diagram of Rindler space with Rindler horizons at $X^{\pm} = 0$ are indicated by $\mathcal{H}^{\pm}$ and Rindler wedges indicated by Roman numerals $I - IV$. The curved lines indicated by $r$ and $r'$ indicate locally inertial frames in Minkowski and Rindler coordinates, respectively. The other symbols have the same meaning as in Minkowski space. Figure adapted from \cite{rindlerfig}.}
			\label{rindlerfig}
		\end{center}
	\end{figure}

	Its conformal diagram is displayed in figure \ref{rindlerfig}, which displays many similarities with the conformal diagram of the Schwarzschild black hole, to be considered next time.
	
	\clearpage

	\section{Penrose diagrams of charged \& rotating black holes}

	\subsection{Penrose diagram for Schwarzschild black hole}
	
	As we saw, Penrose diagrams exhibit the causal structure of space-time. Penrose diagrams are found in two steps:
	
	 A) Choose coordinates that map boundaries of space-time to finite coordinate distance.\\
     B) Conformal rescaling of metric i.e. get rid of divergent part of metric, namely, the conformal factor.

	We treated Minkowski space $\mathbb{R}^{3,1}$, which is geodesically complete, as well as Rindler space, which is not geodesically complete. We will now discuss the Penrose diagram of the Schwarzschild space. We focus on region $I$ i.e. the first outer region of the black hole corresponding to $r>2M$. We transform
	
	$$(r,t) \rightarrow (u,v)~ ~; ~~ u=t-r^*~~,~~~ v=t+r^*$$
	
	Note that $ -\infty \leq t \leq \infty$ and $ -\infty \leq r^* \leq \infty$. The metric is given by
	
	\begin{equation}
	ds^2= -\left( 1-\frac{2M}{r}\right) du dv + r^2 d \Omega^2_{(2)}
	\end{equation}
	
	We have the following null boundaries \par
	\makebox[1.5cm]{$\mathcal{I}^+$ :}  $  u=\text{finite} ~, ~~ v \rightarrow \infty$ ~,\par
	\makebox[1.5cm]{$\mathcal{I}^-$ :}  $ v= \text{finite} ~, ~~ u \rightarrow - \infty$~,\par
	
\vspace{.3cm}
	
	as well as event horizons  \par 
	\makebox[1.5cm]{$H^+$ :}  $u\rightarrow \infty~, ~~ v= \text{finite}$~,\par
	\makebox[1.5cm]{$H^-$ :}  $v\rightarrow -\infty~, ~~ u=\text{finite}$~.\par
	
	\vspace{.3cm}
	
	We define $u\eqqcolon \tan \tilde{U},~ v \eqqcolon \tan \tilde{V}~,~ ~\lvert \tilde{U}\rvert \leq \frac{\pi}{2} , ~\lvert \tilde{V}\rvert \leq \frac{\pi}{2} ~$, which can then be re-expressed in time- and space-like components as 
	
\[ u_k \coloneqq T - R \coloneqq \tan \hat{U} , ~ v \coloneqq T+R \coloneqq \tan{\hat{V}} ~. \] 

The coordinates $(R,T)$ are the Kruskal-Szekeres coordinates we encountered before, in which the metric is expressed as 
	
	\begin{equation}
	ds^2 = \frac{32M^3}{r}e^{-r/2M} \left(dT^2 -dR^2\right) + r^2 d\Omega_{(2)}^2 ~.
	\end{equation}

 This gives the fully extended Penrose diagram, which has the following regions\\
	
	\begin{enumerate}
		\item Region $I$: Exterior of black hole
		\item Region $II$: Interior of black hole
		\item Region $III$: Parallel universe
		\item Region $IV$: White hole
	\end{enumerate} 

\vspace{.6cm}

		\begin{figure}[h!]
		\label{Schwarzschild_penrose}
		\begin{center}
			\includegraphics[width=11cm]{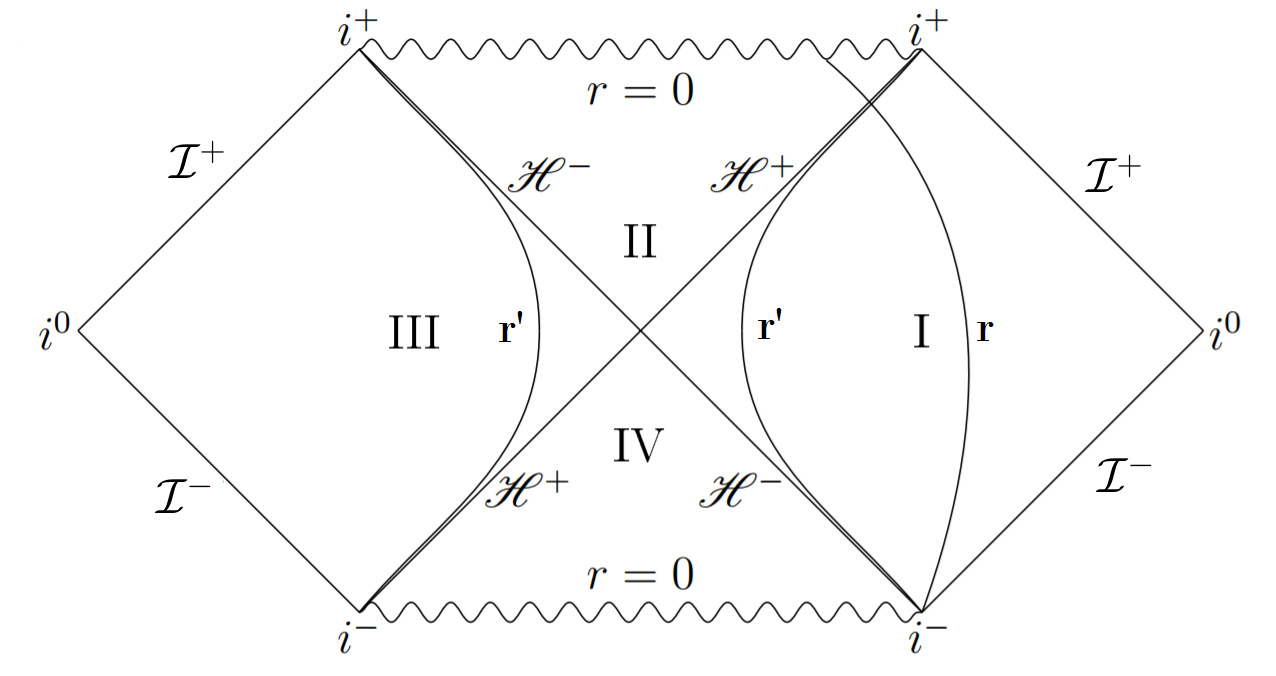}
			\caption{The maximally extended conformal (Penrose) diagram of the Schwarzschild black hole. The curved lines indicated by $r$ and $r'$ indicate locally inertial frames in Kruskal-Szekeres and Schwarzschild coordinates, respectively. Note the difference with Rindler space indicated in figure \ref{rindlerfig}, namely, the requirement that $r>0$ `cuts off' the top and bottom halves of regions $II$ and $IV$, respectively. Figure adapted from \cite{rindlerfig}.}
		\end{center}
	\end{figure}

	\subsection{Charged and rotating black holes}
	
	\subsubsection{Reissner-Nordstr\"{o}m black holes (charged)}
	
	We consider the following action
	
	\[ S=\int d^4x \sqrt{-g} \left( R -\underbrace{\frac{1}{4} g^{\alpha \mu }g^{\nu \beta} F_{\mu \nu} F_{\alpha \beta}}_\text{matter action} \right) ~, \] 
	
	 where $F_{\mu \nu}$ is the electromagnetic field strength tensor. The electromagnetic field equations are $\partial_{[\alpha} F_{\beta\gamma]} = 0$ and $\partial_{\mu} \left( \sqrt{-g} F^{\mu \nu}\right) = -\sqrt{-g} J^{\nu}$, where $J^{\nu}$ is the electromagnetic current which satisfies $\partial_{\nu} \left( \sqrt{-g} J^{\nu} \right) =0 $. By varying the action with respect to the metric, we find that the energy-momentum tensor is given by
	
	\begin{equation}
	T_{\mu \nu} = -F_{\mu \alpha}F^{\alpha}_{\nu} + \left( \frac{1}{4} F_{\alpha \beta} F^{\alpha \beta} - J_{\alpha}A^{\alpha} \right) g_{\mu \nu} ~.
	\end{equation}	

    	For a point-like electric charge, the field strength is given by     	$F_{rt} = \frac{Q}{r^2}$, where $Q$ is the electric charge. The unique closed form solution to $R_{\mu \nu} - \frac{1}{2}g_{\mu \nu}R = 8\pi G T_{\mu \nu}$ is given by
	
	\begin{equation}
	ds^2 = -\left(1-\frac{2M}{r}+ \frac{Q^2}{r^2}\right) dt^2 + \left(1-\frac{2M}{r}+ \frac{Q^2}{r^2}\right)^{-1}dr^2 + r^2 d\Omega_{(2)}^2 ~.
	\end{equation}
	
	This is known as the \textit{Reissner-Nordstr\"om} metric, which is characterized by parameters $(M,Q)$. We again have a curvature singularity (characterized by diverging Kretschmann scalar) at $r=0$. However, we now have two horizons, located at $r=r_{\pm} = M\pm \sqrt{M^2 - Q^2}$. The metric signatures of the Reissner-Nordstr\"{o}m solution are given by

	\[ r>r_+: ~(-,+,+,+)~, ~~~~~~~  r_-<r<r_+~:~(+,-,+,+) ~, ~~~~~~~r<r_-: ~(-,+,+,+)~ . \]
	

 From the metric signatures we can see that that time-like geodesics do not necessarily end at $r=0$ since they are forced to move in the direction of decreasing $r$ only for $r_- < r < r_+ $. The singularity at $r=0$ forms a time-like line and can thus be avoided, as opposed to uncharged (Schwarzschild) black holes, where $r=0$ is space-like. The Reissner-Nordstr\"{o}m metrix has two important limits
	
		\begin{enumerate}
		\item \textbf{Zero charge limit}: $Q \rightarrow 0$, where we retrieve the Schwarzschild solution. Then, $r_+=2M$ and $ r_- = 0$.
		\item \textbf{Extremal limit}: $Q^2 = M^2$, so that $r_- = M = r_+$ .
	\end{enumerate}

	Since $r_+=r_-$ for the extremal black hole, the signature of the metric is $(-,+,+,+)$ everywhere in this space-time. The metric is
	
	\begin{equation}
	ds^2_{\text{Extremal}} = -\left(1-\frac{M}{r}\right)^2 dt^2 + \left(1-\frac{M}{r}\right)^{-2}dr^2 + r^2 d\Omega^2_{(2)}~.
	\end{equation}
	
	Hence we see directly that the metric does not change signature at the horizon at $r=M$. Extremal black holes can be coupled to fermions such that it preserves \textit{supersymmetry}, hence they are often (somewhat erroneously) referred to as a \textit{supersymmetric black holes}. $M^2 = Q^2$ is a limiting case in the sense that we require $M^2 \geq Q^2$. For $M^2 < Q^2$, we get a naked singularity, thus violating the cosmic censorship conjecture. That we have a naked singularity can be seen from the fact that the metric is completely regular for $r\rightarrow 0 $, while we still have a space-time singularity at $r=0$ . A bound such as $M^2 \geq Q^2$ is generally referred to as a \textit{BPS-bound}, we will return to this point in the lectures on black holes in string theory.

	\clearpage
	
	\section{Rotating black holes and black hole mechanics}
	
	In this section, we consider rotating black holes as well as black hole mechanics and start exploring its similarities with thermodynamics.

	\subsection{Rotating (Kerr) black hole}
	
	So far, we considered spherically symmetric black holes. We will now considera black hole with some non-zero angular velocity along azimuthal angle $\phi$. The metric will then depend on $\theta$ in a non-trivial way, but it will not depend on $\phi$ i.e. it will be axially symmetric.
	
	\underline{Remark}: A slowly rotating body at large $r$ has the following metric
	
	$$ ds^2 = \underbrace{ds_0^2}_\text{non-rotating metric}+ \frac{4J}{r} \sin^2 \theta dt d\phi $$
	
	Where $J$ is the angular momentum of the rotating body. We now check that this is reproduced by the full solution, like we did in the case for the metric of a spherical body at large $r$. The full solution is
	
	\begin{align}
	ds^2 = &-\left(1- \frac{2Mr}{r^2+a^2 \cos^2 \theta} \right) \left( du + a \sin \theta d\phi \right)^2 + 2 \left( du +  \sin^2 \theta d \phi\right) \left( dr + a\sin^2 \theta d\phi \right) + \notag \\
	&+  \left(r^2 a^2 \cos^2 \theta \right) \underbrace{\left( d\theta^2 + \sin^2 \theta d \phi^2 \right)}_{d\Omega^2_{(2)}}~,
	\end{align}
	
	where, again $u=t-r -2M \log \left( \frac{r}{2M}-1\right)$. This solution was found by Kerr in 1963. The fact that it took almost 50 years since Einstein published his gravitational equations for someone to discover this solution is probably due to the fact that it has a rather complicated mixing between $t$ and $\phi$. If we redefine our coordinates as $t \to t- 2M \int \frac{r dr}{r^2-2Mr+a^2} ,~ \phi \to - \phi- a \int \frac{r dr}{r^2-2Mr+a^2}$, the metric becomes

	\begin{equation}
	ds^2 = -\left( 1 - \frac{2Mr}{\rho^2}\right) dt^2 - \frac{4Mar \sin^2 \theta}{\rho^2} d\phi dt + \frac{\rho^2}{\Delta}dr^2 + \rho^2 d\theta^2 + \left( r^2 + a^2 + \frac{2Mra^2 \sin^2 \theta}{\rho^2} \right) \sin^2 \theta d\phi^2~,
	\end{equation}

	with $\rho^2 = r^2 + a^2 \cos^2 \theta~~, ~~~ \Delta = r^2 - 2Mr + a^2~~,~~~ a= \frac{J}{M}$.\\
	 
	 
We remark on two limits of parameters $a$ and $M$

	 \begin{enumerate}
	 	\item $a\rightarrow 0$. Then $\rho \rightarrow r$, $\Delta = r^2 - 2Mr$, this gives the Schwarzschild metric.\\
	 \item $M\rightarrow 0 $. This gives flat Minkowski in so-called \textit{oblate spheroidal coordinates}. 
	\end{enumerate}

The Kerr metric does not depend on $t$ or $\phi$ and is invariant under the combination $t\rightarrow -t ~ \& ~ \phi \rightarrow -\phi$. The horizons are located at the points where $g_{rr} \to \infty$ i.e. where $\Delta \to \infty $. We thus find two horizons, located at

\begin{equation}
r_{\pm} = M \pm \sqrt{M^2 - a^2}~.
\end{equation}

As for the Reissner-Nordstr\"{o}m black hole, the metric signature is $(+,-,+,+)$ for $r_- < r < r_+$, and  $(-,+,+,+)$ for $r< r_- $ and $r> r_+$.
We require $M^2 \geq a^2 \sim M^2 \geq \lvert J \rvert$. If this bound is violated the solution is again unphysical since we have a naked singularity. The BPS-saturated (extremal) black hole is given by $M^2 = J^2$.

\subsection{Kerr-Newman black hole}

We can generalize the Kerr metric to include electric charge $Q$ and magnetic charge $P$ by simply replacing $2Mr$ with $2Mr - (Q^2 + P^2)$. We will set $P=0$ here, as is customary. The result is known as the \textit{Kerr-Newman metric}, which is given by

\begin{align}
ds^2 =&  -\left( 1 - \frac{2Mr - Q^2}{\rho^2}\right) dt^2 - \frac{2a \left( 2Mr - Q^2 \right)^2 \sin^2 \theta }{\rho^2} d\phi dt + \frac{\rho^2}{\Delta} dr^2 \rho^2 d\theta^2 + \notag\\
& + \left( r^2 + a^2 + \frac{ (2Mr - Q^2 ) a^2 \sin^2 \theta}{\rho^2} \right) \sin^2 \theta d\varphi^2 ~.
\end{align}

Where $a \coloneqq \frac{J}{M}$, $\rho = r^2 + a^2 \cos^2 \theta$, $\Delta = r^2 + a^2 -2Mr + Q^2 $. For this metric, we have

\begin{equation}
d M = \frac{1}{16\pi}\frac{r_+ - r_-}{r_+^2+a^2}dA + \underbrace{\frac{a}{r_+^2+ a^2 }}_{\Omega_H} dJ + \underbrace{\frac{Q r_+}{r_+^2 + a^2 }}_{\Phi} dQ ~,
\end{equation}

where $\Omega_H = \frac{d\phi}{dt} \bigg \rvert_{r=r_H} = \frac{a}{r_+^2+ a^2 }$ can be interpreted as the angular velocity.

\subsection{Laws of black hole thermodynamics (mechanics)}

We will now compare black hole mechanics with thermodynamics. In particular, we will see during the next lectures that black holes are an objects with a finite temperature that lose energy via thermal (Hawking) radiation. We will see that for a classical black hole $( \hbar = 0)$, $T=0$ and there is no radiation. That black holes emit thermal radiation is thus an inherently quantum-mechanical phenomenon.

\subsubsection{Zero'th law}


Recall the definitions of null-hypersurfaces and Killing horizons. Take a set of hypersurfaces $\{ S\}$ i.e. of 3d achronal submanifolds of 4d space-time characterized by $S\left( x^{\mu}\right) =0$, where $S$ is some function. A vector field normal to $S$ is given by $\ell^{\mu} = N(x) \partial^{\mu} S(x)$, where $N(x)$ gives a normalization. A null hypersurface $\Sigma$ is then characterized by 

\[\ell^2  \bigg\rvert_{\Sigma} = \ell^{\mu} \ell_{\mu}  \bigg\rvert_{\Sigma} = 0~,  \] 

 i.e. its normal vector $ \ell$ is orthogonal to itself. We then consider the \textit{Lie derivative}, which is the derivative of a tensor along a vector field. Namely, the Lie derivative tells us how a space-time tensor field changes when we take an infinetesimal step along a vector field defined in that space-time. The Lie derivative of the metric along a \textit{Killing vector field} $\xi$ vanishes i.e. $ \left( \mathcal{L}_\xi g \right)_{\mu \nu} =0$, with $ \left( \mathcal{L}_\xi g \right)_{\mu \nu}  = \xi^{\lambda } \partial_{\lambda} g_{\mu \nu} + \partial_{\mu } \xi^{\lambda}g_{\nu \lambda} + \partial_{\nu } \xi^{\lambda}g_{\mu \lambda}$. A \textit{Killing horizon} is then given by a null hypersurface $\Sigma$ for which there exists a Killing vector field that is normal to it i.e.  for some $\xi^{\mu } = f(x) \ell^{\mu}(x)$ defined at $\Sigma$, we have $\xi^2 \bigg\rvert_{\Sigma} = f^2 \ell^2 \bigg\rvert_{\Sigma} = 0$. One can show that

 $$\xi^{\sigma}\nabla_{\sigma} \xi^{\mu} \bigg\rvert_{\Sigma} = \kappa \xi^{\mu} ~,$$

where $\kappa$ is the \textit{surface gravity} of the black hole. For the Schwarzschild metric, we have $\kappa = \frac{1}{4M}$. For an asymptotically flat space-time, $\kappa$ gives the acceleration of a static observer at the Killing horizon. Another useful formula is

\[\kappa^2  = - \frac{1}{2} \left( \nabla^{\alpha} \xi^{\beta} \right) \left( \nabla_{\alpha} \xi_{\beta} \right) \bigg\rvert_{\Sigma}~.\]

A shortcut for computing $\kappa$ for Schwarzschild goes as follows. Take Euclidean Schwarzschild i.e. let $t \rightarrow i \tau $ (Wick rotation). The metric is then

\[ ds^2 = \left( 1 - \frac{2M}{r} \right)d\tau^2 + \left( 1 - \frac{2M}{r} \right)^{-1}dr^2 + r^2 d\Omega_{(2)}^2 ~. \]

This is equivalent to considering our system at finite temperature. The periodicity of $\tau$ gives the temperature. We now examine near-horizon region of black hole by introducing $\frac{x^2}{8M } = r-2M$. This gives

\[ds^2 = (\kappa x)^2 d\tau^2 + dx^2 + \kappa d\Omega_{(2)}^2 ~,\]

where $\kappa = \frac{1}{4M}$. This is the metric of $\mathbb{R}^2$ in angular coordinates. $\tau$ thus has the periodicity $\tau \sim \tau + \frac{2\pi}{\kappa} \eqqcolon \tau + \beta $, from which we see that $\beta = T^{-1} = \frac{2\pi}{\kappa}$. 
	
\clearpage

\section{Black hole mechanics and thermodynamics}

In the 1970's, Bekenstein realized that, assuming the second law of thermodynamics holds, a black hole must carry entropy \cite{bekenstein}. This follows from the fact that if we throw an object with some entropy into a black hole, the entropy of the total system may not decrease, hence the entropy of a black hole must grow when we throw an object into it. Such thought experiments can be used to derive the laws of black hole mechanics, which turn out to be profoundly connected to the laws of thermodynamics. In particular, certain parameters from black hole physics will be seen to correspond to thermodynamic quantities. We state the laws of black hole mechanics and the corresponding laws of thermodynamics

\begin{center}
	\begin{tabular}{ |c|c|c| } 
		\label{thermotable}
		\textbf{Law} & \textbf{Thermodynamics} & \textbf{Black hole mechanics} \\ 
		\hline
		0 & T constant in equilibrium & $\kappa=$constant \\  
		1 &   $\delta E = T \delta S$ & $\delta M = \frac{\kappa}{8\pi}\delta A + \Omega \delta J + \Phi \delta Q$ \\
		2 & $\delta S \geq 0 $ & $\delta A \geq 0$   \\
		3 & Cannot let $T \rightarrow 0$ in a finite number of steps & Cannot let $\kappa \rightarrow 0$ in a finite number of steps 
	\end{tabular}
\end{center}

\vspace{.3cm}
During the next few lectures we will further explore the connection between the laws of thermodynamics and black hole mechanics.

	
	\subsection{First law of black hole mechanics}
The \textit{first law of black hole mechanics} tells us how the mass of a black hole changes with its horizon area, charge, and angular momentum. It reads

\begin{equation}
\delta M = \frac{\kappa}{4\pi}\delta A + \Omega \delta J + \Phi \delta Q ~.
\end{equation}

 We will show that this is the correct expression for stationary, axisymmetric, and asymptotically flat space-times. We need our space-time to have these properties so that mass and angular momentum are well-defined. To find an expression for the mass of a black hole, we first consider the example of electromagnetism. The electric charge inside a volume $V$ is given by
	
	\[Q\left(V \right) = \int_V \rho dV = \int_V \nabla \cdot \vec{E} dV = \oint_{\partial V} \vec{E} \cdot  \hat{n} dS~,\]
	
	where $\hat{n}$ is the normal vector to $\partial V$.	The covariant generalization of this expression is
	
	\[  Q\left( V \right) = \int_V d V \sqrt{\gamma} ~ n_{\mu} j^{\mu} = \int_V  d V \sqrt{\gamma} ~n_{\mu} \nabla_{\nu} F^{\mu \nu} = \oint_{\partial V} d S \sqrt{\tilde{\gamma}}~ n_{\mu} \sigma_\nu F^{\mu \nu}~,\]
	
	where $\gamma$ and $\tilde{\gamma}$ are induced three- and two-dimensional metrics with corresponding unit normals given by $n_\mu$ and $\sigma_\nu$, repectively. 
	
	We now consider general relativity. The \textit{Komar integral} associated to a Killing vector $\xi$ is given by
	
	\[ Q_{\xi} \left( V \right) = - \frac{1}{4\pi} \oint_{\partial V} d S \sqrt{\tilde{\gamma}}~ n_{\mu} \sigma_\nu  \left(\nabla^{\mu } \xi^{\nu}\right) _{\mu \nu}~. \]
	
 Comparing this expression with that of electromagnetism, we note that $\nabla^{\mu } \xi^{\nu} = - \nabla^{\nu } \xi^{\mu}$, similar to $F^{\mu \nu} = - F^{\nu \mu}$. We now use $ \nabla_{\nu} \nabla_{\mu} \xi^{\nu} = R_{\mu \nu} \xi^{\nu}$, which holds for any Killing vector $\xi^{\nu}$, to rewrite
	
	\[Q_{\xi} \left( V \right) = -\frac{1}{4\pi} \int_{ V} d V \sqrt{\gamma} ~ n_{\mu} R^{\nu}_{~ \mu} \xi^{\mu} ~. \]
	
	 \underline{Remark}: If $\nabla^a T_{ab} = 0$ i.e. if energy and momentum are conserved, then $T_{ab} \xi^b=0$. From this it follows that $Q_{\xi}$ is a conserved quantity. Hence, the Komar mass i.e. the Komar integral associated with the time-like Killing vector $\xi$ is
	 
	 \begin{equation}
	 M_{\xi} \left( V \right) = - \frac{1}{4 \pi } \oint_{\partial V} d S \sqrt{\tilde{\gamma}}~ n_{\mu} \sigma_\nu \nabla^{\mu} \xi^{\nu} ~,
	 \end{equation} 
	
The mass $M$ is subject to the \textit{positive energy theorem}, due to Witten \& Yau, which reads as follows


\emph{If $T_{ab}$ satisfies the \text{dominant energy condition} i.e. every observer observes strictly positive energy, then $M_{ADM} \geq 0$, and $M_{ADM} = 0$ only for Minkowski space-time. The dominant energy condition entails that $T_{ab} u^a \geq 0$ for any $u^a$ that is a tangent vector to a general future-directed causal curve.}


If, instead, we consider axial Killing vector $\phi^\nu = (\partial_\phi)^\nu$, then the Komar integral gives the angular momentum

\begin{equation}
J_{\phi} \left( V \right) = \frac{1}{8 \pi } \oint_{\partial V} d S \sqrt{\tilde{\gamma}}~ n_{\mu} \sigma_\nu \nabla^{\mu } \phi^{\nu}  ~.
\end{equation}

\subsection{Smarr's formula}

We use $B$ to refer to the interior of a black hole, and $\Sigma$ to refer to the exterior. We assume $\partial B$ is a Killing horizon for $K = \xi  - \Omega_H \phi $, where we take into account possible non-zero rotation of the black hole i.e. $\Omega_H \neq 0 $ for Kerr(-Newman) black holes. This gives

\begin{align*}
M_k\left( S^2_{\infty} \right) & = - \frac{1}{4 \pi }  \int_{S^2_{\infty}} \nabla^{\mu} \xi^{\nu} dS_{\mu \nu}\\
 & = -\frac{1}{4\pi} \int_{ \Sigma} R^{\mu}_{~ \nu} K^{\nu} dS_{\mu} + \frac{1}{4\pi} \int_{ \partial B } \nabla^{\mu} K^{\nu}dS_{\mu \nu} \\
 & = \int_{ \Sigma} 2\left( T^{\mu}_{\nu} - T\delta^{\mu}_{\nu} \right) K^{\nu} dS_{\mu} + 2\Omega_H J + \frac{1}{4\pi} \int_{ \partial B } \nabla^{\mu} \xi^{\nu} dS_{\mu \nu}  ~.
\end{align*}

Using $\xi^{\mu}\nabla_{\mu}\xi^{\nu} = \kappa \xi^{\nu}$, an expression encountered previously in the discussion of Killing vectors, we can rewrite the last term in the last expression to find Smarr's law

\begin{equation}
M = \Phi Q + 2 \Omega_H J + \frac{\kappa}{4\pi} A~,
\end{equation}

where $\Phi$ is the electromagnetic potential between $r= \infty$ and $r=2M$. There are two ways to use Smarr's law to arrive at the first law of black hole mechanics. We will first derive it via the more complicated method. Assuming $Q=0$, we have

\[\delta M = \frac{1}{4\pi} \left( A \delta \kappa + \kappa \delta A \right) + 2\left( \delta \Omega_H J + \Omega \delta J \right) ~.\]

 We also have $- \delta M = \frac{1}{4\pi } \delta \kappa + 2 \delta \Omega_H J$, which one can derive. Combining the two gives the first law of black hole mechanics
 
 \begin{equation}
 \delta M = \frac{\kappa}{8\pi}\delta A + \Omega \delta J ~. 
 \end{equation}

The alternative derivation of the first law goes as follows. We use the uniqueness theorem, which states that the surface area of a black hole is a (non-trivial) function of only the mass, angular momentum, and charge. Since we set $Q=0$, this can be written as $A=A(M,J)$, akin to $M=M(A,J)$. Now note that $A$ and $J$ are both proportional to the $( \text{Komar mass})^2$ from dimensional reasons. We thus have $\alpha M = M \left( \alpha^2 A, \alpha^2 J \right)$ i.e. $M(A,J)$ is a homogeneous function of degree $\frac{1}{2}$. Hence $A\frac{\partial M }{\partial A} + J \frac{\partial M}{\partial J }= \frac{1}{2}M = \frac{\kappa}{8 \pi} A + \Omega_H J$, so that

$$ A\left( \frac{\partial M}{\partial A} - \frac{\kappa}{8 \pi }\right) + J \left( \frac{\partial M}{\partial J} - \Omega_H \right) = 0~.$$

Since $A$ and $J$ are independent, the terms multiplying them in the above expression are both equal to zero. We thus find

\begin{align}
\delta M &= \frac{\partial M}{\partial A} \delta A + \frac{\partial M}{\partial J} \delta J \notag \\
 & = \frac{\kappa}{8 \pi } \delta A + \Omega_H \delta J ~.
\end{align}

\clearpage

\section{Black hole thermodynamics}

\subsection{Previously: Zero'th and first laws} 

During last lecture, we discussed black hole thermodynamics and mechanics. The zero'th law states that surface gravity is constant over an event horizon. For our discussion of the first law we considered Komar quantities. For some surface $S \subset  M$ with volume element $dS_{\mu}$ with boundary $\partial S $ with volume element $dS_{\mu \nu}$ and $\xi$ a Killing vector, we have general Komar quantity

\begin{equation}
Q_k = -\frac{1}{4\pi} \int_{ \partial V = S} d S \sqrt{\tilde{\gamma}}~ n_{\mu} \sigma_\nu \nabla^{\mu} \xi^{\nu}~.
\end{equation}

E.g. for $k \sim \partial_t$ or $\tilde{k} \sim \partial_{\phi}$, we find the Komar mass or angular momentum, respectively

\begin{align}
M &= - \frac{1}{4\pi} \int_{ S} d S \sqrt{\tilde{\gamma}}~ n_{\mu} \sigma_\nu \nabla^{\mu} k^{\nu} ~, \notag \\
J &= - \frac{1}{8\pi} \int_{S}d S \sqrt{\tilde{\gamma}}~ n_{\mu} \sigma_\nu \nabla^{\mu} \tilde{k}^{\nu}~.
\end{align}

The total mass inside some $S$ is then

\begin{equation}
M = \Phi Q + 2 \Omega_H J + \frac{\kappa}{4\pi} A ~.
\end{equation}

Combining these gives the first law

\begin{equation}
dM = \frac{\kappa \left( M,J,Q \right)}{8 \pi}dA + \Omega \left( M,J,Q \right) dJ + \Phi\left(M,J,Q \right) dQ ~,
\end{equation}

where $\Omega$ is the angular velocity at the horizon, and $\kappa$ and $\Phi$ are the acceleration and electric potential at the horizon with respect to a stationary asymptotic observer.

\subsection{Second law}

A natural question concerning black holes is whether there are restrictions on the possible charges, areas, angular momenta, and masses of black holes for a given set of boundary conditions. This question is relevant for matter that collapses to a black hole, or generally for dynamical problems. The answer is that there are indeed restrictions, for example, the BPS-bound gives $J^2 \leq M^2$, and the same for $Q$. Further, we have Hawking's area theorem, which states that

\begin{equation}
\frac{dA}{dt} \geq 0  ~.
\end{equation}

 This expression holds assuming that our space is asymptotically flat, we have cosmic censorship i.e. no naked singularities, and the weak energy condition holds i.e. $T_{\mu \nu} v^{\mu} v^{\nu} \geq 0$, where $v^{\mu}$ is an arbitrary time-like vector.

\underline{Proof} (relies on causality): A \textit{causal curve} is a curve that is nowhere space-like. We define the \textit{causal past} of some surface $S$ as $J^- (S) \coloneqq \{ x \in \mathcal{M} \mid \exists s \in S , x \leq s \} $, where $x \leq s$ means that the time coordinate of $x$ is less than or equal to $x$. The \textit{causal future} is then defined as $J^+ (S) \coloneqq \{ x \in \mathcal{M} \vert ~ \exists s \in S , x\geq s \} $.  We now consider \textit{chronological curves} which are defined to be everywhere time-like. The \textit{chronological past} and \textit{future} are then given by $I^{-} \coloneqq \{ x \in \mathcal{M} \mid \exists s \in S , x < s \}$ and $I^+ (S) \coloneqq \{ x \in \mathcal{M} \mid \exists s \in S , x > s \} $. The boundary of the causal past is $\partial J^-(S) = J^-(s) / I^- (S)$, $\partial J^- (S)$ is generated by a set of null geodesics, which are referred to as the \textit{null geodesic generators} of $\partial J^- (S)$. There is a lemma due to Penrose (1967) which states that in any subset $S \subset M$, a null geodesic generator of $\partial J^- (S)$ cannot have future endpoints on $\partial J^- (S)$. In other words, the area of a trapped surface cannot decrease. Since the event horizon of a black hole is such a trapped surface, Hawking'are theorem follows trivially.



\subsubsection{Consequences for coalescing black holes}

We consider the limits of mass-energy conversion of a black hole collision (such as observed by LIGO). Consider two black holes with masses $M_1$ and $M_2$, respectively. They coalesce to form a third black hole, which has mass $M_3$. During this process, part of the total energy is converted to gravitational waves; this fraction of the total energy is given by $M_1 + M_2 - M_3$. We can then define an \textit{efficiency coefficient}  $\eta \coloneqq \frac{M_1+M_2-M_3}{M_1+M_2} = 1- \frac{M_3}{M_1+M_2}$. Black hole surface area is $A= 16 \pi  M^2$, so Hawking's area theorem tells us that $A_3  = 16 \pi M_3^2 \geq 16 \pi \left( M_1^2 + M_2^2 \right)$, hence $M_3 \geq \sqrt{ M_1^2 + M_2^2}$ i.e. 

\begin{equation}
\eta \leq 1 - \frac{\sqrt{ M_1^2 + M_2^2}}{M_1+M_2} = 1- \frac{1}{\sqrt{2}}~.
\end{equation} 

One then easily sees that a single black hole cannot split up into two separate black holes. The exceptions are BPS black holes, which can freely split up without decreasing total entropy.

\subsection{Third law of black hole mechanics}

The third law of black hole mechanics states that it is impossible to let the surface gravity $\kappa$ go to zero in a finite number of operations. We give this statement without proof, but it can be seen from the fact that we need to let $M \to \infty $ or $(a^2 + Q^2 + P^2) \to M^2$. We thus need to add an infinite amount of (BPS-saturated) matter. We thus see that extremal (BPS) black holes cannot be formed continuously since they have $\kappa =0$. This is related to supersymmetry, since BPS black holes can be made invariant under supersymmetry while the non-BPS black holes cannot.

\clearpage

\section{Black holes and entropy}

So far, we have considered the laws of black hole mechanics and briefly discussed their correspondence to the laws of thermodynamics. We now wish to consider the laws of black hole mechanics as thermodynamical statements. This means that we will define a temperature and an entropy for our black holes. In thermodynamics, we may consider e.g. the grand canonical ensemble, where the system is characterized by quantities $\mu, V, T$, nammely chemical potential, volume, and temperature, respectively. The first law of thermodynamics is then

$$dE = TdS - pdV + \mu dN~,$$

where $p$ and $N$ are pressure and particle number, respectively. The first law of black hole mechanics is  

$$dM = \frac{\kappa}{8\pi} dA + \Omega dJ + \Phi dQ~.$$ 

By comparing the two expressions, we see that $-pdV+\mu dN$ is analogous to $ \Omega dJ+ \Phi dQ$, so that $\frac{\kappa}{8\pi}dA$ is the entropy term. Moreover, Hawking's area theorem tells us that $\frac{dA}{dt} \geq 0$ which further solidifies the anology between $A$ and $S$. The precise relation will involve $\hbar$, signalling the importance of quantum effects. The full expression is Bekenstein-Hawking area law

\begin{equation}
S= \frac{c^3A}{4  G_N \hbar}~.
\end{equation}

Hence $S$ diverges for $\hbar \rightarrow 0$. We can rewrite this as

$$S = \frac{A}{4 \pi L_p^2} \sim \frac{1}{4\pi} \frac{A}{\left(10^{-33}\right)^2}~.$$

Where $L_p$ is the Planck length $ \sim 10^{-33}$  cm. E.g. for a solar mass black hole with $r_s \sim 3$ km, $S \approx 10^{77}$.

In statistical physics, entropy is interpreted as information. The von Neumann entropy is given by $S = -  \sum_n  p_n \log p_n$, where $p_n$ are probabilities satisfying $\sum_n p_n =1$. Entropy is related to information (quantum bits). One naturally wonders what the quantum bits of black holes are i.e. what are the carriers of black hole information. People have speculated that the quantum bits are somehow `distributed' over the horizon, since $S \propto A$. This area law has inspired the principle of holography, which is the idea that gravitational degrees of freedom are dual to degrees of freedom on a holographic `screen' which has one dimension lower than the gravitational system. An explicit example of holography is the AdS/CFT correspondence \cite{adscft}, where the gravitational bulk and the holographic screen are the AdS space and the boundary CFT, respectively.

\clearpage

\section{Hawking and Unruh radiation}

We saw that there are striking similarities between thermodynamics and black hole mechanics as summarized in table \ref{thermotable}. These were considered by many to be mere surface similarities, until Hawking showed that semiclassical black holes emit thermal radiation at inverse temperature $\beta = \frac{2\pi}{\kappa}$ \cite{hawking}. That is, black holes seem to be truly thermodynamic objects which radiate at a well-defined temperature. However, this immediately poses a problem, as thermal radiation is in a mixed qunatum state, which means we need a density matrix to describe it. Hence, if a black hole formed out of a pure state evaporates into mixed thermal radiation, we have a pure-to-mixed state transition, which violates the unitarity postulate of quantum mechanics. There thus seems to be a conflict between quantum mechanics and general relativity, which presents perhaps the most important unsolved problem of contemporary theoretical physics. The remaining sections will look at the origin of this problem as well as some partial solutions that have been proposed thus far.

To investigate Hawking and Unruh radiation \cite{hawking} \cite{unruh} we take a semiclassical approach, which means that we quantize fields on a classical curved background space-time. A full treatment would include the back-reaction of quantized fields on the metric, but such calculations are typically very complicated, if not impossible. Before performing the semiclassical calculation, we can make a simple estimate of the Hawking temperature by using Wien's law $ \lambda = \frac{ \hbar c}{k_B T}$. If we then take $\lambda$ to be the Schwarzschild radius $\lambda = r_s = \frac{2GM}{c^2}$, we find $T = \frac{\hbar c^3 }{2k_B GM }$, which is a factor $4\pi$ larger than the real value. We see that the temperature goes to zero when $\hbar$ goes to zero, signalling that Hawking radiation is a quantum-mechanical effect. The heuristic picture of Hawking radiation is that we have production of a particle-antiplartile pair sufficiently close to the horizon that one of the particles passes through the horizon and falls inward to the singularity. The other particle of the pair is maximally entangled with the first, and as a consequence it propagates outward to radial infinity. The outward propagating particles constitute Hawking radiation, at Hawking temperature $T_H = \frac{\hbar c^3 }{8 \pi G M k_B }$. In Rindler space-time we have Unruh radiation associated to the fact that there is a horizon in Rindler space. We have the corresponding Hawking radiation in Schwarzschild space-time. In both cases the existence of a horizon obscures our notion of the vacuum as well as particle number, as we will see below. We first review a few standard ideas of QFT in Minkowski space. 

\subsection{Free scalar field in Minkowski space}

We have $ds^2=-dt^2 + d\vec{x}^2$ and $S\left[ \phi \right] = -\frac{1}{2} \int d^4 x \left[ \eta^{\mu \nu} \partial_{\mu} \phi(x) \partial_{\nu} \phi(x) + m^2 \phi^2 \right]$. The equation of motion is

 $$ \partial^2_t \phi - \nabla^2 \phi + m^2 \phi =0~.$$

We perform a Fourier transformation by writing $ \phi(t, \vec{x}) = \int \frac{d^3 \vec{k}}{(2\pi)^{3/2}} e^{i \vec{k} \cdot \vec{x}} \phi_{\vec{k}} (t) $, which gives 

$$ \ddot{\phi}_{\vec{k}} + \left( k^2 + m^2 \right) \phi_{\vec{k}} = 0~.$$

We now perform canonical quantization, using $\pi \left( t , \vec{x} \right) = \partial_t \phi \left( t , \vec{x} \right)$ and promoting $\phi$ and $\pi $ to operators $\hat{\phi}$ and $\hat{\pi}$. The equal time commutators are then

\begin{equation}
\label{harmosc}
 \left[ \hat{\pi} \left( t , \vec{x} \right),\hat{\phi} \left( t , \vec{y} \right) \right] = i \hbar \delta \left( \vec{x} - \vec{y} \right)~~,~~~ \left[ \hat{\phi}, \hat{\phi}\right] =0= \left[ \hat{\pi}, \hat{\pi}\right] ~.
  \end{equation}

We make the ansatz that we can write

\begin{equation} 
\hat{\phi} ( \vec{x}, t) = \int \frac{d^3 \vec{k}}{(2\pi)^{2} \sqrt{2\pi \omega}} \left[ v^*_{\vec{k}}(t) \hat{a}^-_{\vec{k}} e^{i \vec{k}\vec{x}} + v_{\vec{k}}(t) \hat{a}^{+}_{\vec{k}} e^{ - i \vec{k}\vec{x}}  \right]~,
\end{equation}

 i.e. that we can split the scalar field into positive and negative frequencies. Plugging this into equation \ref{harmosc} gives

\begin{equation}
\label{wrcond}
\left[ \hat{a}^-_{\vec{k}} , \hat{a}^{+}_{\vec{p}}  \right] =  \delta \left( \vec{k} - \vec{p} \right)~~ , ~~~\left[ \hat{a}^- , \hat{a}^- \right] = 0 = \left[ \hat{a}^+ , \hat{a}^+ \right]~~ , ~~~\dot{v}_k(t) v_k^*(t) - \dot{ v}^*_k(t) v_k(t) = 2i ~ ,
\end{equation}

where the last equality of \ref{wrcond} is known as the \emph{Wronski condition}. We implicitly define the vacuum $\ket{0}$ in terms of the annihilation operators as $\hat{a}^-_k \ket{0} = 0$. The mode functions $v_{\vec{k}}$ are not completely defined by the commutation relations. The operators $\hat{a}^-_{\vec{k}}$ and $\hat{a}^+_{\vec{k}}$ are also not unambiguously defined. The extra condition we require is that the vacuum is the state of lowest energy of the Hamiltonian, which is given by

$$ \hat{H}(t) = \frac{1}{2} \int d^3 k \left[ \left( \dot{v}_k^2 + \omega_k^2 v_k^2 \right) \hat{a}_{\vec{k}}^+ \hat{a}_{-\vec{k}}^+ + \text{h.c.} + \left( \lvert \dot{v}_{\vec{k}} \rvert^2 + \omega_k \lvert v_{\vec{k}} \rvert^2 \right) \left( 2\hat{a}_{\vec{k}}^+ \hat{a}_{\vec{k}}^- + \delta^{(3)}(0) \right) \right] ~, $$

here the $\delta(0)$ gives an infinite contribution to the energy and $\omega_k$ are the frequencies. The vacuum expectation value of $H(t)$ is then

$$ \obket{0}{\hat{H}}{0} =  \int d^3 \vec{k} \left( \lvert \dot{v}_k \rvert^2 + \omega_k \lvert v_k \rvert^2 \right)~. $$

The energy density is therefore $ \epsilon = \lvert \dot{v}_k \rvert^2 + \omega_k \lvert v_k \rvert^2$. A solution of the energy minimization is 

\begin{equation}
v_k(t) = \sqrt{ \frac{1}{\omega_k}} e^{i \omega_k t} ~~, ~~~ \omega_k\text{ = frequencies~.}
\end{equation}

We define the annihilation (creation) operators coresponding to positive (negative) frequency modes. Plugging the solution back into Hamiltonian gives

$$ \hat{H}(t) = \int d^3 k \omega_k \hat{a}_k^+ \hat{a}^-_{- \vec{k}} ~.$$

The momentum operator is given by $p^i = - \int d^3 \vec{x} \pi \partial_i \phi = - \int \frac{d^3 \vec{p}}{(2\pi)^3} \vec{p} a^+_p a^-_{-p}$, excited states with energy $\omega_k $ and momentum $\vec{k}$ are given by $\hat{a}^+_{\vec{k}} \ket{0}$. The \textit{Fock space} is then spanned by

\begin{equation}
\ket{ n_1, n_2, \dots} = \frac{1}{\sqrt{n_1! n_2! \dots}} \left( \hat{a}^+_{\vec{k}_1} \right)^{n_1} \left( \hat{a}^+_{\vec{k}_2}\right)^{n_2} \dots \ket{0} ~.
\end{equation}

For QFT in flat space, the vacuum, particle numbers, and momenta/energies are all well-defined quantities.  We will see that this is no longer the case for curved spaces.


\clearpage

\section{Quantum field theory in curved space-time backgrounds  }

\subsection{Last time: Minkowski space with free scalar field}

The scalar field action is given by $S\left[ \phi \right] = - \frac{1}{2} \int d^4x \left[ \eta^{\mu \nu } \partial_{\mu} \phi \partial_{\nu} \phi + m^2 \phi^2 \right]$. We promote the field $\phi( t, \vec{x})$ to an operator $\hat{\phi} ( t, \vec{x})$ with associated creation and annihilation operators, which we can then make time-dependent as

\begin{align*}
\begin{rcases}
\hat{a}_{\vec{k}}^{+} & \rightarrow \hat{a}_{\vec{k}}^{+}(t) \coloneqq e^{i \omega_k t} \hat{a}_{\vec{k}}^{+} \\
\hat{a}_{\vec{k}}^{-} & \rightarrow \hat{a}_{\vec{k}}^{-}(t) \coloneqq e^{- i \omega_k t} \hat{a}_{\vec{k}}^{-}
\end{rcases}
\text{Dispersion relation:} ~ \omega_{\vec{k}} = c \lvert \vec{k} \rvert ~~,~~~ m=0~.
\end{align*}

The vacuum state $\ket{0_M}$ is implicitly defined by $\hat{a}^-_{\vec{k}} \ket{0_M} = 0$, the Fock space is given by

$$ \ket{ n_1 ,n_2 , \dots} = \frac{1}{\sqrt{n_1!n_2! \dots }} \left( \hat{a}^+_{\vec{k_1}} \right)^{n_1}\left( \hat{a}^+_{\vec{k_2}} \right)^{n_2} \dots \ket{0_M} ~.$$

\underline{Remarks}

\begin{enumerate}
	\item  The vacuum energy is divergent as $E_0 = \sum_{\vec{k}} \frac{1}{2} \omega_{\vec{k}} ,~ \frac{E_0}{V } \int \frac{d^3 \vec{k}}{(2\pi)^3} \frac{1}{2} \omega_k \sim \int dk k^3$. We thus introduce a UV cut-off $k_{max} \sim M_{Planck}$ so that $\frac{E_0}{V} \approx \frac{M_{Planck}}{V} \approx 10^{94} g/cm^3$. The cosmological constant would then be $\Lambda \approx M_{Planck}^4$, which is much higher than the observed cosmological constant which corresponds to $\frac{E_0}{V} \approx 10^{55} g/cm^3$. This is referred to as the \textit{cosmological fine-tuning problem}, which is one of the main open problems in contemporary theoretical physics.

\item In Fock space language, the out-states are related to in-states by a scattering ($S$-) matrix as $\ket{\text{out}} = S \ket{\text{in}}$. In QFT, we require $S$ to be unitary. We will see that a black hole appears to give rise to non-unitarity. In Minkowski space, $\ket{0}_{in} \sim \ket{0}_{out}$. We will see that this changes when we go to curved space-times, where the in-states (out-states) are defined on $\mathcal{I}^-$ ($\mathcal{I}^+$)
 
\item Poincar\'{e} transformations act as $\hat{\phi}(t, \vec{x}) \rightarrow \phi'( t', \vec{x}')$ e.g. for a Lorentz boost, $ \vec{k} \vec{x} - \omega t = \vec{k}'\vec{x}' -\omega' t' $. This induces $a^-_{\vec{k}} \rightarrow b^-_{\vec{k}'} \simeq a^-_{\vec{k}}$ and $ \ket{0_M} = \ket{0'_M}$. This means that all inertial observers in Minkowski space will agree on the number of particles in Minkowski space, i.e. QFT in flat space-time has unambiguously defined vacuum and particle states. This will be markedly different when we go to curved space-times. 
\end{enumerate}
\vspace{.3cm}

In the context of general relativity there is no preferred coordinate system. Hence, if one observer sees well-defined particles with respect to a set of positive and negative frequency modes $a^+$ and $a^-$, respectively, another observer will generally see a different number of particles corresponding to new modes $b^+$ and $b^-$. We will also see that $\ket{0}_a \neq \ket{0}_b$ i.e. different observers will have different corresponding vacuum states. In general, $b$-modes are related to $a$-modes by \textit{Bogoliubov transformations} i.e. $a_k = c_1 b_k^- + c_2 b_k^+$,  $c_1,c_2 \in \mathbb{C}$.

In curved space-time, we have squeezing and rotation of light-cone as a consequence of varying gravitational potential, as well as $x \leftrightarrow t$ when we cross horizon. This indicates that we will mix positive and negative frequency modes. This is relevant in two instances:

\begin{enumerate}
	\item Accelerated observers in Minkowski space-time, who encounter a phenomenon called the \textit{Unruh effect}.
\item Curved space-time, in particular black holes. Here we will see the Hawking effect, the curved space analogue of the Unruh effect. 
\end{enumerate}

This discussion also applies in cosmology e.g. the FLRW space-time and particle-antiparticle creation at the big bang. The relevant ground state here is the so-called \textit{Bunch-Davies vacuum}. 

\subsection{Unruh effect}

We start from two-dimensional Minkowski space, with inertial observer corresponding to the metric $ds^2 = -dt^2 + dx^2$. We now go to Rindler space with velocity $ u^{\alpha} (\tau)= \frac{dx^{\alpha}}{d\tau} = \left( \dot{t}(\tau) , \dot{x}(\tau) \right) $ and constant acceleration $a$ given by $a^{\alpha}  = \dot{u}^{\alpha}$. Again, we will see that the notion of particles and the definition of positive and negative frequency modes depends on the observer. Namely, for the inertial observer, the modes are defined with respect to $t$, while for the accelerated (co-moving) observer, the modes are defined with respect to $\tau$. We will denote the latter observer by $\xi_0$. We now compare in 3 steps:

\vspace{.2cm}

\begin{enumerate}
	\item Determine the trajectory of the  accelerated observer
\item Define a new accelerated coordinate system, which is co-moving with respect to the accelerated observer. We will refer to these coordinates as Rindler coordinates.
\item Solve the wave equation of the scalar particle in both coordinate systems and compare corresponding vacua and modes.
\end{enumerate}

\vspace{.2cm}

These steps are performed in the following fashion:
\vspace{.1cm}

\begin{enumerate}
	\item We go to light-cone coordinates given by $u=t-x$, $v=t+x$, so that $ds^2 = -du dv$. For a Lorentz boost, our light-cone coordinates transform as $u\rightarrow u' = \alpha  u$, $v \rightarrow v'= \frac{1}{\alpha} v$. The trajectory of the accelerated observer is given by $u(\tau) = - \frac{1}{a} e^{-a\tau}$, $v(\tau) = \frac{1}{a} e^{a\tau}$ corresponding to $t(\tau) = \frac{v+u}{2} = \frac{1}{a} \sinh a\tau~~,~~~ x(\tau) = \frac{1}{a} \cosh a\tau$.

\item The Rindler coordinates $\tilde{u}, \tilde{v}$ are then implicitly defined by $u  = -\frac{1}{a} e^{-a \tilde{u}} ~~,~~~ v= \frac{1}{a} e^{a\tilde{v}}$. Then $ds^2 = -du dv = - e^{a(\tilde{v} - \tilde{u})} d\tilde{u} d\tilde{v} = - e^{2a \xi^1} \left( - \left(d\xi^0\right)^2 + \left(d\xi^1\right)^2 \right)$, where $\xi^0 = \frac{\tilde{u}+ \tilde{v}}{2} ~~,~~~ \xi^1 = \frac{\tilde{u}- \tilde{v}}{2}$. However, $(\xi^0, \xi^1)$ is not a complete coordinate system i.e. it does not cover all of Minkowski space. We refer to figure \ref{rindlerfig}, where the curve indicated by $r'$ has constant $\xi^1$. 

\item We now introduce quantum fields on our space-time. We will see that we will find different particle numbers for different observers.  $t$ and $\xi^0$ are related in a non-trivial way, which implies that positive frequency modes with respect to $t$ will be a superposition of positive and negative frequency modes with respect to $\xi^0$. 
\end{enumerate}

\vspace{.2cm}

Consider again the scalar field action

\[S\left[ \phi \right] = -\frac{1}{2} \int  d^2x \sqrt{-g} ~g^{\alpha \beta } \partial_{\alpha}\phi\partial_{\beta} \phi ~.\]

 Conformal transformations, i.e. transformations of the form $g_{\alpha \beta } \rightarrow \tilde{g}_{\alpha \beta } = \Omega^2(x) g_{\alpha \beta}$, leave our action invariant. This is easy to see since $\sqrt{ - \tilde{g}} = \Omega^2 \sqrt{-g}$, $\tilde{g}^{\alpha \beta} = \Omega^{-2} g^{\alpha \beta}$. Hence, we can see that going to Rindler space i.e. to coordinates $( \xi^0, \xi^1)$ simply corresponds to a conformal rescaling. We have 
 \[S= -2 \int du dv ~ \partial_u \phi \partial_v \phi  = -2 \int d\tilde{u} d\tilde{v}~ \partial_{\tilde{u}} \phi \partial_{\tilde{v}} \phi ~ .\] 
 
 The field equations are thus given by $\partial_u \partial_v \phi = 0 = \partial_{\tilde{u}} \partial_{\tilde{v}} \phi$, which is solved by
 
 \[\phi(u,v) = A(u) + B(v)~~,~~~ \phi( \tilde{u} , \tilde{v} ) = \tilde{A}(\tilde{u}) +  \tilde{B}(\tilde{v}) ~.\]

 The modes that make up $A(u)$ ($B(v)$) is referred to as right-moving (left-moving). In the following we will focus on the right-moving modes. In Minkowski space, $\phi(u) \sim e^{-i\omega u } = e^{-i\omega(t-x)}$, while in Rindler space we have $\phi(\tilde{u}) \sim e^{-i\Omega \tilde{u}} = e^{-i\Omega( \xi^0 - \xi^1)}$. Namely, $\omega$ is the Minkowski frequency and $\Omega$ is the Rindler frequency. We consider the domain where $x > \lvert t \rvert$, corresponding to the Rindler wedge where both our Rindler and Minkowski coordinates are well-defined. We thus have
 
 \begin{equation}
\hat{\phi} = A(u) = \int \frac{d\omega}{(2\pi)^{1/2}} \frac{1}{\sqrt{2\omega}} \left( e^{-i\omega u } \hat{a}^-_{\omega} + e^{i\omega u } \hat{a}^+_{\omega}\right) = \int \frac{d\omega}{(2\pi)^{1/2}} \frac{1}{\sqrt{2\Omega}} \left( e^{-i\Omega u } \hat{b}^-_{\Omega} + e^{i\Omega u } \hat{b}^+_{\Omega}\right) = \tilde{A}(\tilde{u})~.
 \end{equation}
 
  This is easy to see since the definition of our quantum field cannot depend on our choice of coordinate system. We have two sets of modes with commutation relations
 
 \[\left[ \hat{a}^-_{\omega} , \hat{a}^+_{\omega'} \right] = \delta \left( \omega - \omega' \right) ~~ \text{(Minkowski) and } \left[ \hat{b}^-_{\Omega} , \hat{b}^+_{\Omega'} \right] = \delta \left( \Omega - \Omega' \right) ~~\text{(Rindler)}~. \]
 
  Correspondingly, we have two vacua, $\ket{0_M}$ and $\ket{0_R}$, which satisfy 
  
  \[ \hat{a}^-_{\omega} \ket{0_M} = 0 = \hat{b}^-_{\Omega} \ket{0_R}~.\]
  
   Next time we will compare these two vacua and derive the Unruh effect.

\clearpage

\section{Unruh and Hawking effects}

We express a quantum scalar field in Minkowski and Rindler space as

 $$ \hat{\phi} = \int_0^{\infty} \frac{d\omega}{(2\pi)^{1/2}} \frac{1}{\sqrt{2\omega}} \left[ e^{-i\omega u } \hat{a}^-_{\omega} +e^{i\omega u } \hat{a}^+_{\omega} \right] = \int_0^{\infty} \frac{d\Omega}{(2\pi)^{1/2}} \frac{1}{\sqrt{2\Omega}} \left[ e^{-i\Omega \tilde{u} } \hat{b}^-_{\Omega} +e^{i\omega \tilde{u} } \hat{b}^+_{\Omega} \right]$$

In Minkowski space, $u$ are light-cone coordinates, $\omega$ are frequencies, $\hat{a}^+_{\omega}, ~\hat{a}^-_{\omega}$ are creation and annihilation operators, $\ket{0_M}$ is the vacuum. $\tilde{u}$ and $\Omega$ are the coordinates and frequencies for Rindler space, which has $\ket{0_R}$ as its vacuum state. We are naturally led to ask what is the `correct' or `true' vacuum. The answer is that different observers have different vacuum states, hence there is no single `true' or otherwise preferred vacuum state. For example, if a particle detector (observer) is accelerated, its correct vacuum is $\ket{0_R}$, if it is not being accelerated the correct vacuum is $\ket{0_M}$. With respect to $\ket{0_R}$, $\ket{0_M}$ contains infinitely many excited states. We now calculate the relation between $\ket{0_M}$ and $\ket{0_R} $ i.e. between $a^{\pm}$ and $b^{\pm}$. We use the ansatz that they are related by a \textit{Bogoliubov transformation}, which is of the form

$$\hat{b}^-_{\Omega} = \int_0^{\infty} d\omega \left[ \alpha_{\Omega \omega} \hat{a}^-_{\omega} - \beta_{\Omega \omega} \hat{a}^+_{\omega}\right] ~.$$

Note that there does not exist an inverse Bogoliubov transformation since Rindler space covers only half of Minkowski space. From the commutation relations for $b_{\Omega}$, we find  

$$\int_0^{\infty} d\omega \left( \alpha_{\Omega \omega}\alpha^*_{\Omega ' \omega } - \beta_{\Omega \omega}\beta^*_{\Omega ' \omega } \right) = \delta\left( \Omega - \Omega'\right)~.$$

This leads to $\frac{1}{\sqrt{\omega}} e^{-i\omega u } = \int^{\infty}_0 \frac{d\Omega'}{\sqrt{\Omega'}} \left( \alpha_{\Omega' \omega} e^{-i \Omega'\tilde{u}} - \beta^*_{\Omega' \omega} e^{i \Omega' \tilde{u}} \right) $. Multiplying this with $e^{ \pm i \Omega \tilde{u}}$ and integrating over frequencies $\Omega$ gives

\begin{align}
\alpha_{\Omega \omega} (\beta_{\Omega \omega} )= \int_{ -\infty}^{\infty} e^{\pm i \omega u + i \Omega \tilde{u}} d\tilde{u} 
&= \pm \frac{1}{2\pi} \sqrt{\frac{\Omega}{\omega}} \int^{\infty}_{-\infty} (-au)^{-i\frac{\Omega}{a} -1} e^{\mp i \omega u } du \notag\\
&= \pm \frac{1}{2\pi} \sqrt{\frac{\Omega}{\omega}} \exp \left( \frac{i\Omega}{a} \ln \frac{\omega}{a} \right) \Gamma \left( - \frac{i\Omega}{a} \right)
\end{align}

and 

\begin{equation}
 \lvert \alpha_{\Omega \omega} \rvert^2 = e^{2 \pi \Omega/a} \lvert \beta_{\Omega \omega} \rvert^2~,
\end{equation}

where $a$ is the acceleration. We now go to the Rindler frame i.e. the accelerated observer and we compute the occupation number of Rindler states in the Minkowski vacuum. The expectation value of the occupation number, $ \hat{N}_{\Omega} = \hat{b}^+_{\Omega} \hat{b}^-_{\Omega} $, is

\begin{align}
\langle \hat{N}_{\Omega} \rangle & = \obket{0_M}{\hat{b}^+_{\Omega} \hat{b}^-_{\Omega}}{0_M} \notag \\
&= \obket{0_M}{ \int d\omega \left[ \alpha^*_{\omega \Omega}  \hat{a}^+_{\omega} - \beta^*_{\omega \Omega} \hat{a}^-_{\omega} \right] \int d\omega' \left[ \alpha_{\omega' \Omega}  \hat{a}^-_{\omega'} - \beta_{\omega' \Omega} \hat{a}^+_{\omega'} \right]  }{0_M} \notag\\
&= \int d\omega \lvert \beta_{\Omega \omega}\rvert^2 ~.
\end{align}

Normalization condition with $\Omega = \Omega'$ is then given by

\begin{equation}
\int d\omega \left( \lvert \alpha_{\Omega \omega}\rvert^2 - \lvert \beta_{\Omega \omega}\rvert^2 \right) = \delta(0)~.
\end{equation}  

Hence

\begin{equation}
\langle N_{\Omega} \rangle = \left[ \exp \left( \frac{2\pi \Omega}{a} \right) -1 \right]^{-1} \delta(0)~.
\end{equation}

 This is divergent due to $\delta(0)$, which signals that we are considering a space of infinite volume. We compute instead the particle density, which is given by

\begin{equation}
n_{\Omega} = \frac{\langle N_{\Omega} \rangle}{V} = \frac{1}{\exp \left( \frac{2 \pi \Omega}{a} \right) -1} ~.
\end{equation}

Where $V$ is the volume of our space-time. This expression gives the massless particles detected by an accelerated observer in the Minkowski vacuum. We see that it obeys Bose-Einstein statistics, which shows that it corresponds to a thermal bath at temperature

\begin{equation}
T_{Unruh} = \frac{a}{2\pi} ~.
\end{equation}

Reintroducing the constants that we previously set to zero gives

\begin{equation}
T_{Unruh} = \frac{\hbar a}{2 \pi k_B c} ~.
\end{equation}

We see that the Unruh temperature goes to zero if we let $\hbar$ go to zero, signalling its quantum-mechanical origin. If we take $a \approx \frac{c}{\text{micrometer}}$, $T_{unruh} \approx 2K$ i.e. the Unruh temperature is typically very low and thus Unruh radiation is very hard to detect. 


\clearpage

\subsection{Hawking Radiation}

A famous result due to Hawking states that black hole emit a thermal spectrum of particles. Hawking radiation, which was discovered before Unruh radiation, was rather unexpected, since it was previously believed that particles can only be produced in non-static gravitational fields. We will see that particles with positive and negative frequencies can be produced near the horizon of a static black hole. We do not go through this derivation in great detail since it is similar to that of Unruh radiation. Indeed, by the equivalence principle, inertial and gravitational acceleration are locally indistinguishable, so that the corresponding particle creation should be locally indistinguishable as well. 

The first coordinate system is the Schwarzschild space-time in tortoise coordinates. These contain a coordinate singularity at the horizon. These coordinates are analogous to the Rindler coordinates in the derivation of the Unruh effect. The quantity analogous to the Rindler acceleration $a$ is the surface gravity $\kappa$, which is the acceleration at the event horizon with respect to an asymptotic observer. The tortoise coordinate is given by $r^*(r) = r + r_s \ln \left( \frac{r}{r_s} -1 \right) $, with $r_s = 2M$, the Schwarzschild radius. We see that $r^* \rightarrow \infty$ for $r \rightarrow r_s$ i.e. we are considering one external region of the black hole corresponding to a single Rindler wedge. The coordinate range is $ r_s \leq r \leq \infty$ corresponding to $ - \infty \leq r^* \leq \infty$. Going to light-cone coordinates $\tilde{u} = t- r^*$ and $\tilde{v} = t+ r^*$, the two-dimensional tortoise coordinates are of the form

\begin{equation}
ds^2_{r^*} = - \left( 1 - \frac{ r_s}{r(\tilde{u}, \tilde{v} )} \right)d\tilde{u} d\tilde{v} ~.
\end{equation}

This is conformally equivalent to a flat metric. We see that it is singular at $r=r_s$, hence these coordinates only cover region $I$.

The other coordinate system are the Kruskal-Szekeres (KS) coordinates, corresponding to a free falling observer. These coordinates are non-singular at the horizon, hence a locally inertial observer in these coordinates should not notice anything special at the horizon. This coordinate system is analogous to the Minkowski coordinates for the Unruh effect, since they cover all four coordinate wedges of Rindler or Schwarzschild space and are non-singular everywhere except at $r=0$. We go to coordinates $ u = -2r_s \exp \left( - \frac{\tilde{u}}{2 r_s} \right) ,~ v = 2 r_s \exp \left( \frac{\tilde{v}}{2 r_s} \right)$, which gives the following metric

\begin{equation}
ds^2_{KS} = - \frac{ r_s}{ r(u,v)} \exp \left[ 1 - \frac{ r(u,v) }{r_s} \right] du dv ~.
\end{equation}

This metric is also conformally equivalent to a flat metric, so that it is regular at $r=r_s$ and covers coordinate wedges $I - IV$. We now introduce the 2-dimensional scalar field theory action

\begin{equation}
S \left[ \phi \right] = - \frac{1}{2} \int g^{\alpha \beta} \partial_{\alpha} \phi \partial_{\beta} \phi \sqrt{-g} d^2 x ~.
\end{equation}

Note that this action is conformally invariant. We can expand in both bases, giving $\phi \left( \tilde{u} , \tilde{v} \right) = \phi( u,v)$. $\phi \left( \tilde{u} , \tilde{v} \right)$ corresponds to the \textit{Boulware vacuum} $\ket{0_B}$ with corresponding creation and annihilation operators, $\hat{b}_{\Omega}^-$ and $ \hat{b}_{\Omega}^+$, while $\phi( u,v)$ corresponds to Kursal-Szekeres vacuum $\ket{0_{KS}}$ and operators $ \hat{a}_{\omega}^-, \hat{a}_{\omega}^+$. We use similar notation to the one we used for the Unruh effect to signal that the Boulware vacuum $\ket{0_B}$ is analogous to the Rindler vacuum and \textit{Kruskal-Szekeres} vacuum $\ket{0_{KS}}$ is analogous to the Minkowski vacuum.

Consider an observer, A, at constant $r^*$ i.e. constant $r$, which corresponds to the accelerated Rindler observer, and another observer, B, at non-constant $r^*$ which is an inertial observer. What is the particle spectrum seen by A in the Kruskal-Szekers vacuum $\ket{0_{KS}}$? Analogous to our previous computation, we find that

\begin{equation}
\langle \hat{N}_{\Omega} \rangle = \obket{ 0_{KS}}{ \hat{b}^+_{\Omega} \hat{b}^-_{\Omega} }{0_{KS}} = \frac{1}{ \exp\left( \frac{2\pi \Omega}{\kappa} \right) -1 } \delta(0) ~.
\end{equation}

We thus find a thermal spectrum at Hawking temperature

\begin{equation}
T_H = \frac{\kappa}{2\pi} = \frac{1}{8\pi M} = \frac{\hbar c^3}{8 \pi G k_B  M}~.
\end{equation}

Where we introduce all constants of nature at the last equality. Note again that $T \to 0 $ when $ \hbar \to 0 $, signalling the quantum-mechanical mechanical origin of Hawking radiation as it did for Unruh radiation.

\clearpage

\section{Information loss paradox}

The results of the last few lectures can be summarized as follows: \\

\begin{center}
	\begin{tabular}{ |c|c|c| } 
\textbf{Phenomenon}:	& \textbf{Unruh effect} & \textbf{Hawking radiation} \\ \hline
Origin: &	Accelerated coordinate systems & Gravitational background of a black hole\\  
Vacuum of full space: &	  $\ket{0_M}$  & $\ket{0_{KS}}$  \\
Vacuum of wedge:	&	$\ket{0_R}$  & $\ket{0_B}$   \\
Temperature: &	$a$ =  acceleration & $\kappa$ = surface gravity \\
Coordinates: & $u = - a^{-1} e^{ - a \tilde{u}}$, $v = - a^{-1} e^{ a \tilde{v}}$  & $u = - \kappa^{-1} e^{ - \kappa \tilde{u}}$, $v = - \kappa^{-1} e^{\kappa \tilde{v}}$\\
	\end{tabular}
\end{center}
\vspace{.3cm}


Hawking and Unruh radiation are found by calculating the occupation number 

\[\obket{0}{ b^+_{\Omega } b^-_{\Omega} }{0} = \left[ \exp\left( \frac{2 \pi \Omega}{a (\kappa)} \right) -1 \right]^{-1} \delta(0)~,\]

 where $b^+_{\Omega },~ b^-_{\Omega}$ are the Boulware (Rindler) modes and $\ket{0}$ is the KS (Minkowski) vacuum.

\underline{Remarks}:

\begin{enumerate}
	\item  For a black hole in thermal equilibrium with an external heat bath with temperature $T_H$, the black hole emits and absorbs particles at the same rate. 

\item For a black hole in empty space, the black hole only emits particles, hence it will evaporate, $\delta M<0$, and it disappears within finite time. 

\item Recall that $dM = \frac{1}{8 \pi M} d\left( \frac{A}{4} \right) = T_H dS$, where $T_H = \hbar \frac{\kappa}{2 \pi}$, so that $ S=\frac{A}{4 \hbar} \xrightarrow{ \hbar \rightarrow 0} \infty$. 

\end{enumerate}
\vspace*{.3cm}

Hawking radiation is thermal, which signals the loss of information regarding the initial matter state that formed the black hole. For example, presume a black hole is formed from highly energetic muons, schematically, as $\mu^+ \mu^- \rightarrow$`black hole'$\rightarrow$`gravitons+electrons+photons+etc'. This schematic process is meant to show that we completely lose information about the fact that the black hole here was formed by muons, since the emitted Hawking radiation consists of many different particles. More precisely, in quantum mechanics, we have a time evolution operator which we write as $S(t)=\exp \left( \frac{iHt}{\hbar} \right)$, where $H$ is our Hamiltonian. In the context of QFT, $S$ is referred to as the S-matrix. If we know $\ket{ \psi( t_0 )}$, then $\ket{\psi(t)} = S(t) \ket{ \psi( t_0 )}$. We can generalize this using the \textit{quantum density matrix}, which we can write as

\begin{equation}
\hat{\rho} = \sum_i c_i \ket{\psi_i} \bra{\psi_i} ~.
\end{equation}

with $\ket{\psi_i} \in \mathcal{H}$, $c_i \in \mathbb{C}$, and $tr \hat{\rho} =1$. When $\hat{\rho}^2 = \hat{\rho}$, $\hat{\rho}$ describes a \textit{pure state}. For $\hat{\rho}^2 \neq \hat{\rho}$, $\hat{\rho}$ describes a \textit{mixed state}. Hawking asserted that due to the thermal nature of Hawking radiation, the outgoing particles are always described by a mixed state. Namely, the corresponding density matrix is $\rho_H = \sum_n e^{- \beta E_n} \ket{n}\bra{n}$, where $\ket{n}$ are energy eigenstates with energy $E_n$. Hence, we seem to go from a pure state to a mixed state, which signals that we break unitarity. Hawking thus introduced the non-unitary \textit{dollar matrix} \cite{hawdol}, which relates $\rho^{\text{final}}$ to $\rho^{\text{initial}}$ as $ \rho^{\text{final}} = \$ \rho^{\text{initial}}$, as an ad hoc non-unitary substitute for the usual S-matrix.

\subsection{Possible solutions to the information problem}

Solutions to the information loss paradox that can be found in the literature include:

\begin{enumerate}
	\item Black holes violate quantum mechanics. This claim isno longer very popular.
	\item Information stays inside the black hole. This requires one to stop Hawking radiation artificially so that black hole remnant remains, since Hawking temperature increases when the black hole shrinks. This is not generally considered very plausible.
	\item Information goes somewhere else, e.g. region $II$ or some (other) parallel universe.
	\item Quantum mechanics is ok - Information comes out with the Hawking radiation. This is typically considered most plausible, hence we will make a few remarks on this possibility.   
\end{enumerate}

So far, we did not consider the back-reaction of infalling matter on the black hole. Furthermore, we did not treat gravity quantum-mechanically. One may hope that taking these effects into account will restore black hole unitarity. This would entail that the Hawking quanta are not truly thermal. Another possibility is that the information is stored on the horizon, so that the microscopic details of Hawking radiation are determined by a holographic principle. The most explicit holographic principle know to date is the AdS/CFT correspondence, which states that the gravitational AdS bulk is dual to a non-gravitational conformal field theory (CFT) at the boundary of the AdS space. Since CFT's undergo unitary time evolution, so should the corresponding AdS spaces, even when they contain a black hole.

Another proposed (partial) solution is called \textit{black hole complementarity}, which in turn gave rise to the firewall paradox. Another proposed picture is the so-called \textit{fuzz-ball}, and yet another picture is the idea that a black hole is a graviton condensate at the quantum critical point. We will now comment on black hole complementarity and the firewall paradox.

\subsubsection{Black hole complementarity (Susskind, Thorlacius, 't Hooft)}

According to black hole complementarity, information is simultaneously reflected and passed through the horizon. The reflected information can be perceived by the outside observer, while the information that passes through can be perceived by the freely falling observer, but no single observer can confirm both pictures at the same time. This gives rise to the notion of the textit{stretched horizon} \cite{strhor}, which is a thin `membrane' with a thickness of the order of one Planck length. According to the external observer, the infalling and reflected information gets `heated up' at the horizon, yet the infalling observer does not see anything special. This picture requires entanglement between the infalling and reflected information.

\subsubsection{Firewall paradox}

In 2012, the AMPS (Almheiri, Marolf, Polchinski, Sully) paper was published \cite{amps}, which is nicely reviewed in \cite{amps2}. AMPS pointed out a flaw in black hole complementarity. They argued that the following three statements cannot be simultaneously true.

\begin{enumerate}
	\item Hawking radiation is in a pure state.
	\item Information is emitted from a region (e.g. the stretched horizon) near the horizon, where an efective description of gravity in terms of GR is correct.
	\item Infalling observer encounters nothing special when crossing the horizon. 
\end{enumerate}

The proposed solution by AMPS is to give up the third statement i.e. the infalling observer will be `burned' at the horizon, hence this is known as the firewall paradox.

\clearpage

\section{Solitons in String Theory}

The next few sections consider the construction of the Tangherlini black hole in string theory as first done by Strominger and Vafa \cite{stromvaf}. We will look at the construction of D- and p-branes, the compactification of superstring theories, and the construction of black holes from charges associated to the branes. Using a stringy version of the electromagnetic duality, we can express the black hole entropy in terms of both the event horizon area and the number of microstates. We will see that the entropy is given by $S=\frac{A}{4} = \ln \Omega$, where $A$ is the event horizon area and $\Omega$ is the number of microstates, which shows the statistical origin of the entropy of stringy black holes. These sections are based on the review found at \cite{strsol}.

\subsection{Review of electrodynamics in Minkowski space}

The vacuum Maxwell's equations are

\begin{align}
d(*F)= 0 ~, \notag\\
dF = 0~. 
\end{align}

Where $F$ is the field strength in form notation and $*F$ is its \textit{Hodge dual}. Generally, the Hodge dual of an (r-d)-form $\omega_{\mu_1 \mu_2 \dots \mu_{r-d}}$ is given by $\left(*\omega\right)_{\mu_1 \mu_2 \dots \mu_r} = \frac{ \sqrt{-g}}{r!(r-d)!} \epsilon_{\mu_1 \dots \mu_r}^{ ~~~~~~ \mu_{r+1} \dots \mu_d} \omega_{\mu_{r+1} \dots \mu_d}$. We then have $d *F = *j_e$, where $j_e$ is the electromagnetic current. We denote $*F$ as $\tilde{F}$, following standard notation. We have the following expression for the electric charge

$$ e = \oint_{S^2} d \vec{n} \vec{E} = \oint_{S^2} *F = \int_{B^3} d*F = \int_{B^3} *j_e ~.$$

The expression $dF=0$ is known as the \textit{Bianchi identity}. This is an identity since we can write $F=dA$, where $A$ is the electromagnetic field in form notation. The Poincar\'{e} identity then tells us that $dF = d^2 A = 0$. Electromagnetism in form notation allows us to easily define a magetic current and charge as follows. We have $dF = j_g$; $F= 2 dA + \omega$, where $d* \omega = 0 $ and $d\omega = j_g$, where $j_g$ is the magnetic current. We then define magnetic charge as

\begin{equation}
g = \int_{ \partial \mathcal{M}} F = \int_{ \mathcal{M}} j_g ~.
\end{equation}

Where $\mathcal{M}$ is some three-dimensional volume in our (3+1)-dimensional Minkowski space. We still have the gauge freedom given by $A \rightarrow A+d \Lambda$, where $\Lambda$ is a 0-form i.e. a scalar. We are now going to use these techniques in higher-dimensional spaces, particularly for solitonic solutions in string theory. We will be considering d-dimensional spaces and (p+1)-form potentials of the form $A^{(p+1)} = A_{\mu_1 \dots \mu_{p+1}} dx^{\mu_1} \wedge \dots \wedge dx^{\mu_{p+1}} $. We then have gauge freedom $A^{(p+1)} \rightarrow A^{(p+1)} + d\Lambda^{(p)}$, our field strength is of the form $F^{(p+1)} = (p+2) dA^{(p+1)} + \omega^{(p+2)}$. The $\omega$-term is typically absent in electromagnetism, except when we introduce magnetic charges by hand. The charges corresponding to $F$ and $*F$ are given by

\begin{equation}
e = \int_{ \mathcal{M}} * j_e^{(d-p-1)}  ~~, ~~~ g= \int_{ \tilde{\mathcal{M}}} j_g^{(p+3)} ~.
\end{equation}

Where $\mathcal{M}$ and $\tilde{\mathcal{M}}$ are (d-p-1)-dimensional and (p+3)-dimensional subspaces, respectively. In analogy with classical electromagneticm, we assume the charges that give rise to $* j_e^{(d-p-1)} $ to be localized in p spatial dimensions. The magnetic objects corresponding to $j_g$ extends $d-p-4$ dimensions. 

\subsection{Dirac quantization}

If we have take an electric charge along some path that encloses a magnetic charge, the phase of the wave function has to change by $2 \pi n$, $n \in \mathbb{Z}$. This means that

\begin{equation}
eg = 2\pi n ~ \Rightarrow ~ e= \frac{2 \pi n}{g} ~.
\end{equation}

That is, we see that the electric charge is quantized due to the presence of a magnetic charge. 

To summarise, the form degree of the potential gives us the dimension of the electric object. From this, we can obtain the dimension of the magnetic object, that is, $\tilde{F}^{(d-p-2)} = d \tilde{A}^{(d-p-3)}$, so that we see that out magnetic object is $(d-p-3)$-dimensional. We will now apply these ideas to string theory. 

\subsection{Supergravity and p-branes}

 We present the massless degrees of freedom of the following superstring theories in a table:\\

\begin{center}
	\begin{tabular}{ |c|c|c|c|c| } 
		\textbf{Model} & \textbf{Potential} & \textbf{Field strength} & \textbf{p$_e$} & \textbf{p$_m$} \\ 
		\hline
Het., II$_A$, II$_B$ & $B^{(2)} = A^{(1+1)}$ & $H $ & Fundamental string & 5 (NS5-brane)  \\  
II$_A$& $ A^{(1)}$, $A^{(3)}$& $F^{(2)}$, $F^{(4)}$ & D-particle, 2-brane & 6-brane, 4-brane \\
		II$_B$ & $A^{(0)}$, $A^{(2)}$, $A^{(4)}$ & $F^{(1)}$, $F^{(3)}$, $F^{(5)}$ & 1-brane , D-string, 3-brane & D7, D5, 3-brane     \\
		M-theory  & $A^{(3)}$ & $F^{(4)}$ & M2-brane & M5-brane
	\end{tabular}
\end{center}

\vspace{.6cm}

We try to find solutions to the field equations corresponding to the SUGRA action, corresponding to the massless degrees of freedom of the superstring theories listed above

\begin{equation}
S^E_{\text{eff}} = \frac{1}{2K^2} \int d^{10}x \sqrt{-G^E} \left( R\left( G^E \right) - \frac{1}{2} \nabla_{\mu} \Phi \nabla^{\mu} \Phi - \sum \frac{1}{2n!} e^{a \Phi} F^{\mu_1 \dots \mu_n} F_{\mu_1 \dots \mu_n} + \text{fermionic part} \right) ~.
\end{equation}

Here, $ R\left( G^E \right)$ corresponds to the graviton and $ \Phi$ is the dilaton. We have $a= -1$, $n=3$ for the universal (NS-NS) sector, and $a = \frac{5-n}{2}$ for the RR-sector. From now on, we will ignore the $\sum_n$-term in the action and consider only one term in the sum. This gives the following equations of motion

\begin{align}
\delta G^E_{\mu \nu }: ~ R^E_{\mu \nu} & = \frac{1}{2} \partial_{\mu} \Phi \partial^{\mu} \Phi + \frac{1}{2(n-1)!} e^{a \Phi} \left( F_{\mu \rho_1 \dots \rho_{n-1}} F_{\nu}^{\rho_1 \dots \rho_{n-1}} - \frac{n-1}{n(d-2)} F^2 G_{\mu \nu} \right) ~, \notag\\ 
	\delta \Phi: ~\partial_{\mu} \partial^{\mu } \Phi & = \frac{a}{n!} e^{a \Phi} F^2~.
\end{align}

We make the ansatz that $x_{\mu }$ with $\mu \in \{ 0, \dots , p \}$ are the coordinates of (p+1)-dimensional charged objects that are invariant under the Poincar\'{e} group $P(1,p)$, and $y_M$ with $ M\in \{p+1, \dots, d-1 \}$ are coordinates of a space with an $SO(d-p-1)$ isometry. We then define $r \eqqcolon \sqrt{ y_M y^M}$. We then make the ansatz that the metric is of the form

\begin{equation}
ds^2 = e^{2A(r)} dx^{\mu}dx_{\mu} + e^{2 B(r)} dy_M dy^M ~.
\end{equation}

Where the $\mu$ and $M$-contractions are performed with a flat metric. We then make the ansatz that the field trength is of the form

\begin{equation}
F^e_{M \mu_2 \dots \mu_n} = \epsilon_{ \mu_2 \dots \mu_n } \partial_M e^{C(r)} ~.
\end{equation}
 
Plugging this into the equations of motion gives

\begin{align}
e^{2A(r)} &= H(r)^{\frac{-4 \tilde{d}}{\Delta (d-2)}} ~, \notag \\
e^{2B(r)} &= H(r)^{\frac{-4 d}{\Delta (d-2)}}~, \notag \\
e^{2C(r)} &= \frac{2}{\sqrt{\Delta}} H (r)^{-1} ~,\notag\\
e^{\Phi(r)} &= \left(H(r)\right)^{\frac{2a}{\zeta \Delta}}~, \notag\\
\partial_M \partial^M H(r) &=0~.
\end{align}

We then introduce sources to find non-trivial solutions to the last equation. These are of the form $H(r) = 1+ \frac{\alpha}{r^{\tilde{d}}} ~ , ~~ \alpha >0$. The symbols in the above expressions are $\tilde{d} = d-p-3$, $\Delta = a^2 + \frac{2(p-1) \tilde{d}}{2}$, $\zeta = \pm 1$, with $+1$ for the electric solution and $-1$ for the magnetic solution, and $\alpha$ fixed by the charge for the electric solution and $\alpha = \frac{g \sqrt{\Delta}}{2\tilde{d}}$ for magnetic solutions.

To summarise, we found solutions to the low energy effective field theory of string theory. They are extended and carry charges and mass. Additionally, they are BPS-saturated, which entails that there is some residual supersymmetry for these particular solutions. We will use this later to count the number of states. 


\clearpage

\section{Brane solutions}

\subsection{Explicit examples of solitonic solutions to type II SUGRA: p-branes}

In the (NS,NS)-sector, we have the explicit solution given by the fundamental string for $d=10$, $n=3$

\begin{equation}
ds_E^2 = \left( 1+ \frac{\alpha}{r^6} \right)^{-3/4} dx_{\mu}^2 + \left( 1+ \frac{\alpha}{r^6} \right)^{1/4} dy_n^2 ~~,~~~ e^{\Phi} = \left( 1+ \frac{\alpha}{r^6} \right)^{-1/2} \sim g_s ~.
\end{equation}

NS5-brane with $n=7$

\begin{equation}
ds_E^2 =  \left( 1+ \frac{\alpha}{r^6} \right)^{1/4} dx_{\mu}^2 + \left( 1+ \frac{\alpha}{r^6} \right)^{3/4} dy_n^2 ~~,~~~ e^{\Phi} = \left( 1+ \frac{\alpha}{r^6} \right)^{1/2} \sim \frac{1}{g_s} ~.
\end{equation}

Here, $n$ is the dimension of the field strength tensor under consideration. The fact that we have $r^6$ comes from the fact that the codimension of the object under considertaion is $6$. These objects are dual to each other, as we can see from the string couplings for the two objects, namely, $g_s$ and $\frac{1}{g_s}$, respectively. For the 3-brane, we have $n=5$ i.e. the object is self-dual. The corresponding metric is

\begin{equation}
ds_E^2 = \left( 1+ \frac{\alpha}{r^4} \right)^{-1/2} dx_{\mu}^2 + \left( 1+ \frac{\alpha}{r^4} \right)^{1/2} dy_n^2 ~~~,~~ e^{\Phi} = 1 ~.
\end{equation}

Lastly, the 5-brane has the same metric as the NS5-brane, given above, but with $\Phi = \frac{1}{\Phi_{NS5}}$.

\subsection{D-branes}

For the remainder of this lecture, we will consider D-branes, which are originally introduced as Dirichlet boundary conditions for open strings. Dirichlet boundary conditions are given by

$$ \delta X^i \bigg\rvert_{\sigma = 0, \pi} = 0 ~. $$

This is necessary for open string to be consistent with T-duality, as we will discuss later. From T-duality, we will see that 

\[ X^i \bigg\rvert_{\sigma = 0, \pi} = c\text{,  for some } c \in \mathbb{C}~. \] 

In general, we call the following set-up a Dp-brane (where p indicates the spatial dimensionality of the brane). We have Neumann boundary conditions in $p$ dimensions i.e. 

\[\partial_{\sigma} X^{\mu} \bigg\rvert_{\sigma = 0, \pi} = 0 ~, ~~ \mu = 0, \dots, p~, \]

 and Dirichlet boundary conditions in $d-p$ dimensions i.e.

\[ \delta X^i \bigg\rvert_{\sigma = 0, \pi} = 0~,~~ i = p+1, \dots, d~. \] 

It may seem as if the D-branes are rigid objects, but we will see during the remainder of the lecture that this is not the case. We will see that they are dynamical objects with a tension $T_p \sim 1/g_s$. Hence, at $g_s \ll1$, $T_p \gg 1$, so that the D-branes appear static. At $g_s \gg 1$, the opposite is true, and it will be easy to see the dynamics of the D-brane, which hints at the fact that D-branes are indeed solitonic (non-perturbative) objects in our string theory. 

\subsection{T-duality for the closed string}

 We compactify our string theory on a circle or torus, and look at the spectrum of our theory. If we consider compactification on a circle of radius $R$, we know that the spectrum is classified by $(\text{momentum, winding number}) = \left( \frac{n}{R}, m \right) ~, ~  m,n \in \mathbb{Z}$. One can show that this theory is dual to a theory compactified on a circle with radius $\frac{\alpha'}{R}$ and a pectrum classified by $\left( m , \frac{n}{R}\right)$ vie T-duality. Note that momentum and winding number are exchanged under T-duality, which maps $X_R$ to $-X_R$ and $X_L$ to $X_L$.

 \subsection{T-duality for the open string}

For the open string, we cannot define a meaningful winding number, hence we have only momentum. Via T-duality, we map to a  theory without momentum and with only winding number, since we fix the end-points of our open string. We again map  $X_R$ to $-X_R$ and $X_L$ to $X_L$. For the open string, we also exchange von Neumann and Dirichlet boundary conditions. A general feature of T-duality along one of the dimensions of our Dp-brane, we go from a Dp-brane to a D(p-1)-brane, whilst if we perform T-duality along one of the dimensions normal to the Dp-brane, we go to a D(p+1)-brane. 

\subsection{Massless spectrum of open string}

The massless states are those without winding. One can show that these are given by $\alpha^{\mu}_{-1} \ket{0}$ and $\alpha_{-1}^9 \ket{0}$. $\alpha^{\mu}_{-1} \ket{0}$ correspond to $U(1)$-gauge bosons in (8+1)-dimensional target space, while $\alpha_{-1}^9 \ket{0}$ corresponds to a scalar in target space. Generalizing this to higher-dimensional tori, the massless states will correspond to $U(1)$-gauge boson in $(p+1)$ dimensions and $(9-p)$ scalars in $(p+1)$ dimensions, which describe the position of the Dp-brane. Since these states do not have any momentum orthogonal to the D-brane, we can see that the gauge theory lives on the world volume of the D-brane. Due to supersymmetry, the gauge bosons share their multiplet with a scalar and a spinor.

\subsection{Generalization to several D-branes}

We now extend to a situation where we have N D-branes, N $\in \mathbb{N}$, we will see that this will give rise to supersymmetric Yang-Mills theories. Namely, if one associates a charge to the end of our open strings, one can show that the string scattering satisfies a super Yang-Mills algebra. The charges are referred to as \textit{Chan-Paton factors}. This introduces a non-abelian structure from which one can find a non-abelian gauge potential. Namely, we will find a gauge field $A_9$ of the form $A_9 = \frac{1}{2\pi R} \text{diag} (\theta_1, \theta_2 ,\dots, \theta_n)$, where $\theta_i \in \mathbb{R}$. Introducing a gauge connection changes our canonical momentum and the bosonic field in the compactified dimension as

$$p = \frac{n}{R} + \frac{\theta_j - \theta_i}{2 \pi R} ~~,~~~ X^9_{(i,j)} = (\sigma, \tau) = C_{(i,j)} = + \left(2L + \frac{\theta_j - \theta_i}{\pi} \right) R_D \sigma + \text{oscillators}~. $$
 
Setting $C_{(i,j)} = \theta_i R_D$, we have $ X^9_{(i,j)} = 2 \pi L R_D  + \theta_j R_D$. Since D-branes break $P(1,9)$, i.e. the Poincar\'{e} group of $\mathbb{R}^{1,9}$, they also break 10d SUSY.

\subsection{Dynamics of D-branes}

We will see now that D-branes are in fact dynamical, as they interact with closed and open strings, they couple to gravity, the dilaton, and RR-fields, and they fluctuate in position and shape. The perturbative degrees of freedom are the open strings. We consider two D-branes with a closed string propagating between them, following a calculation done by Polchinski. Depending on how we choose our time direction, this process can either be seen as an open string vacuum amplitude, or a closed string exchange. Matching the amplitudes for these processes is expressed as

\begin{equation}
 \obket{0}{e^{-2 \pi^2 \left(L_0 + \bar{L}_0 -2 \right)/t }}{0}_{\text{tree}} = \mathcal{A} = \obket{0}{ e^{-2t \left(L_0^D -1 \right)  }}{0}_{\text{one-loop}} ~.
 \end{equation}
	
	Here, $L_0^D$ is the $L_0$ associated to the D-brane. Further, the length of the closed sting is $l = 2\pi^2/t$, whereas the open string has length $t$. One can show that this amplitude is zero if the D-branes are BPS-saturated, which means that there is no force between the parallel D-branes as the attractive force exerted by gravity and the dilaton equals the repulsive force from the RR-fields. The result of this calculation is
	
	\begin{equation}
	\lim_{r \to \infty} \mathcal{A} = V_{p+1 } (1-1) 2\pi (4 \pi^2 \alpha') \Delta_{(q-p)} (r^2) ~.
	\end{equation}
	
	Here, $\Delta_{(q-p)} (r^2)$ is the propagator in $(q-p)$-dimensional subspace with $r$ the distance from the D-brane. The $+1$ and $-1$ contributions come from the gravity+dilaton and the RR-fields, respectively. We now use this expression to find the tension $T_p$ and the charge density $\rho_p$ by comparing this result with the one obtained from the low energy effective action

	\begin{equation}
	S_{\text{eff}}^{II} = \frac{1}{2\kappa^2} \int d^{10}X \left[ \sqrt{-G} \underbrace{\left( R(G) + \frac{1}{2} \left(d\Phi\right)^2 + \frac{1}{12} e^{-\Phi} \left(dB\right)^2\right)}_{\text{Universal part (NS,NS)}} +  \underbrace{ \sum_p \frac{1}{2 (p+2)! } e^{\frac{3-p}{2} \Phi} \left( dA^{(p+1)} \right)^2  }_{\text{(R,R)}}     \right] ~.
	\end{equation}

	The effective action on the Dp-brane is

	\begin{equation}
	S^D_{\text{eff}} = T_p \int_{ \mathcal{M}^{(p+1)}} d^{p+1} \xi e^{\frac{3-p}{4} \Phi} \sqrt{ -\det\left( \Pi*G + \Pi*B + 2\pi \alpha' F \right) } ~,
	\end{equation}
	
	where $\Pi^*$ is the pull-back from 10-dimensional target space onto the D-brane. This is known as the \textit{Dirac-Born-Infeld action}. Lastly, we will need the coupling of the Dp-brane with charge $\rho_p$ to the RR-field $A_{\mu_1 \dots \mu_p}^{(p+1)}$, which is given by.
	
	\begin{equation}
	S_{eff}^D = \rho_p \int_{ \mathcal{M}^{(p+1)} } \Pi^* A^{(p+1)} ~.
	\end{equation}
	
	In the next sections, we will use these epressions to construct a stringy black hole state.

    \clearpage
    
    \section{Dimensional reduction and black holes}

    We start from the effective action
    
    \begin{align*}
    S_{\text{eff}}^{II} + S_{\text{eff}}^D+S_{\text{eff}}^{WZ} = &\int d^{10}x    \left[ \frac{1}{2\kappa^2} \left( \sqrt{-G} R+ \frac{1}{2} \left(d \Phi \right)^2 + \frac{1}{2(p+2)!} \left( dC^{(p+1)} \right)^2 \right) +  \right. \\
     & \qquad  \left. ~~ ~+ \sum_{i=1,2} \left(T_p \delta^{(9-p)} (x_{\perp} - a_i) \right) \left( - \frac{p-3}{4} \Phi + \sqrt{-G} \right) + \rho_p C^{(p+1)} \delta^{(9-p)} (x_{\perp} - a_i) + \dots \right] ~.
    \end{align*}
	
	
	Dropping all coupling and higher order terms, since we are only interested in the long range behaviour, we find
	
	\begin{equation}
	\ln Z_{vac} = 2 V_{p+1 } \kappa^2 \left( \rho_p^2 - T_p^2 \right) \Delta_{(q-p)}(p^2) \overset{!}{=} 0 ~~, ~~~ \rho_p^2 = T_p^2 = \frac{\pi}{\kappa^2} \left(4 \pi^2 \alpha' \right)^{3-p} ~.
	\end{equation}
	
	The last expression tells us that the charge of a Dp-brane is equal to its tension. We now consider Dirac-Zwanziger quantization. We move a Dp-brane around its magnetic dual, which is a D(6-p)-brane, which gives 
	
	
	$$ \rho_p \rho_{6-p} = \frac{\pi n}{\kappa^2} ~.$$

	This gives charge quantization, analogous to the Dirac quantization condition in electromagnetism. Instead of calculating $e \oint_{\mathcal{C}} A$ for a closed curve around a monopole an requiring the expression gives a multiple of $2\pi$, we now calculate $\rho_p \oint_{\mathcal{M_{p+1}}} C^{(p+1)}$, to give the expression above. This shows that the D-branes are states with minimal RR-charge, hence we call them elementary.
	 
	 In short, introducing D-branes in this way breaks half of our original supersymmetry (as we found for the p-branes). We find that $\rho_p = T_p$ and $\rho_p \rho_{6-p} = \frac{\pi n}{\kappa^2}$. The low energy limit is a SUGRA p-brane solution. In this picture, our perturbative degrees of freedom are massless open string states ending on D-branes.

	 \subsection{Black holes in string theory}
	 
	 Our strategy for constructing and analyzing string black holes will follow the following steps:
	 
	 \begin{enumerate}
	 	\item Construct black hole via dimensional reduction of p-brane solutions. The result will be an extremal black holes in 5d, which has a temperature equal to zero so that Hawking radiation is absent.
	 	\item Use D-brane description of the p-branes (i.e. relating strongly and weakly coupled limits of our theory) to count microstates $N$, calculate entropy as $S_{\text{stat}} = \log N$
	 	\item Compare to $S_{\text{BH}} = \frac{A}{4}$	 	
	 \end{enumerate}
	
	\subsubsection{Dimensional reduction of effective action}
	
The string frame action in terms of $R$ and $\phi$ reads

\begin{equation}
S= \frac{1}{2\kappa_D^2} \int d^D x \sqrt{-G} e^{-2 \phi_D} \left(R+ 4\partial_{M} \phi \partial^{M} \phi_D \right) ~~, \text{take}~ x^M = (x^{\mu} ,x) ~,~~  \text{with} ~ x \sim x+2\pi R n,~ n\in \mathbb{Z} ~.
\end{equation}

We perform a metric anzats of the form

\begin{equation}
G_{MN} = \left( \begin{array}{c|c}
\bar{G}_{\mu \nu} + e^{2 \sigma} A_{\mu } A_{\nu} & e^{2 \sigma } A_{\mu} \\ \hline
e^{2 \sigma} A_{\nu} & e^{ 2 \sigma}
\end{array} \right) ~.
\end{equation}


This corresponds to considering only massless states, since we assume $A_{\mu}$ to be independent of $x$, the coordinate of our compactified dimension. Here, $\bar{G}_{\mu \nu}$ is the $(D-1)$-dimensional metric in string frame, $A_{\mu}$ is a Kaluza-Klein gauge field, and $\sigma$ is the Kaluza-Klein scalar. We have $ (- G)^{1/2} = e^{\sigma} ( - \bar{G})^{1/2}$, hence the geodesic length $\rho$ of our compactified dimension is related to the parametric length $R$ of the compactified dimension as 

\begin{equation}
2 \pi \rho = 2 \pi R e^{ \langle \sigma \rangle} ~.
\end{equation}

We will see that $\langle \sigma \rangle$ is not fixed by the equations of motion. We thus define a new dilaton in our $(D-1)$-dimensional theory

\begin{equation}
\bar{\phi}_{D-1} = \phi_D - \frac{\sigma}{2}~.
\end{equation}

We then find

\begin{equation}
S= \underbrace{\frac{1}{2\kappa_{D-1}^2}}_{ =\frac{ \pi R}{\kappa_D^2} } \int d^{D-1} x \sqrt{ - \bar{G}} e^{-2 \bar{\phi}_{D-1}} \left( \bar{R} + 4 \partial_{\mu} \bar{\phi}_{D-1} \partial^{\mu} \phi_{D-1} - \partial_{\mu} \sigma \partial^{\mu} \sigma  - \frac{1}{4} e^{2 \sigma} F_{\mu \nu} F^{\mu \nu} \right)  ~.
\end{equation}

The massive states in this theory (which we ignore here) are charged with respect to the $U(1) \simeq S^1$ gauge theory on the compactified circle with charges $Q \sim \frac{n}{R}$. The gauge coupling depends on the modulus and the dilaton field. In the full effective action of type $II_A$ and $II_B$, there are also tensor fields which have to be dimensionally reduced.

\subsubsection{Dimensional reducation of p-branes}

We have two ways types of dimensional reduction.

\begin{enumerate}
	\item Compactify along world volume directions of the p-brane. We then go from a p-brane in D dimensions to a (p-1)-brane in (D-1) dimensions
	\item We compactify along directions transverse to the p-brane. This gives rise to a p-brane in (D-1) dimensions. The transverse directions are not isometry directions of the p-brane, but we can use the no-force property of BPS to construct a periodic array of p-branes.
\end{enumerate}

We consider the second of the dimensional reduction constructions. We now split off the the transverse directions as $\vec{x} = (\bar{x},x) $. At all $x= 2\pi Rn$, $n \in \mathbb{Z}$, we have

\begin{equation}
H = 1+ \sum_{-\infty}^{\infty} \frac{Q}{ \lvert \vec{x} - \vec{x}_n \rvert^{D-p-3}} ~~ ~,~~ \text{with } \vec{x}_n = (0 , 2\pi n R) ~.
\end{equation}

Through some rather involved calculations, one can show that

\begin{equation}
H = 1 + \frac{Q}{R \lvert x \rvert^{D-p-4}} + \mathcal{O} \left( e^{ - \frac{\lvert \bar{x} \rvert }{R}} \rvert \right)~.
\end{equation}

One says that the p-brane is delocalized or `smeared' along the circle. This state is invariant under rotations of the circle.

\subsubsection{Interlude: Tangherlini black hole}

The Tangherlini black hole is the 5-dimensional generalization of the extremal Reissner-Nordstr\"{o}m black hole solution, it is of the form

\begin{equation}
ds^2_E = - H^{-2} dt^2 + H \left( dr^2 + r^2 d \Omega_{3}^2 \right) ~,
\end{equation}

where $H$ is a function that is harmonic with respect to the transverse coordinates, i.e. $H =  1+ \frac{Q}{r^2} $, where $Q$ is the electric charge of the black hole. In these coordinates, the event horizon is at $r=0$, where $H$ diverges. The area of the event horizon and corresponding is then

\begin{equation}
A= 2\pi^2 \lim_{r \rightarrow 0 } \left( r^3 H^{3/2} \right) = \underbrace{2 \pi^2}_\text{Area of $S^3$} Q^{3/2} ~~, ~~~ S_{\text{BH}} = \frac{A}{4} = \frac{\pi^2}{2} Q^{3/2} ~.
\end{equation}

\subsection{Dimensional reduction of the D1-brane}

We now try to construct a 5-dimensional black hole by dimensional reduction of a 1-brane, starting from an expression from last lecture for the string frame action

\begin{equation}
ds^2_{SF} = H_1^{-1/2} \left( dt^2 + dy^2 \right) + H_1^{1/2} \left( dx_1^2 + \dots + dx_8^2 \right) ~~, ~~~ e^{-2 b_{10}} = H_1^{-1} ~~, ~~~ H_1 = 1+ \frac{Q_1^{(10)} }{r^6} ~.
\end{equation}

We now compactify along $x_4, x_5, \dots, x_9 = y$. $x_9$ is a direction along our D-brane, i.e. compactifying along this dimension reduces the dimensionality our our p-brane to a (p-1)-brane, according to the first of the two methods of compactification outlined above. This gives the metric

\begin{equation}
ds^2_{SF} = - H_1^{1/2} dt^2 + H_1^{1/2} \left( dx_1^2 + \dots + dx_4^2 \right)~~, ~~~ H_1 = 1+ \frac{Q_1^{(5)}}{r^2}~.
\end{equation}

Where $Q_1^{(5)}$ can be calculated from $Q_1^{(10)}$ in the original string frame action. Additionally, the dilatons are related as $e^{-2 \phi_5} = e^{-2 \phi_{10}} \sqrt{ G_{internal}} = H_1^{-1/4} $. We now go from the string frame solution to the Einstein frame solution, which reads

\begin{equation}
ds_E^2= H_1^{-2/3} dt^2 + H_1F^{1/3} \left( dr^2 + r^2 d\Omega_3^2 \right) ~.
\end{equation}

We then find for the Bekenstein-Hawking entropy \cite{stromvaf}

\begin{equation}
A= 2\pi^2 \lim_{r\rightarrow 0 } \left( r^3 \sqrt{\frac{Q_1}{r^2}} \right) = 0 \Rightarrow S_{\text{BH}} = \frac{A}{4} = 0~.
\end{equation}

The solution does not have an event horizon. This is due to the Kaluza-Klein scalars $\sigma$ which become singular at the event horizon and at spatial infinity. This signals that these solutions don't make sense as solutions to the lower-dimensional theory. In our case, the toriodal radii behave as $R_{5,6,7,7} \rightarrow \infty$, $R_9 \rightarrow 0 $ when we approach the horizon. We are only interested in regular solutions, so we have to look further. Finding such regular solutions is typically referred to as \textit{stabilization of the moduli}. The generic method for stabilization is to employ ratios of harmonic functions, i.e. we consider not just $H_1$ but ratios of harmonic functions.

\subsection{Solution: p-brane superposition}
We are interest in a superposition of p-branes of different kinds, where preserver part of the original supersymmetry We consider a Dp-Dp'-system which are either parallel or perpendicular in all dimensions. For such a BPS-solution, the dimensionality of the mutual transverse directions is $4k$, $k \in \mathbb{N}$. In these directions, we either (N,D) or (D,N) boundary conditions, i.e. on one brane we have Neumann and on the other brane we have Dirichlet boundary conditions.

\clearpage

\section{Black holes in string theory from p/D-branes}

\subsection{p-branes}

p-Branes are p-dimensional extended solutions of 10-dimensional (super)string theory, particularly supergravity i.e. the massless sector, coupled to n-form gauge fields. If our 10-dimensional target space is parametrized by $(t, x_1, \dots , x_9)$, then e.g. a 5-brane can by parametrized by $(t, x_1, \dots, x_5)$, where we set the transverse coordinates $(x_6, \dots x_9)$ to zero. The metric is a harmonic function associated to some charge $Q$, then 

\begin{equation}
H(r) = 1 + \frac{Q}{r^{\Delta}} ~~,~~~ \Delta ~\text{ depends on the codimension of the brane.}
\end{equation}

these solutions look similar to black holes, in particular they can have a horizon, hence they are sometimes referred to as black branes. The charge $Q$ characterizes the coupling of the p-brane to a (p+1)-form gauge field.

\begin{enumerate}
	\item p=0: particle (when we look at it from a distance) coupled to a 1-form field $A^{(1)}$. This describes a black hole
	\item p=1: string coupled to $A^{(2)}$.
\end{enumerate}

\subsection{D-branes}

D-branes are hypersurfaces in 10-dimensional space-time, on which we define boundary conditions for open strings. Namely, we fix the endpoints of the open strings fixed on the D-brane i.e. we define Dirichlet boundary conditions for open strings on the D-brane. If we take the D-brane to be p-dimensional, we refer to it as a Dp-brane. Polchinski discovered that there exists a duality between D-branes and p-branes, namely

The picture of a black hole in string theory is either as

\begin{enumerate}
	\item A bound state of p-branes for $g_s \sim 1$, where the entropy can be found in terms of the horizon area as $S_{\text{BH}} \sim A$.
	\item A bound state of D-branes for $g_s \sim 0$, where the entropy can be found in terms of the number of microstates as $S_{\text{BH}} \sim \ln N$. 
\end{enumerate}

The former picture is valid in the regime where the string coupling $g_s \sim \mathcal{O}(1)$, while the latter picture is valid when $g_s \rightarrow 0$. We can interpolate between the two pictures by employing supersymmetry. The entropy will then depend on the charges $\{Q_i \}$ of the branes, but not on $g_s$. We can thus compute the entropy in the two pictures and compare the expressions. In picture I, we compute the entropy from the area of the horizon, namely $S_{\text{BH}} = S_{\text{BH}} (Q_1 , \dots, Q_n)$. In picture II, we count the degeneracies of the open strings attached to the D-branes, namely $S_{\text{Microscopic}} = S(Q_1 , \dots, Q_n) = \log (\# \text{states})$. We will later see that these two expressions agree, thus validating this picture. The number of D-branes will be seen to correspond to the charges $\{ Q\}$. In this lecture we will focus picture I. We will construct a 5-dimensional black hole since these are easier to construct. 


 

 
 
 

 
 \subsection{10-dimensional gravity} 
 
 We choose
 
 $$ \mathcal{M} = \mathbb{R}^{1,4} \otimes T^5 ~, $$

i.e. we compactify 5 of the 10 dimensions. We have

\begin{enumerate}
	\item Double-dimensional reduction, where one wraps certain world-volume directions of the p-brane around n-dimensional cycle of $T^5$. This gives $p'= p-n$
	\item Normal dimensional reduction, where the world volume of the p-brane is in $\mathbb{R}^{1,4}$. This gives $p'= p$
\end{enumerate}

We consider three examples.

\subsubsection{D1-brane}

This case was briefly consdiered in the previous section. The D1-brane is parametrized by the coordinates $(t,y_1)$, and $T^5$ is parametrized by $(y_1,\dots , y_5)$, i.e. we wrap one dimension of our D1-brane on $T^5$ to give a 0-dimensional black hole-like object.
 
 For a D1-brane, we have
 
 \begin{equation}
 ds_{\text{string}}^2 = H_1^{-1/2} \left( -dt^2 + dy^2 \right) + H^{1/2} \left( dx_1^2 + \dots + dx_8^2 \right) ~~, ~~~ H_1 = 1 + \frac{Q_1}{r^6} ~.
 \end{equation}
 
 The dimension $\Delta= 6$ is given by the codimension D1-brane.
 
We have the dilaton $e^{-2 \phi_{10}} = H_1(r)^{-1}$, i.e. we have to rescale the metric by $e^{-2 \phi_{10}}$ to find the metric in Einstein frame

\begin{equation}
ds_E^2 = H_1^{-2/3} dt^2 + H^{1/3} \left( dr^2 + r^2 d\Omega^2_{(3)}\right) ~.
\end{equation}

This is the metric for the (1+4)-dimensional uncompactified space. The area is once again computed as

\begin{equation}
A = 2\pi^2 \lim_{r \rightarrow 0 } \left( r^3 \sqrt{\frac{Q_1}{r^2}} \right) =0 ~.
\end{equation}

We see that this is not the black hole we are looking for. This follows from the fact that this is a $\frac{1}{2}$-BPS object i.e. we have 16 supersymmetry generators in our (4+1)-dimensional target space. We will see that we need to further reduce supersymmetry by a factor 4. 

\subsubsection{D1, D5 - system} 

We now have a D1-brane in the same configuration as in the previous example. We also have a D5-brane, of which we compactify all 5 dimensions on $T^5$ so that we again recover a 0-dimensional object in target space.

The computation of the metric requires so called \textit{superposition rules}, which we do not present here. Heuristically, we need to superimpose the harmonic functions $H_1(r)$ and $H_5(r)$ associated to $Q_1$ and $Q_5$, which is possible since we are considering BPS objects. Both the D1-brane and the D5-brane have SUSY $\frac{1}{2}$ i.e. they are $\frac{1}{2}$-BPS objects, which means we further reduce the number of target space SUSY generators by a factor 2 so that the resulting target space has 8 SUSY generators. This gives the following expression for the metric

\begin{equation}
ds_E^2 = - \left( H_1 H_5 \right)^{-2/3} dt^2 + \left( H_1 H_5 \right)^{1/3} \left( dr^2 + r^2 d\Omega_{(3)}^2 \right) ~~, ~~~ H_1 = 1 + \frac{1}{r^6} ~~,~~~ H_5 = 1 + \frac{Q_5}{r^2}~.
\end{equation}

Then

\begin{equation}
A= (2 \pi)^2 \lim_{r \rightarrow 0 } \left( r^3 \sqrt{ \frac{Q_1 Q_5}{r^4} } \right) = 0~.
\end{equation}

\subsubsection{D1, D5 , pp-wave configuration}

This configuration has charges $Q_1$, $Q_5$ associated to the D1- and D5-brane, respetively, as well as $Q_k$, which is the momentum in the D1-direction. The associated harmonic functions are $H_1$, $H_5$, and $H_k$, the resulting configuration is $\frac{1}{8}$ SUSY i.e. we have $\frac{32}{8} =4$ SUSY generators in target space. We then have the metric

\begin{equation}
ds_E^2 = - \left( H_1  H_5 H_k \right)^{-2/3} dt^2 + \left( H_1 H_5 H_k \right)^{1/3} \left(  dr^2 + r^2 d\Omega_{(3)}^2 \right) ~.
\end{equation}

The area is then

\begin{equation}
A = 2 \pi^2  \lim_{r \rightarrow 0 } \left( r^3 \sqrt{\frac{Q_1 Q_5 Q_k}{r^6}}\right) = 2\pi^2 \sqrt{Q_1 Q_5 Q_k} ~.
\end{equation}

Hence

\begin{equation}
S_{\text{BH}} = \frac{\pi^2}{2} \sqrt{ Q_1 Q_5 Q_k}~.
\end{equation}

We see that this agrees with the entropy of the Tangherlini black hole when we set $Q_1 = Q_5 = Q_k$. One caveat is that this formula for $S_{\text{BH}}$ only holds for large $Q$. Next time, we will derive the same expression by counting string microstates.


\clearpage

\section{Black hole microstate counting}

The D-brane configuration under consideration consists of $Q_5$ D5-branes, $Q_1$ D1-branes, and $Q_k$ quanta of light-like (conventionally) left-moving momentum along the common compactified dimension (similar to Kaluza-Klein modes). This is an excited BPS state. The statistical entropy is given by the number of distinct ways in which we can distribute the total momentum on the available excited states of the system.  Since we have light-like momenta, we have to look for massless excitations. We use the BPS nature of our configuration to go to another point in parameter space which facilitates the calculation of the number of excited states. We consider the picture where $N g_s \ll 1$, while our calculation of the entropy in the previous lecture was performed at $N g_s \gg 1$. Here, $N$ is the number of branes, and $g_s$ is the string coupling constant. The fact that $N g_s \ll 1$ allows us to use string perturbation theory, while $N g_s \gg 1$ gives non-perturbative solutions. In the former picture, which we will use today, the D-branes will be used as boundary conditions for open strings. 

We compactify on the five torus $T^5 = \left(S^1\right)^5$; we proceed to make one of these circles $S^1$ large. The D5-branes wrap around all five copies of $S^1$ in $T^5$, while the D1-branes wraps only around the large $S^1$. We thus have internal momentum flowing along the large circle in the picture presented in the previous lecture. We thus have the following situation:\\
 
\begin{center}
	\begin{tabular}{ |c|c|c|c|c|c|c|c|c|c } 
		0 & 1 & 2 & 3 & 4 & 5 & 6 & 7 & 8 & 9\\
		\hline
	D5 & x &  &  &  & x & x & x & x & x\\
	D1 & x &  &  &  &  &  &  &  & x
	\end{tabular}
\end{center}

\vspace{.4cm}

where `x' indicates that the D-brane extends into this dimension. As we did previously, we compactify dimensions 5-8 on $T^4$, and we compactify the 9$^{\text{th}}$ dimension on the large copy of $S^1$. After dimensional reduction on $T^4$, the D-brane system is 1+1-dimensional with compact space direction. At low energies, the effective world volume theory of $Q_p$ D-branes is a $U(Q_p)$ super Yang-Mills (SYM) theory. In our case, we thus obtain a 2d SYM theory with $\mathcal{N} = (4,4)$ SUSY and gauge group $U(Q_1) \otimes U(Q_5)$. We thus have two kinds of light-like excitations.

\begin{enumerate}
	\item Excitations which start and end on the same D-brane stack, e.g. those on start and on the D1-brane. These form a vector multiplet.
	\item Excitations which start and end on different D-brane stacks. These form a hypermultiplet i.e. a supersymmetric multiplet.
	\end{enumerate}

States in the vector multiplet transform under the adjoint representation of $U(Q_1) \otimes U(Q_5)$, while states in the hypermultiplet transform under the (anti)fundamental representation of $U(Q_1)$ $\left(U(Q_5) \right)$, or under the (anti)fundamental representation of $U(Q_5)$ $\left(U(Q_1) \right)$. Here, the former case corresponds to strings that start at the D1-brane stack, while the latter correspond to strings that end there. To identify massless excitations, we have to find all flat directions of the scalar potential. This potential has a complicated valley structure with two main branches:

\begin{enumerate}
	\item Coulomb branch, named such since it corresponds to a non-zero VEV of the scalar field in the vector multiplet.
	\item Higgs branch, named such since it is characterized by a non-zero VEV of the scalar in the hypermultiplet, which couples to the gauge fields in a way that is reminiscent to the Higgs effect.
\end{enumerate}

If we are in the Coulomb branch, the scalars of the hypermultiplet become massive. Conversely, in the Higgs branch, the scalars of the vectormultiplet become massive. Hence, the two branches are mutually exclusive. We are interested in the branch with the largest number of available states i.e. which maximizes the entropy. In the Coulomb branch, the gauge group is spontaneously broken down to $U(1)^{Q_1} \otimes U(1)^{Q_5}$. The only string states that remain are those that start and end on the same brane, since those that stretch between different branes have a non-zero mass. The number of massless states we end up with is thus $Q_1 + Q_5$, which are the Cartan directions of our gauge group. In the Coulomb branch, the branes are moved away from each other so that they do not form a bound state. In the Higgs branch, the gauge group is broken to $U(1)$. Since the VEV of the scalars of the vectormultiplet vanish along the Higgs branch, all the branes stay on top of each other, hence they form a bound state. In this picture, the $U(1)$ gauge freedom parametrizes overall translations along the large copy of $S^1$.

Via a rather involved analysis of the system, one can show that the potential has $4 Q_1 Q_5$ flat directions. Heuristically, the factor 4 arises as follows. We can have strings going from the D1-stack to the D5-stack, namely, the states correspond to $Q_1 \otimes \bar{Q}_5$ and $\bar{Q}_1 \otimes Q_5$, where $\bar{Q}_i$ corresponds to the antifundamental representation. Additionally, we have a complex scalar. The complex scalar is the supersymmetric partner of two Weyl spinors, which contributes another factor of $Q_1 Q_5$ two our microstates. If we increase the size of our large $S^1$, we can reduce the energy carried by individual excitations. We can then use the infrared limit of the effective theory of massless modes.

The $\mathcal{N} = (4,4)$ SUSY implies that the infrared fixed point of the theory is a superconformal sigma model with a hyper-K\"{a}hler target space. This entails that the central charge of the theory is that of a free theory, which is $c=1$ for a boson field and $c=1/2$ for a Majorana-Weyl fermion field, so that the total central charge is 

\begin{equation}
c_{\text{tot}} = \frac{3}{2}\hspace{.1cm} 4Q_1 Q_5 = 6 Q_1 Q_5~.
\end{equation}

 We can then use Cardy's formula for the asymptotic number of states $N(E)$ for a system with total energy $E$ in a 2d CFT with compact support

\begin{equation}
N(E) = \exp \sqrt{ \pi c_{\text{tot}} E L/3}   ~.
\end{equation}

Here, $L$ is the volume of space. Note that this formula holds only when $E \gg 1$. We thus find

\begin{equation}
S_{\text{stat}} = \ln N(E) = \sqrt{ \pi c_{\text{tot}} E L/3 } = \sqrt { 2\pi Q_1 Q_5 \frac{Q_k}{R} 2 \pi R } = 2\pi \sqrt{ Q_1 Q_5 Q_k} =S_{\text{BH}}~.
\end{equation}

Where we used the fact that $E = \frac{Q_k}{R}$ and $L= 2 \pi R$. As indicated above, this matches the result of the Bekenstein-Hawking entropy we derived in the previous lecture. Note that we need the fact that the black hole is in a BPS state to consistently compare the black hole picture and the string microstate picture as presented in the two previous lectures.

A possible improvement to the calculation presented above is to consider a 4d-example, where we use D2-brane and D6-brane stacks, an NS5-brane, and a pp-wave. There, the result is

\begin{equation}
S_{\text{stat}} = 2\pi \sqrt{ Q_2 Q_6 Q_5 Q_k} ~.
\end{equation}

Where, in this case $Q_5$ is the charge of the NS5-brane. One can also extend this calculation to the non-extremal case, where our black hole has a finite temperature. This requires we also include right-moving momenta. In this picture, open strings from the left-moving and right-moving sector combine to create closed strings, which constitute Hawking radiation. Oddly enough, there is no in-moving closed string that corresponds to the antiparticle partner of the outgoing Hawking radiation. Since this picture is not BPS-saturated, the states we consider are not protected, so it is not clear whether this calculation is consistent.

\clearpage

\section{Asymptotic symmetries in general relativity and black hole hair}


\subsection{Introduction}

We consider asymptotic symmetries, which characterize the infrared (IR) structure of general relativity. In a general relativistic space-time, these asymptotic symmetries were first considered in the 1960's by Bondi, van der Burg, Metzner, and Sachs \cite{bbm} \cite{s}. Informed by the notion that flat space-times are generally invariant under the Poincar\'{e} group, they expected to find this as the asymptotic symmetry group of asymptotically flat space-times as well. However, the asymptotic symmetry group will turn out to be an infinite-dimensional extension of the Poincar\'{e} group. These asymptotic symmetries, knwon as BMS-transformations, are special kinds of diffeomorphisms which can be subdivided into \textit{supertranslations} and \textit{superrotations}. Asymptotic symmetries are also present in gauge theories such as QED and QCD, where they are related to so-called \textit{large gauge transformations}, namely those that do not go to zero at infinity \cite{qedinfr}. An important point is that these asymptotic symmetries act as global symmetries despite the fact that they are constructed from gauge symmetries i.e. even though they act non-trivially on the Hilbert space of the system. 

Although the work by B(B)MS dates from the 1960's, it received rather little attention until recently, when the corresponding BMS-charges were derived \cite{btbc} and the Ward identities corresponding to BMS-charge conservation were shown to be equivalent to Weinberg's soft graviton theorem \cite{bmswein}, as well as a recently discovered subleading soft graviton theorem \cite{srstsc} \cite{srstw}. Since we have an infinite number of diffeomorphisms with associated conserved charges, we could use these to store an infinite amount of (classical) information, including information that seems to disppear into a black hole. The remaining section will look at the BMS-symmetries as well as their recently found counterpart at the event horizon \cite{don} \cite{don2} and the role they could play in the resolution of the black hole informaion paradox. 

\subsubsection{Classical picture}

Classically, BMS transformations lead to an infinite class of space-time metrics for one particular space-time geometry (manifold), which are nevertheless physically distinct. A BMS transformation acts as

\begin{equation}
\delta_{\small{BMS}} :  g_{\mu \nu} \left( x^{\mu } \right) \mapsto \tilde{g}_{\mu \nu} \left( x^{\mu} \right) ~~~,~~ \text{$g$ and $\tilde{g}$ are physically distinct. }
\end{equation}

This construction can be used for asymptotically flat space-times, such as Minkowski space, but also for black hole space-times. The difference between $g$ and $\tilde{g}$ can be measured by the \textit{gravitational memory effect}. 

\subsubsection{Quantum picture}

Quantum-mechanically, we say that the metric describes a particular state in the Hilbert space i.e. we describe the metric $g_{\mu \nu}$ as some ket $\ket{g_{\mu \nu}}$. The variation $\delta_{BMS}$ is now promoted to an operator with associated charges $\hat{Q}_{BMS}$. These charges satisfy an algebra, which is only known for particular cases such as AdS spaces. In the quantum picture, the variation of the metric is written as
	
	\begin{equation}
	\hat{Q}_{\small{BMS}} \ket{g_{\mu \nu}} = \ket{ \tilde{g}_{\mu \nu}}~.
	\end{equation}

	We thus find an (almost) degenerate set of vacua, where each vacuum is characterized by a metric. They are almost degenerate since we might get slightly different values for physical parameters after applying the BMS transformation as above, which we will see later. This degeneracy of the vacuum that follows from the set of BMS-transformations is then related to the entropy of a given space-time geometry. The BMS transformation transforms a metric to a different metric, which entails that the BMS symmetry group is spontaneously broken. This gives rise to Goldstone modes, which are massless particles corresponding to fluctuations in the `flat' directions in the potential. 
	
	\subsection{The interpretation of the Goldstone particles}
	
	The Goldstone particles that arise due to the spontaneous breaking of the BMS group are soft gravitons, which appear in any gravitational scattering amplitude (even those at tree level). This is the gravitational construction analogous to work done on QED by Kulish and Faddeev in the 1970's.
	
	Consider a four point graviton amplitude, with momenta $p_1, p_2, p_3, p_4$. Such amplitudes are generically IR-divergent, i.e. they diverge when we let $\{ p_i \}$ go to zero. According to Weinberg, such IR divergences are cancelled by the inclusion of infinitely many soft gravitons which `dress' the final state of the scattering process i.e. the final state includes an infinite cloud of soft particles. This is called the \textit{soft theorem}. Schematically, the final state of our four-point ($2 \rightarrow 2$) scattering process is given by
	
	\begin{equation*}
	\ket{\text{final}} = \ket{\text{ two-particle state }} \otimes \ket{\text{ coherent cloud of soft gravitons }}~.
	\end{equation*}  		
	
\begin{figure}[h]
	\begin{center}
		\includegraphics[width=10cm]{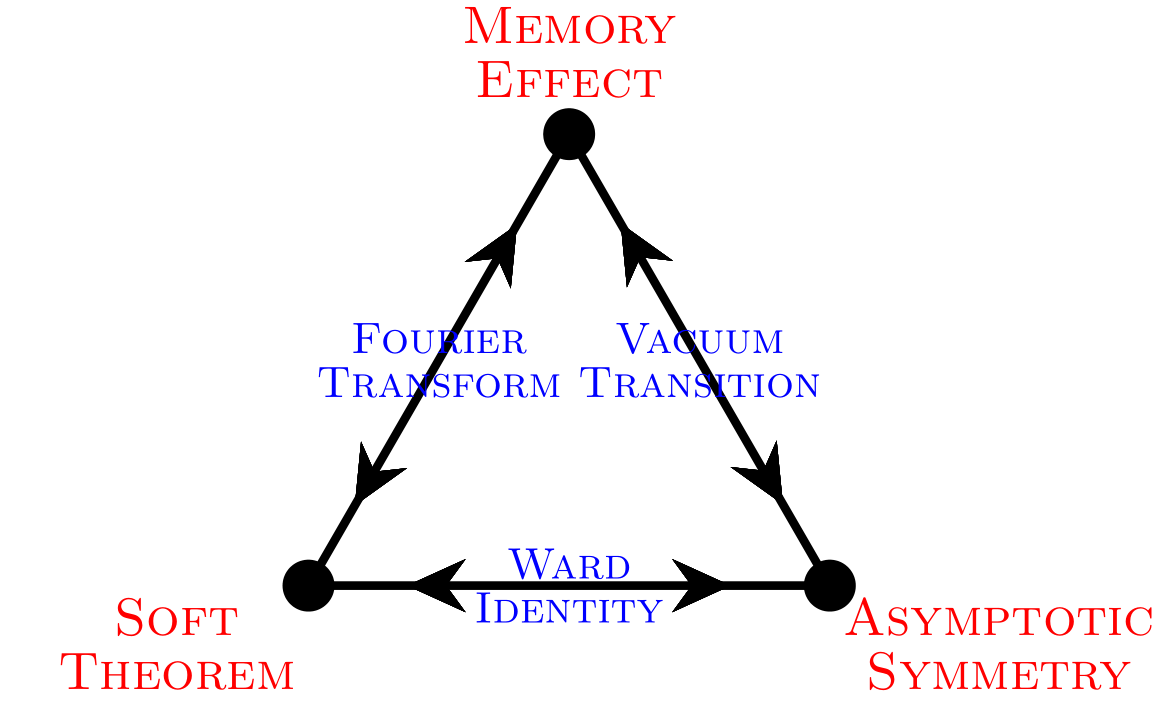}
		\caption{A schematic summary of the relation between soft gauge bosons, asymptotic symmetries, and the memory effect. The remainder of this lecture series is devoted to elucidating these three concepts and exactly how they are related. Figure taken from \cite{strominfr}. \label{strtr}}
	\end{center}
\end{figure}

This is expressed by the so-called BMS Ward identity, as recognized by Strominger. The meaning of the BMS Goldstone modes is thus that the action of $\hat{Q}_{BMS}$ creates a cloud of soft gravitons. An amplitude with $n$ soft modes is related to an amplitude with $n+1$ soft modes by a BMS transformation. The relation between soft gauge bosons (e.g. gravitons), asymptotic symmetries, and the memory effect (to be discussed later) is schematically expressed in figure \ref{strtr}.

	The BMS group provides a possible solution to the BMS information puzzle, posed very recently by Strominger in \cite{stromdec}. Consider a black hole scattering process, where we start from an initial state which is pure. The final state of this scattering process are Hawking modes, which are in a mixed state. \\
	

	Hawking modes have an energy ($\sim$ temperature) $T_H = \frac{1}{8 \pi M} > 0$ , which we consider to be \textit{hard}. Recall that the spectrum of the Hawking modes is $ n(\omega) = \frac{1}{e^{\omega/T_H} -1}$. The final state is then given by some density matrix $\rho = \sum_{\alpha} \rho_{\alpha} \ket{H_{\alpha}} \bra{ H_{\alpha}}$. Strominger argued that this picture is not complete, since each Hawking quantum has to be dressed by an infinite number of soft modes. He then proposed that the soft modes purify the final state due to their entanglement with the hard Hawking modes $H_{\alpha}$.  The full quantum process is then given by a unitary S-matrix as
	
	\begin{equation}
	S \ket{ \psi_{\text{initial}}} = \sum_{\alpha, \beta } \ket{H_{\alpha}} \ket{S_{\beta}} =  \ket{\psi_{\text{\text{final}}}}~,
	\end{equation}
	
	wo that both $\psi_{\text{initial}}$ and $\ket{\psi_{\text{final}}}$ are pure states. We will not comment further on this proposal; interested readers are encouraged to have a look at the literature.

	\subsection{Asymptotic BMS tranformations -  supertranslations and superrotations}
	
	 We investigate the aymptotic symmetries by following these steps:
	 
	 \begin{enumerate}
	 	\item Assume the space-time is asymptotically flat
	 	\item Expand the metric around spatial infinity
	 	\item Consider a special class of diffeomorphisms, namely those that leave the asymptotic structure and boundary conditions at infinity invariant. This entails that the resulting metric is also asymptotically flat
	 \end{enumerate}
	
	\subsubsection{Expansion of the metric around spatial infinity}
	
	We now introduce so-called \textit{Bondi coordinates}, which are often used to describe asymptotic symmetries. However, this are not the only possible choice; their common usage is probably due to historical reasons. We have lightcone coordinates $u$ and $v$ which parametrize $\mathcal{I}^+$ and $\mathcal{I}^-$, respectively. We then use complex coordinates $z$ and $\bar{z}$ to parametrize the two-sphere for a particular value of $u$ or $v$. The \textit{retarded} and \textit{advanced} Bondi coordinates thus given by $(u,r,z,\bar{z})$ and $(v,r,z,\bar{z})$, respectively. We have $z = i \cot (\theta/2 ) e^{i \phi}$, so that the spherical metric is given by
	
	\begin{equation}
	ds^2_{S^2} = d\theta^2 + \sin^2 \theta d \phi^2 = 2 \gamma_{z \bar{z}} dz d\bar{z} ~~,~~~ \gamma_{z \bar{z}}= \frac{2}{\left(1 + z \bar{z}\right)^2}  ~.
	\end{equation}
	
	For example Minkowski space is given by
	
	\begin{equation}
	ds^2 = -dt^2 + d \vec{x}^2 = - du^2 - 2du dr + 2r^2 \gamma_{z \bar{z}} dz d\bar{z}~.
	\end{equation}
	
	We will continue our derivation of the aymptotic symmetries next time.

	\clearpage

	\section{Asymptotic symmetries of 4D space-time geometries}
	
	BMS transformations are related to the emission of soft gravitons, which play the role of Goldstone bosons of spontaneously broken BMS symmetry. In this lecture, we will look more closely at the BMS group of asymptotic symmetries, in particular supertranslations.
	
	\subsection{Supertranslations}
	
	Supertranslations are one component of the asymptotic symmetries of gravity in asymptotically flat space-times, i.e. space-times with asymptotic boundaries $\mathcal{I}^{\pm}$. The derivation of the asymptotic symmetry group is summarized as follows.
	
	\begin{enumerate}
		\item 	We use the retarded BMS coordinates $(u,r,x^A)$ on $\mathcal{I}^+$, where $x^A$ are spherical coordinates. This corresponds to the symmetry group BMS$^+$. There are also the advanced BMS coordinates, where $u$ is replaced by $v$; this gives to BMS$^-$ which acts at $\mathcal{I}^-$. 
		
		\item We use \textit{Bondi gauge}, where we fix the local diffeomorphisms by setting $g_{rr} = 0 = g_{rA}$ and $\partial_r \det \left( \frac{g_{AB}}{r^2} \right) = 0$. This gives the metric
		
		\begin{equation}
		ds^2 = -U du^2 - e^{2 \beta} du dr + g_{AB} \left( d\theta^A + \frac{1}{2} U^A du\right) \left( d\theta^B + \frac{1}{2} U^B du\right)~.
		\end{equation}
		
		\item We go to asymptotically flat space-times, write $ x^A = (z, \bar{z})$, $z= \cot(\theta/2) e^{i \varphi}$, and expand the metric around infinity in $\frac{1}{r}$. This gives

			\begin{align}
		ds^2  = & \underbrace{- du^2 - 2 du dr + 2r^2 \gamma_{z \bar{z}} d z d \bar{z} }_{\text{Minkowski metric}} + \notag \\
		& + \frac{2 m_B}{r} du^2 + r C_{zz} dz^2 +  r C_{\bar{z} \bar{z}} d\bar{z}^2 + D^z C_{zz} du dz + D^{\bar{z}} C_{\bar{z} \bar{z}} du d\bar{z}   + \notag\\
		& + \frac{1}{r} \left( \frac{4}{3} ( N_z + u \partial_z m_B ) - \frac{1}{4} (C_{zz} C^{zz} ) \right)  du d z + \text{c.c.}  + \dots  
		\end{align}
		
	\end{enumerate}
		
		We now have three different functions: $C_{zz} ,~ m_B, ~ N_z$, which are functions of $u,~ z,~ \bar{z}$, but not of $r$. \\
		
		\begin{itemize}
		
	\item	$m_B$ is called the \textit{Bondi mass aspect}. We can integrate it as $ \int_{S^2 } m_B (z ,\bar{z}) dz d\bar{z}$ to give the total Bondi mass, which equals the ADM mass in the case of a black hole.
		
	\item	$N_z$ is called the \textit{angular momentum aspect}. We can integrate it as $ \int_{S^2 } N_z V^z (z ,\bar{z}) dz d\bar{z}$ to give the total angular momentum.
		
	\item	$C_{zz}$ and $C_{\bar{z}\bar{z}}$ are gauge potentials of a gravitational wave, akin to (electromagnetic) vector potentials. 
		
	\item	$N_{zz} \coloneqq \partial_u C_{zz}$ is called the \textit{Bondi news}. $N_{zz}$ and $N_{\bar{z}\bar{z}}$ correspond to the two helicity components of a spin-2 gravitational wave. This quantity is similar to the electromagnetic field strength.
		
	\end{itemize}
	
	\vspace{.2cm}
	
The supertranslations are diffeomorphisms that keep the Bondi gauge and the asymptotic structure of the metric invariant. These are residual large gauge transformations, as they have a non-trivial action at the boundary of space-time. One could naively expect that the only residual symmetries are the Poincar\'{e} group, the finite-dimensional symmetry group of the Minkowski metric. This was indeed the expectation of B(B)MS when they started their calculations. Surprisingly, however, BMS$_{\pm}$ turns out to be infinite-dimensional and contains the Poincar\'{e} group as a subgroup. Hence, general relativity does not simply reduce to special relativity at large radial distances and weak fields. This implies that there is a large space of degenerate metrics (vacua), even in the case of Minkowski space.

More concretely, consider diffeomorphisms which are given by some vector $\xi^{\mu} ( u ,r ,z , \bar{z})$, which we will relate to some $\xi^{\mu} (z , \bar{z})$. We have $ \xi^u, ~ \xi^r \sim \mathcal{O}(1)$, $ \xi^z $ and $ \xi^{\bar{z}} \sim \mathcal{O} \left(\frac{1}{r} \right)$. Then

\begin{align}
 \left( \mathcal{L}_{\xi} g \right)_{ur}   & = - \partial_u \xi^u + \mathcal{O} \left( \frac{1}{r} \right) \notag \\ 
 \left( \mathcal{L}_{\xi} g \right)_{zr} & = r^2 \gamma_{z \bar{z}} \partial_r \xi^{\bar{z}} - \partial_z \xi^u + \mathcal{O} \left( \frac{1}{r} \right) \notag \\
 \left( \mathcal{L}_{\xi} g \right)_{z \bar{z}} &= r \gamma_{z \bar{z}} \left[ 2 \xi^r  + r D_z \xi^z + r D_{\bar{z}} \xi^{ \bar{z}} \right] + \mathcal{O}(1) \notag\\
  \left( \mathcal{L}_{\xi} g \right)_{uu} & = -2 \partial_u \xi^u - 2 \partial_u \xi^r + \mathcal{O} \left( \frac{1}{r} \right)~.
	\end{align}
	
	Where $D_z$ and $D_{\bar{z}}$ are covariant derivatives with respect to $\gamma_{z \bar{z}}$ i.e. they are covariant derivatives on the \underline{unit} sphere. The solution to these equations is

	\begin{equation}
	\xi  = f \partial_u + \frac{1}{r} \left( D^z f \partial_z + D^{\bar{z}} f \partial_{\bar{z}} \right) + D^z D_z f\partial_r ~~~, ~~ f= f(z,\bar{z}) ~.
	\end{equation}
	
	This is the generator of BMS supertranslations. It is given in terms of an \underline{arbitrary} function $f(z , \bar{z} )$, hence we have an infinite family of supertranslations.
	
		\begin{figure}[h]
		\begin{center}
			\includegraphics[width=8cm]{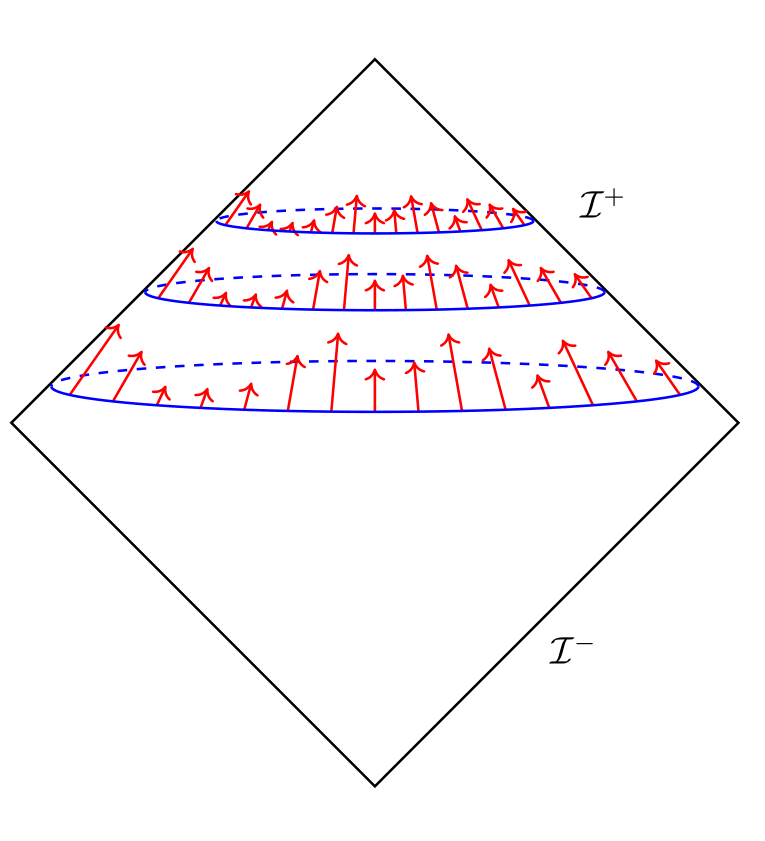}
			\caption{Under the action of a supertranslation in BMS$_+$, $\mathcal{I^+}$ undergoes a translation in $u$ that is a meromorphic function of the angular coordinates $(z,\bar{z})$. Image adapted from \cite{strominf}. }
		\end{center}
	\end{figure}

	As seen in the picture, a supertranslation is nothing but an angle-dependent shift of the retarded time at $\mathcal{I}^+$. This is a generalization (localization) of the four Poincar\'{e} translations. If we take $f(z, \bar{z})$ to be constant, we find $u$-translations. If we take $f(z ,\bar{z})$ to be an $ \ell =1 $ harmonic function, this corresponds to a spatial translation. All supertranslations act non-trivially i.e. they transform physically inequivalent metric into each other. This gives a measurable effect, which is known as the \textit{gravitational memory effect}. This is a gravitational analogue of the Aharonov-Bohm effect. We now look at the action of a supertranslation parametrized by some function $f$ on the quantities given above, which is given by
	
	\begin{align}
	\label{BMSqnt}
		\mathcal{L}_f m_B & = f \partial_u m_B + \frac{1}{4} \left[ N^{zz} D^2_z f + z D_{\bar{z}} N^{zz} D_z f + \text{c.c.} \right]  \notag \\
	\mathcal{L}_f N_{zz} & = f \partial_u N_{zz} ~~~~~~~~~~~~~~~~~~~~ , ~~~ \text{field strength} \notag\\
	\mathcal{L}_f C_{zz} & = f \partial_u C_{zz} - 2 D^2_z f(z ,  \bar{z}) ~~ ~, ~~~ \text{potential}~.
	\end{align}
	
	Note that we have an infinite number of choices for $f(z, \bar{z})$, hence we have an infinite family of metrics.
	
	\subsection{Example: Minkowski space}

	In Minkowski space, $m_B = 0$,  $N_{zz} = 0$, and $C_{zz} =0 $, where the last equality holds in the standard Minkowski metric $ds^2 = - dt^2  + d \vec{x}^2$. We now perform a BMS supertranslation given by some $f(z, \bar{z})$, which will result in another metric which we denote by Minkowski$'$. One sees from \ref{BMSqnt} that the result is given by
	
	\begin{equation}
	m_B' =0 ~~, ~~~ N_{zz}' = 0~~, ~~~ C_{zz}' = 2 D_z^2 f(z, \bar{z})
	\end{equation} 
	
	One can easily show that the curvature components will still be equal to zero i.e. $R_{\mu \nu \rho \sigma} = 0$, which is generally the case if $C_{zz}$ and $C_{\bar{z}\bar{z}}$ equal to the second derivative of some function. Hence, for Minkowski space, we get an infinite family of flat Minkowski metrics. The manifold $C$ of flat vacua (similar to a moduli space) is given by
	
	\begin{equation}
	C = \{  C_{zz} \mid C_{zz} = 2 D^2_z \Phi(z,\bar{z}) ~ \}
	\end{equation}

	 Hence, the space of flat vacua is isomorphic to the space of functions $\Phi (z, \bar{z})$ on $S^2$. Namely, a point in $C$ corresponds to a point in $S^2$. A BMS supertranslation transforms $\Phi$ as
	 
	 $$ \Phi(z,\bar{z}) \rightarrow  \Phi(z, \bar{z} ) - 2 f(z, \bar{z})  $$
	 
	 Minkowski space in Minkowski coordinates in Bondi coordinates is
	 
	 $$ ds^2 = -du^2 - du dr + 2r^2 \gamma_{z \bar{z}}dz d\bar{z} ~~, ~~~ \Phi ( z, \bar{z}) =0 ~. $$
	 
	 After a BMS transformation, the metric is of the form
	 
	 $$ ds^2 = -du^2 -2 du dr + 2r^2 \gamma_{z \bar{z}} dz d\bar{z} + r C_{zz} dz^2 + r C_{\bar{z} \bar{z}} d \bar{z}^2~.$$
	 
	 This is still flat Minkowski space, i.e. $R_{\mu \nu \rho \sigma} = 0$, if and only if $C_{zz} = 2 D^2_z f(z , \bar{z})$.

	\subsection{Quantum-mechanical picture}
	
	In the quantum picture, we write these results as
	
	\begin{align}
&	\text{\underline{Classically}} \hspace{4cm} \text{\underline{Quantum mechanically}} \notag \\
	&C_{zz} = D_z^2 \Phi \hspace{3.8cm} \ket{\Phi} \hat{=} \ket{C_{zz}} \notag\\
	&\Phi \rightarrow   \Phi -2f  \hspace{3.68cm} \ket{\Phi} \rightarrow \hat{Q}_f \ket{\Phi} =  \ket{\Phi - 2f}~.
	\end{align}

	The generator $\hat{Q}_f$ does not annihilate the vacuum i.e. the BMS group is spontaneously broken, as mentioned during the last lecture. This gives rise to massless Goldstone modes. Classically, the Goldstone modes are just the variations of the metric of the form $D_z^2 f \sim \mathcal{L}_f g$. We will make this a bit more precise in due time by considering the gravitational memory effect.

	Consider two classical vacua at some initial time $u_i$ and some final time $u_f$, which differ by a BMS supertranslation. We construct an interpolating space-time that brings us from the vacuum at $u_i$ to the vacuum at $u_f$. At $u_i$, we take $C^i_{zz} = D^2_z \Phi(z , \bar{z})$, and at $u_f$, we have $C^f_{zz} = D^2_z \Phi^f(z,\bar{z})$, where $\Phi^i - \Phi^f = 2f$. We now look for an interpolating metric, which is described by a function $C^{\text{int}}_{zz} ( u, z, \bar{z})$ such that  $ C^{\text{int}}_{zz} ( u_i, z, \bar{z}) = C^i_{zz}$
 and $ C^{\text{int}}_{zz} ( u_f, z, \bar{z}) = C^f_{zz}$. The interpolating metric must have some non-trivial time dependence. This gives a gravitational wave (pulse), since $N_{zz}^{\text{int}} = \partial_u C^{\text{int}}_{zz} \neq 0 $. The picture is as follows. If we send out a gravitational wave past $\mathcal{I}^+$ and turn it off, this will have a measurable effect on $\mathcal{I}^+$ which is known as the gravitational memory effect.  

\clearpage

\section{BMS charges}

We are discussing a family of asymptotically flat metrics (which includes Minkowski space). We do this in the Bondi gauge using Bondi coordinates, as presented during the previous lecture. The Cauchy data (or parameters) divide into the following subclasses:

\begin{enumerate}
	\item \textbf{Mass}: Bondi mass aspect $m_B(u,z,\bar{z})$
	\item \textbf{Angular momentum}: Bondi angular momentum aspects $N_{z}$ and $N_{ \bar{z}}$
	\item \textbf{Radiation}: Gauge potentials $ C_{zz} (u,z,\bar{z})$ and $C_{\bar{z}\bar{z}}(u,z,\bar{z})$, which describe gravitational waves
	\item \textbf{Field strength}: Bondi news $N_{zz} = \partial_u C_{zz}$ and $N_{\bar{z}\bar{z}} = \partial_u C_{\bar{z}\bar{z}}$
\end{enumerate}

What are the symmetries that keep the asymptotic form of the metric invariant? These are the asymptotic Killing vectors (AKV's), which are of the form

\begin{equation}
\xi_{f(z,\bar{z})}^{\mu} = (\xi^u,  \xi^z , \xi^{\bar{z}}, \xi^r) = \left( f(z,\bar{z}), \hspace{.1cm} -\frac{1}{r} D^z f(z,\bar{z}), \hspace{.1cm} -\frac{1}{r} D^{\bar{z}} f(z,\bar{z}), \hspace{.1cm} D_z D^z f(z,\bar{z}) \right) .
\end{equation}

Where $f(z,\bar{z})$ are functions on $S^2$. Note that these AKV's keep the asymptotic form of the metric invariant but act non-trivially on the interior of our space-time. Namely, the variation with respect to an AKV given by some function $f(z,\bar{z})$ is given by

\begin{equation}
 \delta_{f(z,\bar{z})} C_{zz}= f(z,\bar{z}) \underbrace{\partial_u C_{zz}}_{N_{zz}} - 2 D_z^2 f(z,\bar{z})~.
\end{equation}

The second term gives rise to an infinite family of supertranslated metrics. Namely, Minkowski space corresponds to $m_B, ~ N_z, ~ N_{zz}$ all equal to zero. The family of Minkowski vacua is then given by

\begin{equation}
C = \{ C_{zz}  \mid C_{zz} = 2 D_z^2 \Phi(z,\bar{z}) \}~.
\end{equation}

A supertranslation then acts as $ \Phi(z,\bar{z}) \rightarrow \Phi(z,\bar{z}) + f(z,\bar{z})$. \\

\underline{Remarks}:

\begin{enumerate} 
	
	\item The soft quanta (Goldstone bosons of broken BMS symmetry) have zero energy. These given by

\begin{equation}
\lim_{\omega \to 0 } \Delta C_{zz} = \lim_{\omega \to 0 } \int_{u_i}^{u_f}  du \hspace{.1cm} e^{i \omega u } \partial_u C_{zz} ( u, z, \bar{z}) =  C_{zz}(u^f, z, \bar{z}) - C_{zz} ( u^i , z, \bar{z}) = \delta_f C_{zz} = \underbrace{ 2 D_z^2 f(z, \bar{z})}_\text{Goldstone mode}
\end{equation}

In field theory language, this is written as 

$$\ket{\text{Goldstone mode}} = \ket{\text{Soft graviton}}~.$$

In words, this means that Minkowski space is equipped with an arbitrary number of soft modes.

\item The field equations are the familiar Einstein equations

$$ R_{\mu \nu } - \frac{1}{2} g_{\mu \nu } R = 8 \pi G T^{M}_{\mu \nu}~.$$

Where $T^{M}_{\mu \nu}$ is the energy-momentum tensor corresponding to the matter content of our space-time. For the asymptotically flat metrics, we obtain the following on-shell condition

\begin{equation}
\label{dmb}
\partial_u m_B = \frac{1}{4} \left[ D_z^2 N_{zz} + D_{\bar{z}}^2 N_{\bar{z}\bar{z}} \right] - T_{uu} ~~ ~,~~  T_{uu} = \frac{1}{4} N_{zz} N^{zz}  + 4\pi G \lim_{r \to \infty} \left[ r^2 T_{uu}^{M}\right] ~.
\end{equation}

\end{enumerate}

\subsection{Asymptotic charges}

The asymptotic charges can also be defined for the bulk but are only conserved at the null-like boundaries of space-time denoted by $\mathcal{I}^{\pm}$. These charges are Noether charges, which are typically derived in Hamiltonian formalism. A Hamiltonian corresponds to a symplectic structure, which is a symplectic form that depends on position and momentum variables $\omega(p,q)$. We will not derive the symplectic structure here, but instead present the result. The charge corresponding to the supertranslations are
 
 \begin{align}
 Q_{f(z, \bar{z})}   =  \omega ( C_{zz} , \underbrace{\delta_f C_{zz}}_{\mathcal{L}_f C_{zz}} ) & =  ~~ \frac{1}{4 \pi G } \int_{\mathcal{I}^{ \pm} \cong S^2 } dz d\bar{z} \hspace{.1cm} \gamma_{z \bar{z}} f(z, \bar{z}) m_B (u ,z ,\bar{z}) \notag\\
  & = - \frac{1}{4\pi G} \int_{\mathcal{I}^{ +}  \cap \mathcal{I}^{ -} \equiv C } du dz d\bar{z} \hspace{.1cm} f(z, \bar{z}) \partial_u m_B \notag\\
  &= ~~ \frac{1}{4 \pi G } \int_C du dz d\bar{z} f(z , \bar{z} ) \left[ T_{uu} - \frac{1}{4} \left( D_z^2 N^{zz} + D_{\bar{z}}^2 N^{\bar{z} \bar{z}} \right) \right] \notag\\
  & = - \frac{1}{8 \pi G  } \int_C du dz d\bar{z} C_{zz} \partial_u \left( \delta_f C_{zz} \right) ~~~~~ , ~~~ \delta_f C_{zz} = 2 D_z^2 f~.
 \end{align}

Here, $\mathcal{I}^{ +}  \cap \mathcal{I}^{ -} \equiv C$ is a Cauchy surface for our space-time. We have an infinite number of charges since $f(z, \bar{z})$ is allowed to be meromorphic. We see that the charges correspond to the $u$-derivative of the Goldstone mode. We can then compute the charge algebra as

\[\{ Q_f, Q_{f'}  \}_{\text{P.B.}} = Q_{f+f'} ~, \] 

 This gives two copies of the Virasoro algebra, each of which is of the form $ \left[  L_m , L_n \right] = (m-n) L_{m+n}$. This tells us that we are simply performing conformal transformations on $S^2$. \\ 

\underline{Remarks}: 

\begin{enumerate}

	\item The charges are related to soft gravitons $\equiv$ Goldstone bosons given by the ket $\ket{GB}$ via the standard low energy theory as follows
	$$ Q \ket{0} = \ket{0} + \ket{GB} = \ket{0'}$$
	
	\item The expectation values for the charges $Q_f$ for static and stationary backrgounds, i.e. $u$-independent backgrounds, are classically zero. In quantum language
	$$ \langle Q  \rangle = \obket{0}{Q}{0} = \bket{0}{0'} = 0$$ 
	
	However, the following expression is not equal to zero
	
	$$ \obket{0}{Q}{GB} = \bket{GB}{GB} \neq 0 $$
	\end{enumerate}

\subsection{Superrotations}

Superrotations are generalizations of standard rotations. We saw before that the supertranslation chagres are expressed in terms of the Bondi mass aspect $m_B$. We will see that the superrotation charges are  expressed in terms of the angular momentum aspect $N_A \equiv \left( N_z, N_{\bar{z}} \right)$. The superrotations are given by functions $Y_A (z ,\bar{z}) = (Y_z , Y_{\bar{z}} )$. The corresponding charges are

\begin{equation}
Q_{Y_{A}} = \frac{1}{8 \pi G } \int_{ \mathcal{I}^{\pm} \cong S^2 }  dz d\bar{z} \hspace{.1cm} \gamma_{z \bar{z}} Y^A ( z , \bar{z}) N_A ( u, z ,\bar{z})
\end{equation}



They destroy asymptotic flatness, in the sense that they generate a conical deficit angle in space-time. This corresponds to the creation of a cosmic string into the bulk. The charges $Q_f$ correspond to generators $T_{(m,n)}$, while $Q_Y$ correspond to Virasoro generators $L_n$. They satisfy the following algebra:

\begin{align}
\left[ L_m, L_n \right] & = (m-n) L_{m+n} ~, \notag \\ 
\left[ L_m , T_{(p,q)} \right] & = -p T_{(p+m, q)}~, \notag \\
\left[ T, T \right] & = 0~.
 \end{align}

The second expression tells us that supertranslations give rise to superrotation charges.

\subsection{BMS-like transformations and charges on the horizon of Schwarzschild space-time}

A similar structure to asymptotic BMS has been found at the event horizons of a black hole, which we denote by $H^{\pm}$ \cite{don}. We follow the same steps as before:

\begin{enumerate}
	\item Expand the metric around $H^+$
	\item Look for those diffeomorphisms that keep the asymptotic expansion invariant, which we write as BMS$^H$. 
	\item Derive a family of black hole metrics $\delta_{ g(z, \bar{z}) } g^{\text{BH}}_{\mu \nu} = g^{\text{BH}'}_{ \mu \nu} $, where $g(z, \bar{z})$ now plays the role that $f(z, \bar{z})$ played before.
		\item Find Goldstone modes which correspond to gravitational waves which pass through $H^+$
		\item Compute the horizon charges $Q_g$. We will find that, for classical, static black holes, $\langle Q_g \rangle = 0$. Quantum mechanically, we will find that the operators $\hat{Q}_g$ corresponding to the charges $Q_g$ will have a non-zero VEV. 
		\end{enumerate}

\clearpage

\section{The gravitational memory effect}

The gravitational memory effect is a change in spatial separation between two points (on $\mathcal{I}$) due to the passing of gravitational radiation, which is characterized by $N_{zz} \neq 0$. Namely, if we start from two points that are separated by a distance $(s^z, s^{\bar{z}})$ and let gravitational waves pass between them, the change in $s^z$ in retarded coordinates is given by

\begin{equation}
\Delta s^z = \frac{\gamma^{z \bar{z}}}{2r} \Delta C_{\bar{z} \bar{z}}s^{\bar{z}} ~.
\end{equation}

We start from

\begin{equation}
C_{zz}  = -2 D_z^2 C(z,\bar{z}) ~,
\end{equation}

for some function $C(z,\bar{z})$. From \ref{dmb}, we then find that 

\begin{equation}
D_z^2 \Delta C^{zz} = 2 \Delta m_B + 2 \int du T_{uu} ~.
\end{equation}

The solution is given by

\begin{equation}
\Delta C(z,\bar{z}) = -\int d^2 w \gamma_{w\bar{w}} G(z,\bar{z} ; w ,\bar{w}) \int du \left( T_{uu}(w,\bar{w})+\Delta m_B \right) ~,
\end{equation}

where the Green's function is given by

\begin{equation}
G(z,\bar{z} ; w ,\bar{w}) = \frac{ \lvert z-w\rvert^2}{\pi (1+z \bar{z})(1+w\bar{w})} \log \left[ \frac{ \lvert z-w\rvert^2}{(1+z \bar{z}) (1+w \bar{w})} \right] ~~~~~, ~~~ D_z^2 D_{\bar{z}}^2 G(z,\bar{z} ; w ,\bar{w}) = \gamma_{z \bar{z}} \delta^{(2)}(z-w) + \dots
\end{equation}

One can rewrite it as \cite{szsoft}

\begin{equation}
G(z,\bar{z} ; w ,\bar{w}) = \frac{1}{\pi} \sin^2 \frac{\Delta \Theta}{2} \log \sin^2 \frac{\Delta \Theta}{2}
\end{equation}

where $\Delta \Theta $ is the angle on the two-sphere at $\mathcal{I}$ between $ (z,\bar{z})$ and $(w,\bar{w})$. Note that the Green's function effect vanishes for small $\Delta \Theta$ and attains its maximum absolute value at $\Delta \Theta = \pi / 2$.

\clearpage

\section{Current research on BMS-like transformations and charges of black holes}

The BMS-like charges we consider here are also referred to as BMS soft hair, since they correspond to soft gravitons and extend the notion of black hole hair. Today we will focus on the classical picture, the next lecture will focus on the quantum model. Consider Schwarzschild geometry in Eddington-Finkelstein coordinates in advanced time $v=t+r^*$. The only difference of this form of the metric with that of the Bondi coordinates the fact that we use $r^*$ instead of $r$

\begin{equation}
ds^2 = - \left( 1- \frac{r_s}{r} \right)^2 dv^2 + 2 dv dr^* + r^2 \gamma_{z \bar{z}}dz d\bar{z}  ~.
\end{equation}

We see that if we let $r_s \rightarrow 0 $, we recover (flat) Bondi coordinates. We have a family of metrics

\begin{equation}
ds^2 = - \left( 1 - \frac{r_s}{r} \right)^2 dv^2 + 2 dv dr + r^2 \gamma_{z \bar{z}} dz d\bar{z} + \underbrace{\frac{r_s  r}{r - r_s}  C^H_{zz} (v, z, \bar{z}) dz^2 + \frac{r r_s}{r-r_s} C^H_{\bar{z} \bar{z}} (v , z ,\bar{z}) d \bar{z}^2    }_\text{radiation through horizon}  ~.
\end{equation}

We now consider diffeomorphisms that preserve the gauge choice and the form of the metric in the limit where we approach the event horizon. The corresponding asymptotic Killing vectors are given by an infinite set of functions $g(z, \bar{z})$

\begin{equation}
\chi_g^{\mu} = g(z, \bar{z}) \partial_v - \frac{r_s -r}{r r_s} \left( D_z^2 g \hspace{.1cm} \partial_z + \text{h.c.} \right) + D^z D_z g(z ,\bar{z}) \partial_r  ~.
\end{equation}

We refer to the corresponding modes as H-modes, which will later be contrasted with A-modes, to be defined below. The gauge potentials transform as

\begin{equation}
C_{zz}^H \rightarrow C_{zz}^H - 2 D_z^2 g(z, \bar{z})  ~.
\end{equation}

This transformation has the same form as the one we found in previous lectures for $f(z, \bar{z})$. We thus get a family of equivalent Schwarzschild vacua characterized by

\begin{equation}
C^H = \{ C_{zz}^H \mid C_{zz}^H = 2 D_z^2 \Phi(z, \bar{z})\}  ~.
\end{equation}

Which is of the same form found previously at $\mathcal{I}$. One can then show that the mass of the black hole is unchanged. Additionally, we want to check if the family of metrics also correspond to vacuum solutions. Namely, if we start from some $g_{\mu \nu}^0$ and vary with respect to an AKV, we wish to check that the following holds

\begin{equation}
g_{\mu \nu}^g = g^0_{\mu \nu} + \delta_{ \chi} g_{\mu \nu } ~~~~,  ~~~ R_{\mu \nu}\left( g_{\mu \nu}^g \right) = 0  ~.
\end{equation}

One can show that the angular components of the Ricci tensor $\left( R_{\theta \phi}, \dots \right)$ vanish on the horizon. The other components of the Ricci tensor vanish at the linearized level, which corresponds to the fact that such a perturbation corresponds to a gravitational wave. This picture is equivalent to the so-called \textit{membrane paradigm} \cite{penmem}.

\subsection{Soft gravitons}

We have the field strengths analogues $N_{zz}^H = \partial_v C_{zz}$ and $N_{\bar{z} \bar{z}}^H = \partial_v C_{\bar{z} \bar{z}}$, corresponding to the two polarizations of a (soft) graviton. Then

\begin{equation}
N_{zz}^{H, w=0} = \lim_{ \omega \rightarrow 0} \int_{v_i}^{v_f} e^{i \omega v} \partial_v  C_{zz}^H = \hspace{-3cm} \underbrace{2 D_z^2 g(z,\bar{z})}_\text{\hspace{2.5cm}Goldstone modes of Schwarzschild space-time}  ~,
\end{equation}

which makes explicit the fact that we are dealing with a soft $(\omega \to 0)$ limit. The supertranslation charges at the horizon are then given by 

\begin{align}
Q^H & = ~ \int_{I_{\text{BH}} \simeq H^{\pm}  \cup \mathcal{I}^{\pm}} \hspace{-1cm} dv dz d\bar{z} \hspace{0.1cm} \partial_v C_{zz}^H D^2_z g(z ,\bar{z})  \notag\\
& = - \int_{I_{\text{BH}} } dv dz d\bar{z} \hspace{0.1cm} C_{zz}^H \partial_v \left(D_z^2 g \right) \notag\\
& = - \frac{1}{2} \int_{ I_{\text{BH}}}  dv dz d\bar{z} \hspace{0.1cm} C_{zz}^H \partial_v \left( \delta_g C_{zz}^H \right)  ~.
\end{align}

For static (stationary) black hole metrics, $Q^H = 0$, but $ Q^H \neq 0 $ for a collapsing shell of matter.

\subsection{A-modes}

We see in the H-modes that we have $\frac{r_s- r}{r_s r } = \frac{1}{r} - \frac{1}{r_s}$. The first term corresponds to the familiar BMS transformations for flat (Minkowski) metrics, while the second term is new and arises from the presence of a black hole \cite{adgl,adgl2,diet}. We now consider

\begin{equation}
\chi_g'= g(z ,\bar{z}) \partial_v + \frac{1}{r_s } \left( D_z^2 g \partial_z + \text{h.c.} \right) +  D_z^2 g \partial_z   ~,
\end{equation} 

in the context of the metric

\begin{equation}
ds^2 = - \left(1- \frac{r_s}{r} \right) dv^2 + 2 dv dr + r^2 \gamma_{z \bar{z}} dz d\bar{z} + r_s C_{zz}^A dz^2 + r_s C_{\bar{z} \bar{z}}^A d \bar{z}^2  ~.
\end{equation}

This corresponds to $C_{zz}^A \rightarrow C_{zz}^A - 2 D_z^2 g (z ,\bar{z})$, so that we can write

\begin{equation}
\delta_{\chi_g} = \begin{pmatrix}
 0& 0& 0& 0 \\
 0& 0& 0& 0 \\
 0 & 0& \alpha & \beta\\
 0 & 0& \beta & \gamma
\end{pmatrix}  ~,
\end{equation}

where

\begin{align}
 \alpha &= - 2 \frac{r^2}{r_s} \frac{ \partial^2 g}{ \partial \theta^2} ~, \notag\\
 \beta&= -2 \frac{r^2 }{r_s} \left( \frac{ \partial^2 g }{\partial \theta \partial \varphi}  \right) ~, \notag \\
 \gamma&=-  2 \frac{r^2 }{r_s^2} \left( \frac{\partial^2 g}{ \partial \varphi^2} + \sin \theta \cos \theta \frac{\partial g}{\partial\theta} \right) ~.
\end{align}

The A-modes are the Goldstone modes of the family of black hole space-times. We make the following remarks:

\begin{enumerate}
	\item $Q^A =0 $ for static or stationary solutions. 
	 \item Since the Goldstone modes are $\omega =0 $ gravitational waves, they are gapless modes i.e. they carry no energy and leave the mass of the black hole invariant.
	 
\item The wavelength of the A-mode is finite, namely $\lambda_A \approx r_s$. That this is consistent with the fact that they have no energy is due to the infinite redshift at the horizon.

\item A-modes correspond to gravitational waves that travel along the horizon. This is reminiscent of open strings travelling along the black hole, whose degeneracy gives rise to black hole entropy in string theory. 
\item Clasically, we have an infinite number of charges i.e. infinitely many Goldstone modes. We can expand $g( z \bar{z})$ in angular coordinates as

$$g ( \theta ,\varphi) = \sum_{\ell= 0 }^{\infty} \sum_{m=-\ell}^{ + \ell} b_{\ell m } Y_{\ell m} (\theta , \varphi) ~.$$

We thus have one charge $b_{\ell m}$ for each pair $(\ell,m)$ subject to $-\ell \leq m \leq \ell$. These charges are referred to as soft hair.
\end{enumerate}

The classical entropy $S= \frac{A}{4 \hbar G_N } \xrightarrow{\hbar \to 0} \infty $, but this entropy needs an infinite amount of time to be released i.e. it is not classically accessible.

\subsection{Quantum picture of a Schwarzschild black hole}

The physical picture considered here has been developed by Dvali and Gomez \cite{dvgom}. We will see that the classical picture presented above corresponds to a large N-limit of the quantum picture, in a way that is reminiscent of AdS/CFT correspondence. The black hole seems to correspond to a bound state of N gravitons, akin to a Bose-Einstein condensate. We refer to the Goldstone modes as Bogoliubov modes for reasons that will become clear in due time. These modes are the collective excitations of the graviton bound state.  Schematically, we will rename $ \ket{\text{Goldstone}} \equiv \ket{\text{Bogoliubov}}$. The collective mode is formed at the quantum critical point, where the gravitons condense. We have a graviton condensate which we refer to as $\phi$, such that $\langle \phi \rangle \sim \langle gg \rangle$, i.e. the gravitons form a condensate corresponding to a chage $ \hat{Q}^A$. Then, $\hat{Q}^A  \ket{\text{BH}} = \ket{\text{BH'}}$.

\subsubsection{Toy model}

In a condensed matter toy model one can calculate that $M = \sqrt{N} M_p + \mathcal{O} \left( \frac{1}{N}\right)$ and $r_s = \sqrt{N} L_p + \mathcal{O}\left( \frac{1}{N} \right)$, so that 

$$ S \sim r_s^2 = N+ \mathcal{O}(1)  \xrightarrow{N \to \infty}  \infty  ~. $$

There is a collective coupling, given by $\lambda = \alpha N $, where $\alpha$ is the coupling of individual gravitons. The critical point is then at $\alpha = \frac{1}{N} \sim \frac{1}{r_s^2}$ i.e. $\lambda =1$, which is familiar from condensed matter physics. This toy model corresponds to bosons on a circle. We see that $\omega_i = 0 + \mathcal{O} \left( \frac{1}{N} \right)$ at $\lambda = 1$ i.e. the non-degenerate spectrum collapses to a point where it becomes completely degenerate. The $\mathcal{O} \left( \frac{1}{N} \right)$ term signals a breakdown of degeneracy, i.e. the Goldstone modes will get masses of $\mathcal{O} \left( \frac{1}{N} \right)$, so that they become \textit{pseudogoldstone modes}.  The limit $\lambda \to 1$ corresponds to infinite redshift, so that the event horizon can be seen to correspond to the quantum critical point of the BEC.


\clearpage

\section{Quantum hair and quantum black hole vacua}

We have seen that there is an infinite family of classical metrics that describe black hole, particularly Schwarzschild. We transform the metric as follows

\begin{equation}
\delta_{ \chi} g_{\mu \nu}(x) + g_{\mu \nu} ~.
\end{equation}

This gives rise to an infinite number of Goldstone modes. The BMS groups at radial infinity and at the event horizon are related as follows\\

\begin{center}
\begin{tabular}{c|c}
	\textbf{BMS$_{\infty}$} & \textbf{BMS$_H$ } \\ 
	\hline 
\vspace{.1cm}	$\omega = 0$, $\lambda = \infty$ ($ k=0$) & $\omega = 0$, $\lambda = r_s$ \\
	$\omega = ck$ & Infinite redshift at horizon, $ \omega = g_{rr} c k$
\end{tabular}
\end{center}

\vspace{.6cm}

Let us now assume that a black hole is a bound state of $N$ gravitons for $N \to \infty$. The individual coupling between gravitons goes as $\alpha(E) = \frac{E^2}{M_{Pl}^2} = \frac{L_p^2 }{r^2}$. The \textit{collective coupling} is then given by $\lambda = \alpha(E) N = \frac{L_p^2 }{r^2 } N ~ (= \lambda(r))$. At $\lambda(r) =1$, we have quantum criticality i.e. the gravitons condensate. $\lambda(r)$ can be seen as the \textit{holographic coupling}, namely, $\lambda$ gives rise to an emergent (radial) coordinate in a renormalization type picture. Then, $\lambda=1 $ is the fixed point of our renormalization group flow. A similar picture arises in AdS/MERA correspondence. For $\lambda=1$, $r_s^2 = N$, hence $M= \sqrt{N} M_p$ and $r_s = \sqrt{N} L_p$, and $S \sim A \sim r_s^2  \sim N$, so that $N$ can be viewed as the number of microstates. 

\subsection{Collective excitations}

The collective excitations are called Bogoliubov modes. These are the quantum counterparts of the Goldstone modes. Schematically, this is written as

$$\ket{\text{Bogoliubov mode}} = \ket{\text{Goldstone boson}} ~.$$

The Bogoliubov mode has some $ 1/r_s$ and frequency $\Delta \omega \sim \frac{1}{r_2} = \frac{1}{N}$. $\Delta \omega $ is the first correction to the energy degeneracy at $\lambda =1$. Classically, we have seen that the charges for superrotations are zero. In the quantum picture, we write

\begin{equation}
\hat{Q}= \in \partial_v C_{zz}^H \delta_{ \chi_g} g_{\mu \nu} = \int \partial_v \delta_{ \chi_g} g_{\mu \nu}(x)~.
\end{equation}

We then have

\begin{align}
  \text{Classically:     }\delta_{ \chi_g}g_{\mu \nu} &= \sum_ {m, \ell}  b^{\ell m}_{\mu \nu}(r, \theta , \varphi) Y^{\ell m}(\theta , \varphi) \notag \\
\text{Quantum mechanically: } \delta_{ \chi_g} g_{\mu \nu} &= \sum_ {m, \ell} \frac{1}{\sqrt{\omega_{\ell,m}}} \left[ \hat{b}^{\ell m}_{\mu \nu}(r, \theta , \varphi) Y^{\ell m}(\theta , \varphi) e^{-i v \omega_{m \ell}} +  \text{h.c.}  \right]     ~. 
\end{align}

The expression for the charge is 

\begin{equation}
\hat{Q} = \int \partial_v \delta_{ \chi} \hat{g}_{\mu \nu} (v, \theta, \varphi) = -i \sqrt{\omega_{\ell,m}} \left( b_{m \ell} e^{-i \omega_{\ell,m} v} - b_{m \ell}^{\dagger} e^{i \omega_{\ell,m} v} \right)  ~.
\end{equation}

We see that for $\omega_{\ell m } \to 0$, $\frac{1}{N} \to 0$, hence $\hat{Q}$ goes to zero as $\frac{1}{\sqrt{N}}$. We use a coherent state ansatz, which satisfies

$$ \hat{Q} \ket{N} = \ket{N}' ~. $$

This state $\ket{N}$ is given by

$$ \ket{N} = \exp \left( \sum_{m, \ell} n_{\ell m } \left( b_{\ell m} - b_{\ell m}^{\dagger} \right) \right) \ket{0} ~.$$

Here, $n_{\ell m } = \langle b^\dagger_{\ell m} b_{\ell m} \rangle$ i.e. the occupation number of the individual Bogoliubov modes. The VEV of the charge is then

\begin{equation}
\obket{N}{\hat{Q}_{\ell m} }{N} = \sqrt{\omega_{\ell m }n_{\ell m}} \approx \sqrt{\omega_{\ell m }} ~~, ~ \text{for } n_{\ell m}= 0,1  ~.
\end{equation}

The approximation made in the above expression gives

$$
\langle \hat{Q}_{\ell m} \rangle = \frac{1}{\sqrt{N}} \xrightarrow{N \to \infty} ~. 0
$$

We will take $\ell_{\max} \sim r_s = \sqrt{N}$, since for higher angular momenta we cannot ignore the back-reaction. For $\ell \leq m \leq \ell$, we find that

$$ \# \text{modes} \sim \ell_{max}^2 \sim r_s^2 \sim N = \# \text{charges } =  Q_{\ell m} ~.$$

We then assume that each mode corresponds to some qubit with values $n=0,1$. Then, the number of states is $d = 2^N$, so that $S = \log d = N$. We can summarize the picture in the following table\\

\begin{center}
\begin{tabular}{c|c}
	\textbf{Quantum picture}  & \textbf{Geometric picture}\\
	 \hline
	$N=$finite &  $N=\infty$\\
	Quantum criticality & Emergence of the horizon \\
	Bogoliubov modes & A-modes $\delta g_{\mu \nu}$ \\
	Coupling $\alpha \sim \frac{1}{N}$ & $\alpha = 0$ \\
	Energy gap $\omega \sim \frac{1}{N}$ & Massless modes $\omega = 0$ 
	\end{tabular}
\end{center}

\clearpage

\bibliographystyle{unsrt}

\bibliography{lib}


 \end{document}